\documentclass[11pt,a4paper,preprintnumbers]{article}
\usepackage{jheppub}
\usepackage{amsmath,amsfonts,mathrsfs,graphicx,bbm}
\usepackage{fixltx2e}
\usepackage{cancel}

\usepackage{enumitem}

\usepackage{upgreek}
\usepackage{t4phonet}
\usepackage{pifont}

\usepackage[font=small]{caption}

\usepackage{lmodern}

\usepackage{graphicx}
\usepackage{booktabs}
\usepackage{dsfont}

\usepackage{mathrsfs}
\usepackage{amsmath,amssymb}
\usepackage{mathtools}
\usepackage{bbm}
\usepackage{slashed}
\usepackage{braket}

\usepackage{hyperref}

\usepackage{xspace}

\usepackage{color}

\definecolor{grey}{rgb}{0.4,0.4,0.5}
\definecolor{darkgreen}{rgb}{0,0.5,0}
\definecolor{darkred}{rgb}{0.6,0.0,0}
\definecolor{lightbrown}{rgb}{1,0.9,0.8}
\definecolor{brown}{rgb}{0.6,0.3,0.3}
\definecolor{darkblue}{rgb}{0,0,0.8}
\definecolor{darkmagenta}{rgb}{0.5,0,0.5}

\newcommand{\bp}{\bar{p}}
\newcommand{\bq}{\bar{q}}

\numberwithin{equation}{section}

\makeatletter
 \let\old@startsection=\@startsection
 \let\oldl@section=\l@section
 \renewcommand{\@startsection}[6]{\old@startsection{#1}{#2}{#3}{#4}{#5}{#6\mathversion{bold}}}
 \renewcommand{\l@section}[2]{\oldl@section{\mathversion{bold}#1}{#2}}
\makeatother

\renewcommand{\leq}{\leqslant}
\renewcommand{\geq}{\geqslant}

\def\XXint#1#2#3{{\setbox0=\hbox{$#1{#2#3}{\int}$}
    \vcenter{\hbox{$#2#3$}}\kern-.5\wd0}}

\def\be{\begin{equation}}
\def\ee{\end{equation}}

\newcommand{\bea}{\begin{eqnarray}}
\newcommand{\eea}{\end{eqnarray}}
\newcommand{\bei}{\begin{itemize}}
\newcommand{\eei}{\end{itemize}}
\newcommand{\bee}{\begin{enumerate}}
\newcommand{\eee}{\end{enumerate}}
\newcommand{\bal}{\begin{equation}\begin{aligned}}
\newcommand{\eal}{\end{aligned}\end{equation}}

\newcommand{\ads}{${\rm  AdS}_5\times {\rm S}^5\ $}

\def\la{\label}

\def\d {\delta}

\def\p{\phi}

\def\eps{\epsilon}

\def\G{\Gamma}

\def\pa {\partial}

\newcommand{\sfrac}[2]{{\textstyle\frac{#1}{#2}}}

\preprint{{\flushright{ZMP-HH/16-19\\DCPT-16/29\\}}}
\title{Resurgence of the dressing phase for ${\rm AdS}_5\times {\rm S}^5$ }

\author[a,1]{Gleb Arutyunov,}
\note{Correspondent fellow at Steklov
Mathematical Institute, Moscow.}
\author[b]{Daniele Dorigoni}
\author[a]{and Sergei Savin}
\affiliation[a]{ Institut f\"ur Theoretische Physik, Universit\"at Hamburg, Luruper Chaussee 149, 22761 Hamburg, Germany\\
Zentrum f\"ur Mathematische Physik, Universit\"at Hamburg, Bundesstrasse 55, 20146 Hamburg, Germany
}
\affiliation[b]{Centre for Particle Theory \& Department of Mathematical Sciences,
Durham University,\\ Lower Mountjoy 
Stockton Road, Durham DH1 3LE, UK}
\emailAdd{gleb.arutyunov@desy.de, daniele.dorigoni@gmail.com, sergei.savin@desy.de}

\abstract{We discuss the resummation of the strong coupling asymptotic expansion of the dressing phase of the ${\rm AdS}_5\times {\rm S}^5$ superstring. The dressing phase proposed by Beisert, Eden and Staudacher can be recovered from a modified Borel-Ecalle resummation of this asymptotic expansion only by completing it with new, non-perturbative and exponentially suppressed terms that can be organized into different sectors labelled by an instanton-like number. We compute the contribution to the dressing phase coming from the sum over all the instanton sectors and show that it satisfies the homogeneous crossing symmetry equation. 
We comment on the semiclassical origin of the non-perturbative terms from the world-sheet theory point of view even though their precise explanation remains still quite mysterious. }

\begin{document}

\maketitle
\flushbottom

\newpage
\section{Introduction}
In recent years a lot of progress has been achieved in the spectral problem of the gauge-string correspondence by using 
ideas and methods from the theory of integrable models \cite{Arutyunov:2009ga,Beisert:2010jr}. For strings on \ads the corresponding light-cone sigma model is quantum 
integrable which allows one to obtain its spectrum by means of Thermodynamic Bethe Ansatz (TBA) \cite{Arutyunov:2009zu}-\cite{Gromov:2009bc} or the modern incarnation of
the latter known as Quantum Spectral Curve \cite{Gromov:2013pga}.  

 We recall that the construction of the TBA is essentially based on the asymptotic S-matrix for scattering of string world-sheet excitations in the 
 uniform light-cone gauge. This S-matrix is determined by symmetries of the \ads light-cone sigma model up to an overall scalar factor called the dressing factor \cite{Arutyunov:2004vx}.
 Thus, determination of the latter quantity and investigation of its properties constitutes an important part of the spectral problem to which many studies has been devoted 
 in the recent past. In this paper we will undertake an effort to complete the existing considerations and clarify some issues related to a perturbative expansion of the dressing factor (phase)
 at strong coupling.

Before we pass to the discussion of our approach, we briefly recall what is known about the dressing factor $\sigma=e^{i\theta}$, where $\theta$ is the dressing phase. 
The functional form of $\sigma$ as a perturbative power series in the inverse string tension $g$ with coefficients written in terms of local conserved charges was conjectured in \cite{Arutyunov:2004vx} by discretising equations that encode finite-gap solutions of the 
classical string sigma model. Since $g$ is related to the 't Hooft coupling $\lambda$ as $g=\sqrt{\lambda}/2\pi$, from the point of view of gauge theory this power series represents a strong coupling expansion of $\sigma$. 
Further, the asymptotic S-matrix appears to be compatible with crossing symmetry which implies a non-trivial functional equation for the dressing factor -- the crossing equation \cite{Janik:2006dc}.
The found leading (AFS) \cite{Arutyunov:2004vx} and sub-leading (HL)  \cite{Hernandez:2006tk} terms in the strong coupling expansion of $\sigma$ were shown to satisfy the crossing equation \cite{Arutyunov:2006iu}
and an all-order asymptotic solution of the latter was obtained in \cite{Beisert:2006ib}. The weak coupling expansion for $\sigma$ was conjectured in \cite{Beisert:2006ez} (BES) as a sort of analytic continuation 
of the corresponding strong coupling expansion. In opposite to the latter, the weak coupling expansion of $\theta(x_1,x_2)$ has a finite radius of convergence and defines a function which admits an integral representation (DHM) well defined in a certain kinematical region of particle rapidities $x_1,x_2$ and for finite values of $g$ \cite{Dorey:2007xn}. Analytic continuation of the dressing phase to other kinematical regions compatible with crossing symmetry has been constructed in  \cite{Arutyunov:2009kf},
which in fact provides verification of the crossing equation for finite $g$. Finally, under some  assumptions on the analytic structure the minimal solution of the crossing equation has been found and cast precisely in the DHM  form \cite{Volin:2009uv,Vieira:2010kb}.
Let us also note that the dressing phase admits a representation in terms of a single integral (rather than double integral representation of DHM) which proved to be useful for numerical construction of solutions of the TBA equations \cite{Frolov:2010wt}.

It was soon realised \cite{Gomez:2006mf} that a non-perturbative resummation prescription must be implemented if we want to extract the weak coupling expansion of $\sigma$ from the strong coupling data. After a particular non-perturbative prescription to resum the leading order dressing phase contribution at strong coupling, the authors of \cite{Gomez:2006mf} were able to expand it in a suitable weak coupling regime. In this way they found a connection between the strong and weak coupling coefficients of the dressing phase, reminiscent of the analytic continuation conjectured in \cite{Beisert:2006ez}.
Similarly in \cite{Fioravanti:2007un} the authors expanded the dressing phase of \cite{Beisert:2006ez} reproducing precisely the asymptotic strong coupling coefficients. To obtain the strong coupling regime, these authors did not use the contour integral type of argument utilized in \cite{Beisert:2006ez}, but rather implemented a suitable ad hoc regularization procedure to expand the integrand of the Beisert-Eden-Staudacher dressing phase and obtained back the formal asymptotic expansion studied in \cite{Beisert:2006ib}.
Although the results of both \cite{Gomez:2006mf} and \cite{Fioravanti:2007un} suggest that the dressing phase proposed in \cite{Beisert:2006ez} (that we will call BES in what follows) has the correct properties to interpolate between the weak and strong coupling regime, to our mind no rigorous and complete treatment of the resummation procedure of the full strong coupling asymptotic expansion of $\sigma$ exists so far.

In this paper we would like to present the discussion of the strong coupling expansion of the dressing phase, and its resummation, in the modern context of resurgence \cite{Ecalle:1981}.
We show how to resum the strong coupling expansion by using a modified version of the well-known Borel transform method. Our main result is that, in order to reproduce the dressing phase of \cite{Beisert:2006ez}, we have to modify the perturbative strong coupling expansion studied in \cite{Beisert:2006ib} to what is called a transseries expansion by adding new, non-perturbative terms of the form $e^{-4\pi g\,n}$ with $n\geq1$ integer. 
These exponentially suppressed terms can be associated with ambiguities related to the resummation procedure of the purely perturbative expansion.
Having modified the purely perturbative coefficients we need to check once again that the new strong coupling dressing phase satisfies the crossing symmetry equation and indeed we show that these new non-perturbative contributions to $\sigma$ solve the homogenous crossing symmetry equation.

According to our findings the leading non-perturbative correction to the dressing phase comes with an exponentially suppressed factor $e^{-4\pi g}$ multiplied by an infinite perturbative expansion starting from three-loops, {\textit{i.e.}} with the factor $g^{-2}$.
From the purely perturbative point of view, the three-loop coefficient is also distinguished because only starting from three-loops the odd coefficients produce contributions to the dressing phase which satisfy the homogeneous crossing equation, while this is not the case for the one-loop perturbative coefficients or the even ones.
For the non-perturbative contributions it might be that there is a protection mechanism based on vanishing of the zero mode contributions, forcing perturbation theory on top of these new non-perturbative saddles to start from three-loops, in the same spirit to what has been observed for the case of the instanton corrections for the anomalous dimension of the Konishi operator \cite{Arutyunov:2000im,Alday:2016tll}.

The origin of these new, non-perturbative effects in the dressing phase is quite mysterious.
This story is analogous to the non-perturbative effects \cite{Basso:2007wd,Basso:2009gh} emergent in the strong coupling expansion, $g\to \infty$, of the cusp anomalous dimension of $\mathcal{N}=4$. Similarly to the dressing phase, the cusp anomaly has a transseries expansion at strong coupling \cite{Aniceto:2015rua,Dorigoni:2015dha} and, perhaps surprisingly, these exponentially suppressed terms have a semiclassical origin that can be understood from the string theory side.
In the dual, weakly coupled description the calculation of the cusp anomaly translates into the computation of the spectrum of folded spinning strings on \ads, the so-called GKP-strings \cite{Gubser:2002tv}. At low energies we can describe the world-sheet theory in terms of an effective sigma model, containing an $O(6)$ factor \cite{Alday:2007mf}, with a non-trivial strongly coupled IR dynamics.
In a suitable regime \cite{Dunne:2015ywa}, this $2$-d quantum field theory contains non-perturbative objects, {\textit{i.e.}} finite action solutions to the classical equations of motion, that, in the semiclassical approximation, give rise to exponentially suppressed contributions to the energy levels hence explaining the presence of non-perturbative terms in the cusp anomaly expansion at strong coupling, on the gauge theory side. How precisely these non-perturbative objects translate into the full string theory remains however to be understood. 

In the case of the dressing phase the weakly coupled dual side can be most conveniently studied via a different stringy solution: the BMN string \cite{Berenstein:2002jq}.
The S-matrix computed from the sigma model perturbation theory has been shown (see {\textit{e.g.}} \cite{Klose:2006zd,Bianchi:2013nra}) to reproduce the well known first few orders of the dressing phase expansion.
For this reason and from the presence of non-perturbative terms in the dressing phase transseries expansion, we predict the existence of new non-perturbative objects in the world-sheet sigma model theory\footnote{We thank Lorenzo Bianchi for useful discussions on this problem.} (or possibly a suitable complexification thereof) that hopefully one can construct more easily in one of the Pohlmeyer reduced versions of the world-sheet theory \cite{Hoare:2014pna}.
Note also that the leading non-perturbative effect presents in the cusp anomalous dimension \cite{Basso:2007wd,Alday:2007mf} takes the form $e^{-\pi g}$, or $e^{-\sqrt{\lambda}/2}$ in terms of the 't Hooft coupling, while the leading correction we find in the dressing phase is of the form $e^{-4\pi g}$, or $e^{-2\sqrt{\lambda}}$. This stresses once more that these new non-perturbative corrections we find in the dressing phase should have a different semi-classical origin compared to the cusp anomaly ones.

From a mathematical point of view we perfectly understand why these non-perturbative terms must be incorporated in order to represent a very particular analytic function, {\textit{i.e.}} this BES dressing phase, in terms of a transseries expansion, but from a physical point of view it is a very important question to understand the semi-classical origin of these exponentially suppressed contributions in terms of non-perturbative strings configurations.

Finally, we mention the universality of the methods developed in the present paper. Similar results about non-perturbative sectors of the dressing phase might be expected also for the 
case of $q$-deformed theories \cite{Delduc:2013qra}-\cite{vanTongeren:2013gva} and for lower dimensional examples of AdS/CFT, like for instance for ${\rm AdS}_{\rm 3}/{\rm CFT}_2$, see {\it e.g.} \cite{Borsato:2013hoa}-\cite{Borsato:2016xns}. 
Furthermore similar type of methods can be applied also to different observables within the context of AdS/CFT correspondence, for example it was realized in \cite{Heller:2013fn} that the hydrodynamic gradient series for the strongly coupled $\mathcal{N} = 4$ super Yang-Mills plasma is only an asymptotic expansion leading to the works \cite{Heller:2015dha}-\cite{Aniceto:2015mto} dealing with resurgence and resummation issues in the fluid context of ${\rm AdS}_{\rm 5}/{\rm CFT}_4$.

The paper is organized as follows. In Section~\ref{sec:Dress} we review some known facts about the dressing phase and its strong coupling expansion while in Section~\ref{sec:ModBor} we introduce a modified version of the Borel transform to resum the perturbative coefficients. We prove in Section~\ref{sec:Resum} that the Borel-Ecalle resummation of our proposed transseries expansion matches perfectly the BES dressing phase. The exact form of the non-perturbative terms is related to the ambiguity in the resummation of the perturbative expansion, which is computed explicitly in Section~\ref{sec:disc} and then expanded at strong coupling in Section~\ref{sec:Strong}.
In Section~\ref{sec:DispRel} we use a standard dispersion-like argument to show how the perturbative coefficients of the non-perturbative sectors can be reconstructed from the large order behaviour of the purely perturbative ones and finally, in Section~\ref{sec:NPdress}, we use precisely these coefficients to obtain new, non-perturbative contributions to the dressing phase, solutions to the homogeneous crossing symmetry equation.
Because of the involved algebraic manipulations, many of the more technical results obtained in this paper are relegated to the appendices.

\section{The dressing phase}
\label{sec:Dress}
Here we collect some known facts about the dressing phase which we need to our further discussion. 
The S-matrix is determined up to an overall scalar function - the dressing factor $\sigma(x_1^{\pm},x_2^{\pm})$, which satisfies a non-trivial functional equation - the crossing equation. It turns out to be convenient to write the dressing factor in the exponential form $\sigma(x_1, x_2) = e^{i\theta(x_1,x_2)}$. Here the dressing phase
\bea
\theta_{12}\equiv   \theta(x_1^+, x_1^-, x_2^+, x_2^-) = \sum\limits_{r=2}^{\infty} \sum_{\substack{s>r \\ r+s={\rm odd} }} c_{r,s}(g) [q_r(x_1^{\pm})q_s(x_2^{\pm}) - q_s(x_1^{\pm})q_r(x_2^{\pm})]\label{eq:dressing}
\eea
with 
\bea
  q_r(x_k^-,x_k^+) = \frac{i}{r-1} \left[\Big(\frac{1}{x_k^+}\Big)^{r-1} - \Big(\frac{1}{x_k^-}\Big)^{r-1}\right], \label{eq:magcharge}
\eea
where $x^{\pm}$ are subject to the relation 
$$
x^++\frac{1}{x^+}-x^--\frac{1}{x^-}=\frac{2i}{g}\, .
$$
Here $g$ is related to the 't Hooft coupling $\lambda$ as $g=\sqrt{\lambda}/2\pi$.

The phase $\theta_{12}$ can be written as 
\begin{align}
\theta_{12}=&\notag+\chi(x_1^+,x_2^+)-\chi(x_1^+,x_2^-)-\chi(x_1^-,x_2^+)
+\chi(x_1^-,x_2^-)\\
&\label{eq:thetachi}-\chi(x_2^+,x_1^+)+\chi(x_2^-,x_1^+)+\chi(x_2^+,x_1^-)-\chi(x_2^-,x_1^-)\,,
\end{align}
where the function $\chi$ obtained from (\ref{eq:dressing}-\ref{eq:magcharge}) is
\bea
\chi(x_1,x_2) = \sum_{r=2}^\infty\sum_{s=r+1}^\infty \frac{-c_{r,s}(g)}{(r-1)(s-1)}\frac{1}{x_1^{r-1}x_2^{s-1}}\,.\label{eq:Chi}
\eea

The coefficients $c_{r,s}(g)$ admit an asymptotic large $g$ expansion
\bea\label{c_rs}
  c_{r,s}(g) = g^2 \sum\limits_{n=0}^{\infty} c_{r,s}^{(n)} g^{-n-1} ,~~~g\gg 1,
\eea
where the numerical coefficients are given by 
\bea\label{c_rsn01}
  c_{r,s}^{(0)} =\frac{1}{2} \delta_{r+1,s}, ~~~~c_{r,s}^{(1)} = - \frac{1-(-1)^{r+s}}{\pi}\frac{(r-1)(s-1)}{(s+r-2)(s-r)} ,
\eea
and for $n\geq2$ by
\bea\label{c_rsn}
  c_{r,s}^{(n)} = \frac{ (-1)^n \zeta(n)}{2\pi^n\Gamma[n-1] } (r-1)(s-1) \frac{ \Gamma[\frac{1}{2}(s+r+n-3)]\Gamma[\frac{1}{2}(s-r+n-1)] }{ \Gamma[\frac{1}{2}(s+r-n+1)]\Gamma[\frac{1}{2}(s-r-n+3)] } .
\eea
Note that for  $n=0,1$ this expression is formally 0/0 , but nevertheless \eqref{c_rsn01} can easily be recovered from \eqref{c_rsn}.
At any given order in the asymptotic $1/g$ expansion the double series defining $\chi$ is
convergent for $|x_{1,2}|>1$.

The series \eqref{c_rs} is divergent and of Gevrey-1 type\footnote{A series $\{c_n\}_{n\in\mathbb{N}}$ is of Gevrey type $1/m$ if the large orders asymptotic terms are bounded by $\vert c_n\vert <\alpha \,C (n!)^m$ for some constants $\alpha$ and $C$.} since the coefficients \eqref{c_rsn} grow as $c_{r,s}^{(n)} \sim n!$, 
for this reason we can thus perform a Borel resummation of series  \eqref{c_rs}. 

The crossing equation satisfied by the dressing phase has the form
\bea
i\theta(x_j,x_k)+i \theta(1/x_j,x_k) = 2 \log h(x_j,x_k)\,,\label{eq:crossing}
\eea
where the function $h$ is
\bea
h(x_j,x_k) =\frac{x_k^-}{x_k^+} \frac{(1-\frac{1}{x_j^- x_k^-})(x_j^- - x_k^+)}{(1-\frac{1}{x_j^+ x_k^-})(x_j^+ - x_k^+)}\,.
\eea
Here we have chosen to uniformize $x^{\pm}$ in terms of a single variable $x$ via \cite{Arutyunov:2006iu}
\bea
x^{\pm}(x) = x\sqrt{1-\frac{1}{g^2(x-\frac{1}{x})^2}}\pm \frac{i}{g} \frac{x}{x-\frac{1}{x}}\, .
\eea

\section{Modified Borel transform}
\label{sec:ModBor}
We start with recalling that the standard Borel transform of a divergent series 
\bea
  \sum\limits_{n=0}^{\infty} c_n z^{-n-1} 
\eea
with coefficients $c_n$ growing as $n!$ is defined as 
\bea\label{eq:stdBorel}
  {\cal B}_0:~~\sum\limits_{n=0}^{\infty} c_n z^{-n-1} ~\mapsto~ \sum\limits_{n=0}^{\infty} \frac{c_n}{n!}\xi^n .
\eea
The standard Borel image is now convergent to some function $\sum\limits_{n=0}^{\infty} \frac{c_n}{n!}\xi^n = \hat{\varphi}(\xi)$ and the initial series can be resummed through the ``inverse'' of the standard Borel transform which is the Laplace transform
\bea
 \varphi(z) =\mathcal{L}[\hat{\phi}](z)= \int_0^\infty  {\rm d}\xi\,e^{-z\xi} \, \hat{\varphi}(\xi)\sim  \sum\limits_{n=0}^{\infty} c_n z^{-n-1}\, ,\label{eq:Laplace}
\eea
where $\sim$ means asymptotic in the standard sense.
Typically,  $\hat{\varphi}(\xi)$ has singularities which lead to ambiguities in the resummation procedure associated with the choice of integration contour in the Laplace transform as we will discuss in full details later on.

Here, to remove an additional Riemann-zeta factor, we consider a modified (similarly to \cite{Russo:2012kj}) Borel transform\footnote{Note that the summation extends from $n=2$ because $\zeta(1)=\infty$.} 
\bea
  {\cal B}:~~ \sum\limits_{n=2}^{\infty} c_n z^{-n} ~\mapsto~ \sum\limits_{n=2}^{\infty} \frac{c_n}{\zeta(n)\Gamma(n+1)}\xi^n = \hat{\varphi}(\xi) \,,
\eea
which on a monomial acts as 
\bea
{\cal B}[z^{-n}] = \frac{\xi^n}{\Gamma(n+1) \zeta(n)}\,, \qquad \mbox{for}~~\,n\geq 2\,,\label{eq:ModBorel}
\eea
where $\zeta(n)$ denotes the Riemann zeta function.

This transform can be easily inverted by noticing that the momenta of the measure 
\bea
{\rm d}\mu  = \frac{1}{4 \sinh^2( \xi/2)} {\rm d}\xi
\eea
are precisely 
\bea
\langle \xi^n\rangle = \int_0^\infty {\rm d}\mu \, \xi^n=\int_0^\infty {\rm d}\xi\, \frac{\xi^n}{4\sinh^2(\xi/2)} =\Gamma(n+1)\zeta(n)  \qquad \mbox{for}\,n\geq 2\, .
\eea
As seen before, the ``inverse'' can be given via 
\bea \label{inverse1}
  \varphi(z) = z \int_0^\infty \frac{d\xi}{4\sinh^2( \xi\,z/2)}\hat{\varphi}(\xi) \sim  \sum\limits_{n=0}^{\infty} c_n z^{-n-1}\, .
\eea
According to (\ref{c_rs})  the variable $z$  in (\ref{inverse1}) should be identified with $g$. 

Applying this technique to \eqref{c_rs} we can sum up the modified Borel image
\bea
\sum_{n=2}^{\infty} \frac{ c_{r,s}^{(n)} }{ \G(n+1)\zeta(n) } \xi^n\, = \hat{\varphi}_{r,s}(\xi)  ,\eea
where
{\small
\bea
\hat{\varphi}_{r,s}(\xi)&=&\frac{1}{48\pi^3} (r-1)(s-1)\xi^2 \times \\
&\times& \Bigg[ 12\pi ~{}_{4}F_{3}\Big( \left\{ \sfrac{3}{2}-\sfrac{r}{2}-\sfrac{s}{2},\sfrac{1}{2}+\sfrac{r}{2}-\sfrac{s}{2},\sfrac{1}{2}-\sfrac{r}{2}+\sfrac{s}{2},-\sfrac{1}{2}+\sfrac{r}{2}+\sfrac{s}{2} \right\}, \left\{ \sfrac{1}{2},\sfrac{3}{2},2 \right\},\left(\sfrac{\xi}{4\pi}\right)^2 \Big) +  \nonumber\\
 &+&(r-s)(r+s-2)\, \xi ~{}_{4}F_{3}\Big( 
 \left\{ 2-\sfrac{r}{2}-\sfrac{s}{2},1+\sfrac{r}{2}-\sfrac{s}{2},1-\sfrac{r}{2}+\sfrac{s}{2},\sfrac{r}{2}+\sfrac{s}{2} \right\}, \left\{ \sfrac{3}{2}, 2, \sfrac{5}{2} \right\},\left(\sfrac{\xi}{4\pi}\right)^2  
  \Big)  
  \Bigg] \nonumber\eea
  }
\noindent
 with  ${}_{p}F_{q}(\{a_1,\dots,a_p \},\{b_1,\dots,b_q \}, z)$ being the generalised hypergeometric function. 
\smallskip
 
Recalling that $s+r$ must be odd and $r \geq 2$, $s \geq r+1$, we can change to $s+r=2p+1$, $s-r=2q+1$, where the integers $p,q$ are $q\geq0$, $p\geq q+2$, and, with the definition $\xi \equiv 4\pi x$, the modified Borel transform 
\bea\hat{\varphi}_{r,s}(\xi)\equiv \hat{\phi}_{p,q}(x)\eea  takes the form
{\small
\bea\nonumber
   \hat{\phi}_{p,q}(x) : &=& \frac{4}{3} (p-q-1)(p+q)x^2 \times \\
   &\times&\Big[3~{}_{4}F_{3}\left( \left\{ 1-p, p, -q, 1+q \right\}, \left\{ \sfrac{1}{2}, \sfrac{3}{2}, 2 \right\}, x^2 \right) - \label{phihat}\\
  &&~~~~~~    (2p-1)(2q+1)\, x~{}_{4}F_{3}\left( \left\{ \sfrac{3}{2}-p, \tfrac{1}{2}+p, \tfrac{1}{2}-q, \tfrac{3}{2}+q \right\}, \left\{ \tfrac{3}{2}, 2, \tfrac{5}{2} \right\}, x^2 \right) \Big]. \nonumber
\eea
}
\vskip -0.5cm
\noindent
In terms of the variables $p$ and $q$ the perturbative coefficients $c_{r,s}^{(n)}$ acquire the form 
\bea\label{c_pq}
 c_{p,q}^{(n)}
=(-1)^n\zeta(n)\frac{(p+q)(p-q-1)}{2\pi^n \Gamma(n-1)}\frac{\G(\frac{n}{2}+p-1)\G(\frac{n}{2}+q)}{\G(-\frac{n}{2}+p+1)\G(-\frac{n}{2}+q+2)} \, .
 \eea
As discussed earlier in this section, we can naively resum the asymptotic power series with
coefficients (\ref{c_pq}) via
\bea\label{c_rs_int}
  c_{p,q}(g) = c_{p,q}^{(0)}\cdot g + c_{p,q}^{(1)} + \pi g^2 \int_{0}^{\infty} \frac{ {\rm d} x }{ \sinh^2(2\pi g x) } \hat{\phi}_{p,q}(x)\,.
\eea
To understand the region of analyticity of the function $c_{p,q}(g)$ in the complex coupling constant $g$-plane, we need first to understand the analytic properties of the modified Borel transform (\ref{phihat}) in the complex Borel $x$-plane.

To begin, we note that the first hypergeometric function in (\ref{phihat}) is a simple polynomial of degree $2 q$ in $x$. This contribution to the full modified Borel transform is an entire function of $x$ because is coming from the coefficients $c^{(n)}_{p,q}$ with $n$ \textit{even} which are only finitely many in number: from the explicit expression 
(\ref{c_pq}), we see that $c^{(2m)}_{p,q}=0$ for any $m\geq q+2$. 

The second hypergeometric function in (\ref{phihat}) , which we denote as 
\bea\label{eq:Omega}
\Omega(z): = ~{}_{4}F_{3}\left( \left\{ \sfrac{3}{2}-p, \sfrac{1}{2}+p, \sfrac{1}{2}-q, \sfrac{3}{2}+q \right\}, \left\{ \sfrac{3}{2}, 2, \sfrac{5}{2} \right\}, z \right)\, ,
\eea
where $z=x^2$, has a cut along the real interval $(1,+\infty)$. Therefore, 
the resummation formula (\ref{c_rs_int}) does not define an analytic function of $g$, unless we specify a contour of integration that dodges the cut. This introduces an ambiguity in the resummation procedure, related to the particular choice of integration contour, {\textit{i.e.}} that is above or below the real line. For the discontinuity of $\Omega(z)$  we found in Appendix~\ref{app:discon} the following formula
\bea
\label{disc_phin_hat}
&&\Omega(z+i\eps)-\Omega( z-i\eps)=-i\frac{3}{(2p-1)(2q+1)(p+q)!}\times \\
&&~~~~\times\frac{1}{\sqrt{z}} \frac{d^q}{dz^q} z^{q-1}\frac{d^{p-2}}{dz^{p-2}}\Bigg[(1-z)^{p+q}z^{p-\sfrac{1}{2}}~{}_{2}F_{1}(\sfrac{1}{2}+p,\sfrac{3}{2}+q,p+q+1,1-z)\Bigg]\, , ~~~|z|>1\, , \nonumber
\eea
so that combining this with (\ref{phihat}) we find
\bea\label{Disc_PhiHat}
&&{\rm Disc}\,\hat{\phi}_{p,q}=i\frac{4(p-q-1)}{(p+q-1)!}\times \\
&&~~~~z \frac{d^q}{dz^q} z^{q-1}\frac{d^{p-2}}{dz^{p-2}}\Bigg[(1-z)^{p+q}z^{p-\sfrac{1}{2}}~{}_{2}F_{1}(\sfrac{1}{2}+p,\sfrac{3}{2}+q,p+q+1,1-z)\Bigg]_{z=x^2}\, , ~~~|z|>1\, . \nonumber\eea
Note that this discontinuity along the cut $(1,+\infty)$ is purely imaginary and also that ${\rm Disc}\,\hat{\phi}_{p,q}(1)=0$, this will shortly be of importance.
\begin{figure} \begin{center}
\hspace {-2cm}
 \includegraphics[scale=0.25]{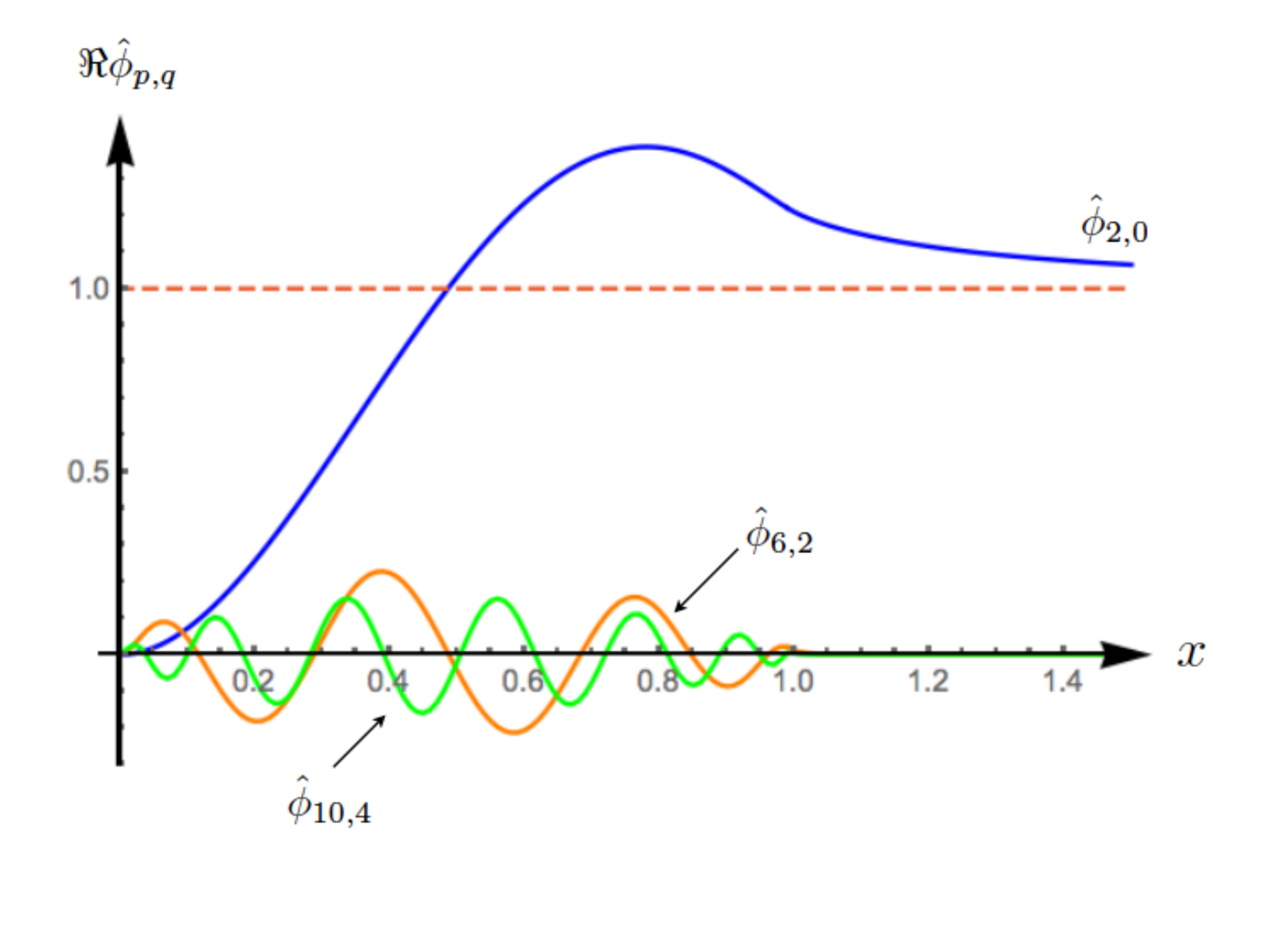}
 \caption{Plot of real part of the function $\hat{\phi}_{p,q}$ for a few values of $p$ and $q$.}\label{fig:hat_phi}
 \end{center}
 \vspace{-1cm}
 \end{figure}
 
One natural way to fix the ambiguity related to the choice of the integration contour is to demand that $c_{p,q}(g)$ must be real for real $g$. Further analysis reveals that  
$\Re\, \hat{\phi}_{p,q}(x)$ has neither pole nor cut on the real line and for generic $p$ and $q$ is a decreasing function as $x\to\infty$, see Figure~\ref{fig:hat_phi}. We thus can define the manifestly real coefficients by 
\bea\label{crs1}
  c_{p,q}(g) = c_{p,q}^{(0)}\cdot g + c_{p,q}^{(1)} + \pi g^2 \int_{0}^{\infty} \frac{ {\rm d} x }{ \sinh^2(2\pi g x) }\Re\,  \hat{\phi}_{p,q}(x)\, ,
\eea
whose strong coupling expansion $g\gg1$ coincide with the original asymptotic formal power series (\ref{c_rs}).
 This prescription for the resummation procedure seems somehow ad hoc but in the next section we will show that it 
corresponds in fact to the {\it median} Borel resummation.

To straightforwardly integrate $\Re\, \hat{\phi}_{p,q}$ is rather difficult because it contains a separate polynomial part. Also the first two terms in (\ref{crs1}) come apart which suggests that they originate from contour integrals around isolated points,
as was explained in \cite{Beisert:2006ez}. Therefore, to proceed, we show that  $\Re\, \hat{\phi}_{p,q}$ admits another but alternative representation through the function 
\bea
\begin{aligned}
\label{BES_c}
\hat{\Phi}_{p,q}(x) = \d_{q,0} + (-1)^{p+q} 2^{5-4p} (p-q-1)(p+q) \frac{\G(2p-2)}{\G(p-q)\G(p+q+1)} \times\\
    \times\, x^{2-2p}\, \cdot{}_{4}F_{3}\left( \{p-1, p-\sfrac{1}{2}, p, p+\sfrac{1}{2}\}; \{2p, p-q, p+q+1\}; x^{-2} \right).
\end{aligned}
\eea
Namely,  both functions $\hat{\phi}_{p,q}$ and $\hat{\Phi}_{p,q}$ share the same real part
\bea\label{main_coin}
\Re\,  \hat{\phi}_{p,q}(x)=\Re \, \hat{\Phi}_{p,q}(x), ~~~~~x>0\, ,
\eea
a statement which is analytically proven in Appendix~\ref{app:phi_to_Phi} in two different ways\footnote{We warn the reader that to verify the coincidence of the real parts of the above functions numerically, for instance, by using {\it Mathematica},
one needs to apply first to the function $\hat{\Phi}$ the command ``FunctionExpand" which renders the answer in terms of complete elliptic integrals of the first and second kind. After that a numerical comparison can be straightforwardly performed.
}. 
At this point it is gratifying to see that (\ref{BES_c}) is essentially the same formula as equation (3.25) in  \cite{Beisert:2006ez}, which has been proposed there to describe a sort of analytic continuation of  the coefficients $c_{rs}$ from strong to weak coupling.
Note that, contrary to $\hat{\phi}_{p,q}$, the new function $\hat{\Phi}_{p,q}$ is an even function of $x$ and this property will be crucial in the next section to extend the integration over the whole real line to implement a Cauchy-like argument . 

\section{Non-perturbative resummation of the coefficients $c_{r,s}(g)$}
\label{sec:Resum}
In this section we prove that the manifestly real resummation (\ref{crs1}) proposed in the previous section does indeed coincide with the coefficients for the BES dressing phase introduced in \cite{Beisert:2006ez}. Furthermore we show that the proposed real resummation (\ref{crs1}) can be understood as the Borel-Ecalle resummation of a particular transseries expansion, generalization of the formal power series (\ref{c_rs}) that we started with.

\subsection{From the Borel sum to the BES dressing phase}
According to the discussion in the previous section, the coefficients $c_{p,q}$ can be represented as 
\bea
\label{crs2}
  c_{p,q}(g) = c_{p,q}^{(0)}\cdot g + c_{p,q}^{(1)} + \sfrac{1}{2}\pi g^2 \int_{-\infty}^{\infty} \frac{ {\rm d} x }{ \sinh^2(2\pi g x) }\Re\,  \hat{\Phi}_{p,q}(x)\, ,
\eea
where the integration was extended to the whole real line since $\Re\,  \hat{\Phi}_{p,q}(x)$ is an even function of $x$.
The rest of the computation follows the same steps as in \cite{Beisert:2006ez} but now for arbitrary values of $r$ and $s$ and, therefore, we outline it here for completeness.

\begin{figure}[t]
\begin{center}
 \includegraphics[scale=0.5]{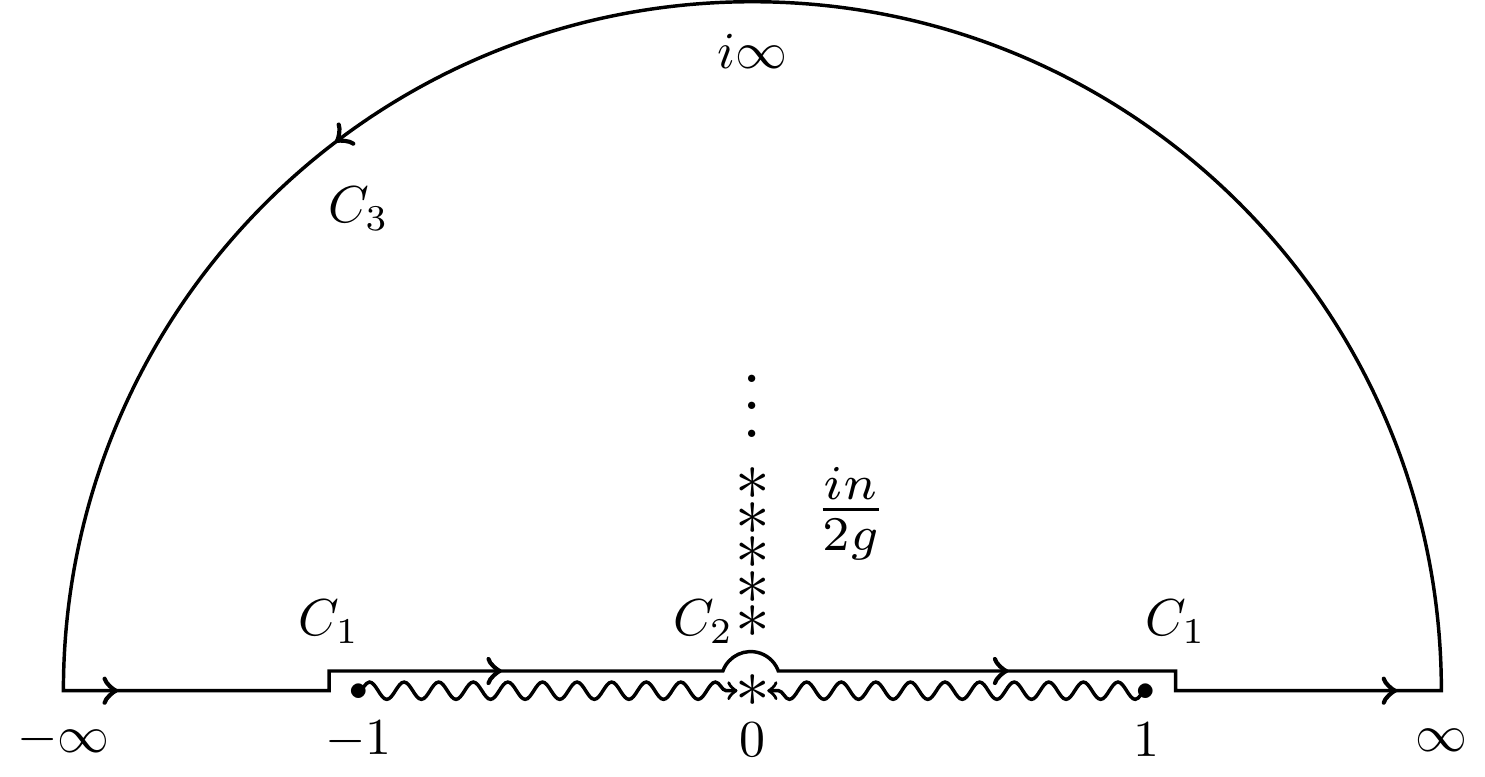}
\end{center}
\vspace{-0.5cm}\caption{Integration contour for the coefficients $c_{r,s}$.}\label{fig:contour}
\end{figure}

The starting point is to pass from integration of $ \hat{\Phi}_{p,q}$ over the real line to integration 
of $ \hat{\Phi}_{p,q}$ along the contour depicted on Figure~\ref{fig:contour}. The function $ \hat{\Phi}_{p,q}$ has a cut on the interval $(-1,1)$ and the integration contour $C_1$
runs just above this cut. Since the kernel $f(z)=\frac{\hat{\Phi}_{p,q}(z)}{ \sinh^2(2\pi g z) }$ is symmetric with respect to $z\to -z$, the contribution 
from two points symmetric around zero amounts to $f(-\bar{z})+f(z)=f(\bar{z})+f(z)=\overline{f(z)}+f(z)=2 \Re f(z)$, because $f(z)$ is real analytic.  
Thus, integration of $f(z)$ above the cut is equivalent to the integration of $\Re{f(z)}$ over the interval $(-1,1)$. One has however to take into account that 
$f(z)$ has a residue at infinity and at $z=0$ which lead to additional contributions. In particular, 
the two isolated terms entering (\ref{crs2}) can be treated (similarly to  \cite{Beisert:2006ez}) as the following contour integrals:
\begin{enumerate}[label=\it{\arabic{*}})~]
\item For the integral around the contour $C_3$, where $|x|\to \infty$ with  $\eps< {\rm arg}\,  x  <\pi-\eps$, a non-trivial contribution occurs only due to the leading term in $\hat{\Phi}_{p,q}(x)=\delta_{q,0}+\ldots$, which is present for $q=0$  only. 
One gets 
{\small
\bea
\hspace{-0.3cm}\sfrac{1}{2}\pi g^2 \int_{C_3}  {\rm d} x\, \frac{\hat{\Phi}_{p,q}(x)}{ \sinh^2(2\pi g x) } =\delta_{q,0} \lim_{|x|\to\infty}\Big[-\sfrac{1}{4}\coth(2\pi g |x|e^{i\theta})\Big|_{\theta=0}^{\theta=\pi}\Big]=
\frac{1}{2}g\delta_{q,0}=c_{r,s}^{(0)}\, .
\eea
}
\item To compute the integral around the contour $C_2$, we have to expand $\hat{\Phi}_{p,q}(x)$ around zero and we find for the leading behaviour 
\bea
\hat{\Phi}_{p,q}(x)=-\frac{16i}{\pi}\frac{(p+q)(p-q-1)}{(2p-1)(2q+1)}x+\ldots \, ,
\eea
{\it i.e.} it is purely imaginary for real $x$.
Note that the case $q=0$ should be treated with care which results in the absence of the leading $\delta_{q,0}$ when $q=0$ in the small $x$ expansion for $\hat{\Phi}_{p,q}(x)$.
Hence, the contribution from the contour $C_2$ is 
\bea
\sfrac{1}{2}\pi g^2 \int_{C_2} {\rm d} x \,\frac{ \hat{\Phi}_{p,q}(x) }{ \sinh^2(2\pi g x) }   &=&\pi g^2 \frac{16i}{\pi}\frac{(p+q)(p-q-1)}{(2p-1)(2q+1)}\int_{C_2} {\rm d}x\,\frac{1}{4g^2 \pi^2\,x}=\nonumber \\
&=&-\frac{2}{\pi}\frac{(p+q)(p-q-1)}{(2p-1)(2q+1)}=c_{p,q}^{(1)}\, .
\eea

\end{enumerate}
Thus, the resummation formula can be written as the following contour integral
\bea
  c_{p,q}(g)=  \sfrac{1}{2}\pi g^2\int\limits_C {\rm d}x\, \frac{ \hat{\Phi}_{p,q}(x) }{ \sinh^2(2\pi g x) }.
\eea
with contour $C=C_1\cup C_2 \cup C_3$ from Figure~\ref{fig:contour}. This is evaluated by Cauchy theorem as 
\bea
   c_{p,q}(g)=i  (\pi g)^2  \sum\limits_{n=1}^{\infty} \text{Res}_{in/2g} \frac{ \hat{\Phi}_{p,q}(x) }{ \sinh^2(2\pi g x) } = \left. \frac{i}{4}\sum\limits_{n=1}^{\infty} \frac{{\rm d}}{{\rm d} x}\hat{\Phi}_{p,q}(x) \right|_{x=\sfrac{in}{2g}  }.
\eea

The derivative is given by
\begin{multline}
  \frac{{\rm d}}{{\rm d} x}\hat{\Phi}_{p,q}(x) = 2^3(-1)^{p+q+1} \frac{\G(2p-1)}{\G(p+q)\G(p-q-1)}\times \\
  \times(4x)^{1-2p}{}_{4}F_{3}\left( \{p-\sfrac{1}{2}, p, p, p+\sfrac{1}{2}\}; \{2p, p-q, p+q+1\}; x^{-2} \right).
\end{multline}
Thus, for the coefficients $c_{p,q}$ we find
\begin{multline}
  c_{p,q}(g) = (-1)^{q} \, 2^{2-2p}\frac{\G(2p-1)}{\G(p+q)\G(p-q-1)}     \times\\
  \times \sum\limits_{n=1}^{\infty} (n/g)^{1-2p}{}_{4}F_{3}\left( \{p-\sfrac{1}{2}, p, p, p+\sfrac{1}{2}\}; \{2p, p-q, p+q+1\}; \left(\frac{2i}{n/g}\right)^2 \right).
\end{multline}
Next, we apply the following identity
\begin{multline}
  z^{-\alpha} {}_{m+k}F_{n}\left( \left\{a_1,...,a_m, \frac{\alpha}{k}, \frac{\alpha+1}{k},..., \frac{\alpha+k-1}{k}\right\}; \{b_1,...,b_n\}; \left(\frac{k\lambda}{z}\right)^k \right) = \\
  = \frac{1}{\G(\alpha)}\int\limits_0^\infty {\rm d}t~ t^{\alpha-1} e^{-z t} {}_{m}F_{n}\left( \left\{a_1,...,a_m\right\}; \{b_1,...,b_n\}; \left( \lambda t \right)^k \right), 
\end{multline}
where we identify $z=n/g$, $\alpha=2p-1$, $k=2$ and $\lambda=i$. Hence, 
\begin{multline}
 c_{p,q}(g) = \frac{(-1)^{q}\,  2^{2-2p}}{\G(p+q)\G(p-q-1)}  \times\\
 \times\sum\limits_{n=1}^{\infty} \int\limits_0^\infty {\rm d}t~ t^{2p-2} e^{-n \,t/g} {}_{2}F_{3}\left( \{ p, p+\sfrac{1}{2} \}; \{ 2p, p-q, p+q+1 \}; -t^2 \right)
\end{multline}
or, going back to the $(r,s)-$representation 
\begin{multline}
 c_{r,s}(g) = 2 (-1)^{(s-r-1)/2} (s-1)(r-1) \frac{1}{\G(r)\G(s)}\times\\
 \times \sum\limits_{n=1}^{\infty}\int\limits_0^\infty {\rm d}t~ t^{r+s-3} e^{-2 n \,t/g} {}_{2}F_{3}\left( \left\{ \sfrac{s+r}{2}, \sfrac{s+r-1}{2} \right\}; \{ r, s, r+s-1 \}; -4t^2 \right)\, .
\end{multline}
Here one can recognise the well-known formula 
\bea
  {}_0F_1(r,-t^2)\,  {}_0F_1(s,-t^2) = {}_{2}F_{3}\left( \left\{ \sfrac{r+s}{2}, \sfrac{r+s-1}{2} \right\}; \{ r, s, r+s-1 \}; -4t^2 \right)
\eea
and use the representation of the Bessel function $J_{\nu}(t)$ via the hypergeometric one 
\bea
  J_{\nu}(2t) = \frac{t^\nu}{\G(\nu+1)}{}_0F_1(\nu+1,-t^2)
\eea
to get
\bea
  c_{r,s}(g) = 2 (-1)^{(s-r-1)/2} (s-1)(r-1) \sum\limits_{n=1}^{\infty} \int\limits_0^\infty \frac{{\rm d}t~ }{t} e^{-2n \,t/g} J_{r-1}(2t) J_{s-1}(2t).
\eea
Summing a geometric series up, one finally gets
\bea\label{eq:BES}
  c_{r,s}(g) = 2 (-1)^{(s-r-1)/2} (s-1)(r-1)\int\limits_0^\infty \frac{{\rm d}t~ }{t(e^t-1)} J_{r-1}(gt) J_{s-1}(gt).
\eea
This formula proves that the median Borel resummed formula for $c_{r,s}$ coincides with the coefficients of the BES dressing phase that first appeared in \cite{Beisert:2006ez}.

\subsection{Non-perturbative ambiguities and median resummation }
\label{subsec:disc}
Let us go back to the initial problem of going from the modified Borel transform to a suitable analytic continuation (\ref{c_rs_int}) of the original asymptotic formal power series (\ref{c_rs}).

To properly define the inverse transform (\ref{c_rs_int}), we need to integrate over a contour where the modified Borel transform $\hat{\phi}_{p,q}(x)$ is not singular. As shown above, the singular directions in the complex $x$ Borel plane, also called {\it{Stokes directions}}, for the case under considerations are $\mbox{Arg} \,x=0$ and $\mbox{Arg} \,x=\pi$.
 
 We can thus introduce the directional Borel resummation via
 \bea\label{c_lat}
  \mathcal{S}_\theta\left[ c_{p,q}\right](g) = c_{p,q}^{(0)}\cdot g + c_{p,q}^{(1)} + \pi g^2 \int_{0}^{e^{i\theta}\,\infty} \frac{ {\rm d} x }{ \sinh^2(2\pi g x) } \hat{\phi}_{p,q}(x)\,,
\eea
which defines an analytic function in the wedge of the complex coupling constant plane given\footnote{
The integral (\ref{c_lat}) is well-defined for $g\in\mathbb{C}$ such
that $\vert \sinh^2(2\pi g x)\vert>1$ for $\vert x \vert$ large
enough. This leads to two disjoint domains of analyticity separated by
the line $\Re (e^{i\theta}g)=0$. We decided to restrict our attention
to the upper domain but one could have directly worked with the union of the two disjoint domains analyzing the discontinuity across them.} by ${\rm{D}}_\theta=\{g\in\mathbb{C}\,\, \vert\,\, \Re (e^{i\theta}g)>0\}$, provided that $\theta$ is a regular direction, {\textit{i.e.}} $\theta\notin\{0,\pi\}$.

For every $\theta$ for which the above integral exists, if we expand for $g\gg1$ we obtain precisely the original asymptotic, formal power series expansion (\ref{c_rs}). Furthermore when $\{0,\pi\} \notin [\theta_1,\theta_2]$ we have that $ \mathcal{S}_{\theta_2}\left[ c_{p,q}\right]$ is the analytic continuation of $ \mathcal{S}_{\theta_1}\left[ c_{p,q}\right]$, {\textit{i.e.}} $ \mathcal{S}_{\theta_1}\left[ c_{p,q}\right](g)= \mathcal{S}_{\theta_2}\left[ c_{p,q}\right](g)$ for every $g\in {\rm{D}}_{\theta_1}\cap{\rm{D}}_{\theta_2}$. This allows us to analytically continue the function $ \mathcal{S}_{\theta_1}\left[ c_{p,q}\right](g)$ on a wider wedge of the complex $g$-plane, {\textit{i.e.}} on the union of the two domains ${\rm{D}}_{\theta_1}\cup{\rm{D}}_{\theta_2}$.

\begin{figure} \begin{center}
 \includegraphics[scale=0.8]{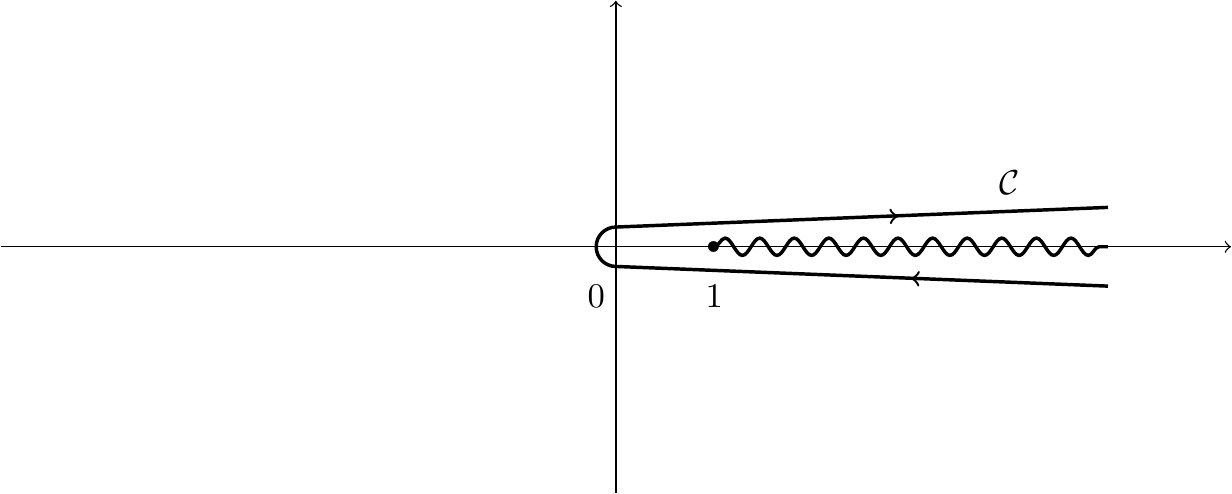}
 \caption{Integration contour in the Borel plane used to compute the difference between lateral resummations.}\label{fig:Hankel}
 \end{center}
 \end{figure}

Due to the presence of singularities in the Borel plane, if we keep on increasing ${\rm Arg} \,g$, or equivalently $\theta$, we will necessarily encounter branch cut singularities for the analytic continuation of the purely perturbative asymptotic power series (\ref{c_rs}). To understand the reason for that, we pick $\epsilon>0$ and small, and consider the two {\it lateral resummations} across the Stokes line $\theta=0$ given by $\mathcal{S}_{\pm\epsilon} \left[c_{p,q}\right](g)$, a similar story holds for the other Stokes line $\theta= \pi$.
These two analytic functions, although having the same asymptotic expansion (\ref{c_rs}), differ from one another on the intersection of their domains of analyticity. Their difference (related to the so called Stokes automorphism) can be written as an integration over the Hankel contour $\mathcal{C}$ shown in Figure~\ref{fig:Hankel}, originating from infinity below the positive real axis, circling the origin and then going back to infinity above the positive real axis:
\bea\label{eq:DeltaS}
&&\Delta S_{p,q}(g)\equiv \mathcal{S}_{+\epsilon} \left[c_{p,q}\right](g) - \mathcal{S}_{-\epsilon} \left[c_{p,q}\right](g)
= \pi g^2 \int_\mathcal{C} \frac{ {\rm d} x }{ \sinh^2(2\pi g x) } \hat{\phi}_{p,q}(x)\\
&&=\notag \pi g^2 \int_1^{\infty}\frac{{\rm d}x}{\sinh^2(2\pi g x)} {\rm Disc}\, \hat{\phi}_{p,q}(x)
= \sum_{n=1}^{\infty}(4\pi n g^2)\,e^{-4\pi n g} \int_0^{\infty}{\rm d}t \,   e^{-4\pi n g t}\,  {\rm Disc}\, \hat{\phi}_{p,q}(t+1)\,,
\eea
where we used the fact that the discontinuity (\ref{Disc_PhiHat}) starts at $x=1$ and, in the last step, we expanded the $\sinh$ for $g\gg1$ to make explicit the exponentially suppressed factor $e^{-4\pi n g}$, benchmark of non-perturbative physics.

We specify the determination of the analytic continuation of (\ref{c_rs}) as follows
\bea
c_{p,q}^{P}(g) = \left\lbrace\begin{matrix}
\mathcal{S}_{-\epsilon} \left[ c_{p,q}\right](g)\,,\qquad0<\mbox{Arg} \,g<\pi\,,\\
\,\,\mathcal{S}_{+\epsilon} \left[ c_{p,q}\right](g)\,,\qquad-\pi<\mbox{Arg}\, g<0\,,\\
\end{matrix}
\right.\label{eq:cPert}
\eea
where the suffix $P$ reminds us that this is only the resummation of the perturbative power series (\ref{c_rs}).
This analytic function has two branch cuts, one for $\mbox{Arg} \, g = 0 $ and the other for $\mbox{Arg} \, g=\pi$.
The two discontinuities are easy to obtain using the formula (\ref{eq:DeltaS})
\bea
 {\rm Disc}_0\,c^P_{p,q}(g) &=&\label{eq:Disc0} -\Delta S_{p,q}(g)\,,\\
 {\rm Disc}_\pi\,c^P_{p,q}(g) &=&\label{eq:DiscPi} -\Delta S_{p,q}(-g)\,,
\eea
where for the discontinuity along the direction $\mbox{Arg} \, g=\pi$, we used the results, proven in the previous section, that the discontinuity of the modified Borel transform is a function of $x^2$ over the Borel plane. We will study in detail these discontinuity in the following section.

Note that using the function $c_{p,q}^{P}(g)$ to obtain the analytic continuation of the perturbative power series (\ref{c_rs}) for real and positive coupling yields two different and complex results, depending whether we reach $g\in\mathbb{R}^+$ from the upper or lower complex half plane. This ``ambiguity'' in our resummation procedure suggests that despite (\ref{eq:cPert}) has the correct asymptotic power series expansion it misses nonetheless crucial non-perturbative contributions and leads to the wrong ({\textit{i.e.}} non-physical) analytic continuation.
 
To obtain an analytic continuation that is real for real coupling we make use of the median resummation \cite{DelPham},  {\textit{i.e.}} the appropriate, unambiguous, analytic continuation that is real for real coupling.

In the case at hand the median resummation of (\ref{c_rs}) is given by
\bea
c_{p,q}^{TS}(g) = \mathcal{S}_{med}\left[ c_{p,q}\right](g)=\left\lbrace\begin{matrix}
\mathcal{S}_{-\epsilon} \left[ c_{p,q}\right](g)+\frac{1}{2} \, \Delta S_{p,q}(g)\,\,,\qquad0<\mbox{Arg} \,g<\pi/2\,,\\
\,\,\mathcal{S}_{+\epsilon} \left[ c_{p,q}\right](g)-\frac{1}{2} \,\Delta S_{p,q}(g)\,\,,\qquad-\pi/2<\mbox{Arg} \,g<0\,.
\end{matrix}
\right.\label{eq:cNP}
\eea
 
The superscript is to remind us that, upon expansion of the analytic function $c_{p,q}^{TS}(g)$ for $g\gg1$, we do not obtain just the power series (\ref{c_rs}) but rather the full transseries representation\footnote{With a slight abuse of notation we denote the median resummation (\ref{eq:cNP}) with the same symbol as its transseries representation (\ref{eq:TS}) having in mind that they both uniquely define the one and the same analytic function.} (see \cite{Edgar})
 \bea
 c_{p,q}^{TS}(g) = c_{p,q}(g)+ \frak{s} \sum_{n=1}^\infty (4\pi n g^2)\,e^{-4\pi n g} \,{\tilde{\Phi}}^{\text{\tiny{ NP}}}_{p,q}(4\pi g n)\label{eq:TS}
 \eea
 where the first term denotes the purely perturbative power series (\ref{c_rs}), and the transseries parameter $\frak{s}  = -i/2$ for $0<\mbox{Arg} \,g <\pi/2$ and  $\frak{s}  = +i/2$ for $-\pi/2<\mbox{Arg}\, g <0$.
 The function $c_{p,q}^{TS}(g)$, when expanded at strong coupling, contains infinitely many exponentially suppressed, {\textit{i.e.}} non-perturbative, terms of the form $e^{-4\pi n g} $. Each of these non-perturbative contributions is multiplied by a formal power series ${\tilde{\Phi}}^{\text{\tiny{ NP}}}_{p,q}(4\pi g n)$ whose standard Borel transform (\ref{eq:stdBorel}) can be extracted easily from
  (\ref{eq:DeltaS})
  \bea
 {\hat{\Phi}}^{\text{\tiny{ NP}}}_{p,q} (t) = \mathcal{B}_0 \left[{\tilde{\Phi}}^{\text{\tiny{ NP}}}_{p,q}\right](t) =+i\, {\rm Disc}\, \hat{\phi}_{p,q}(t+1)\,.
  \eea

 The transseries (\ref{eq:TS}) is a formal representation of the analytic function (\ref{eq:cNP}) that encodes all of its monodromies in the complex $g$-plane.  
 The Borel-Ecalle resummation of (\ref{eq:TS}) gives precisely (\ref{eq:cNP}), in particular for real and positive coupling $g$ we have that the two lateral resummation of the transseries coincide  \bea
 c_{p,q}^{TS}(g)  &=& \mathcal{S}_{-\epsilon} \left[ c_{p,q}\right](g)-\frac{i}{2}\sum_{n=1}^\infty (4\pi n g^2)\,e^{-4\pi n g} \mathcal{L}\left[{\hat{\Phi}}^{\text{\tiny{ NP}}}_{p,q}\right](4\pi g n)\\
  &=&  \mathcal{S}_{+\epsilon} \left[ c_{p,q}\right](g)+\frac{i}{2}\sum_{n=1}^\infty (4\pi n g^2)\,e^{-4\pi n g} \mathcal{L}\left[{\hat{\Phi}}^{\text{\tiny{ NP}}}_{p,q}\right](4\pi g n)
 \eea
 where $\mathcal{L}$ denotes the standard Laplace integral (\ref{eq:Laplace}), inverse of the standard Borel transform.
Furthermore, by combining (\ref{eq:DeltaS}) and (\ref{eq:cNP}), the Borel-Ecalle resummation of the transseries gives
 \bea
   c_{p,q}^{TS}(g) = c_{p,q}^{(0)}\cdot g + c_{p,q}^{(1)} + \pi g^2 \int_{0}^{\infty} \frac{ {\rm d} x }{ \sinh^2(2\pi g x) }\Re\,  \hat{\phi}_{p,q}(x)\,,\label{eq:TSresum}
 \eea
 which is precisely the integral form (\ref{crs1}) used in the previous section that we proved coinciding with the coefficients (\ref{eq:BES}) of the BES dressing phase.
 So we learn that the correct strong coupling expansion of the BES coefficients (\ref{eq:BES}) is not simply given by the asymptotic power series (\ref{c_rs}) but rather from the transseries (\ref{eq:TS}) which coincides with (\ref{c_rs}) perturbatively but it contains infinitely many new exponentially suppressed terms.
 
Note that in the present case the median resummation is very simple and ultimately consists in taking the real part of the modified Borel transform of the purely perturbative expansion.
Generically, physical observables are represented with multiple parameter transseries and the actual implementation of the median resummation can be very complicated.
We refer to \cite{Aniceto:2013fka} for a comprehensive discussion on the cancellation of non-perturbative ambiguities and the construction of the median resummation in one- and two-parameters transseries, relevant for more general physical observables than the one discussed in the present paper.

Another important thing to keep in mind is that the problem under consideration is indeed a linear problem which roughly means that each instanton sector does not ``communicate'' in an intricate way with all the others. This is a very lucky case which simplifies dramatically the Borel-Ecalle resummation procedure. One of the central points in Ecalle's works \cite{Ecalle:1981} is precisely the decodification of the complicated set of relations connecting the different perturbative coefficients in different sectors and the deep intertwining between all sectors: perturbative and non-perturbative. In the present case this could go under-appreciated due to the linearity of the problem and perhaps one of the nicest illustrations where the full power of Ecalle's work can be better appreciated is shown in a nonlinear case \cite{Couso-Santamaria:2015wga} within the context of large-$N$ dualities where the authors are also able to obtain a very explicit strong-weak coupling interpolation similar to the one described in our paper.

As already shown in the previous section, equation (\ref{eq:TSresum}) coincides with physical answer given by the coefficients of the BES dressing phase (\ref{eq:BES}),
but in order to obtain (\ref{eq:TSresum}) we had to pass from the formal power series (\ref{c_rs}) to the transseries (\ref{eq:TS}). This amounted to introduce infinitely many non-perturbative contributions and ultimately means that the initial purely perturbative formal power series (\ref{c_rs}) is not enough to reconstruct the physical answer. 

It is worth emphasizing that, due to its asymptotic nature, the strong coupling transseries representation (\ref{eq:TS}) is only a formal object but its Borel-Ecalle resummation defines a perfectly good analytic function in a wedge of the complex $g$-plane. In particular, this means that the weak coupling expansion coefficients, obtainable from the gauge theory side, must be encoded in some intricate way in the strong coupling transseries coefficients (\ref{eq:TS}). 
We do not know how to read this weak coupling expansion directly from the strong coupling transseries, but as proven above, the median resummation of the strong coupling coefficients yields precisely the coefficients of the BES dressing phase (\ref{eq:BES}), which directly allow for a weak coupling expansion that matches precisely the gauge theory results as shown in \cite{Beisert:2006ez}.

\section{Discontinuity of the Laplace transform}
\label{sec:disc}
As we have just seen, the ambiguity in the Borel resummation procedure comes from the discontinuity of the Laplace transform. In this section we therefore compute this discontinuity 
explicitly by analyzing 
\bea
\Delta S_{p,q}(g)=\pi g^2 \int_1^{\infty}\frac{{\rm d}x}{\sinh^2(2\pi g x)} {\rm Disc}\, \hat{\phi}_{p,q}(x)\, ,
\eea
where 
\bea\label{Disc_Omega}
&&{\rm Disc}\,\hat{\phi}_{p,q}=i\frac{4(p-q-1)}{(p+q-1)!}\times \\
&&~~~~z \frac{d^q}{dz^q} z^{q-1}\frac{d^{p-2}}{dz^{p-2}}\Bigg[(1-z)^{p+q}z^{p-\sfrac{1}{2}}~{}_{2}F_{1}(\sfrac{1}{2}+p,\sfrac{3}{2}+q,p+q+1,1-z)\Bigg]_{z=x^2}\, , ~~~|z|>1\, . \nonumber\eea
The important property of the discontinuity is that ${\rm Disc}\,\hat{\phi}_{p,q}(1)=0$.

Since we have 
\bea
\frac{1}{\sinh^2 (2\pi g x)}=\sum_{n=1}^{\infty} 4n e^{-4\pi g n x }\, ,
\eea
we can write
\bea
\Delta S_{p,q}(g)=-g \sum_{n=1}^{\infty}\int_1^{\infty}{\rm d}x \,  (-4\pi n g)\,  e^{-4\pi n g x}\,  {\rm Disc}\, \hat{\phi}_{p,q}(x)\, .
\eea
Substituting here the explicit formula for the discontinuity we get
\bea
\label{DSQ}
\Delta S_{p,q}(g)&=&-2ig\sfrac{(p-q-1)}{(p+q-1)!} \sum_{n=1}^{\infty}\int_1^{\infty}{\rm d}z  \,    (-h_n\sqrt{z})\,  e^{- h_n\sqrt{z} }\times\label{eq:disc} \\
&\times &\frac{d^q}{dz^q} z^{q-1}\frac{d^{p-2}}{dz^{p-2}}z^{p-\sfrac{1}{2}}\Big[(1-z)^{p+q}~{}_{2}F_{1}(\sfrac{1}{2}+p,\sfrac{3}{2}+q,p+q+1,1-z)\Bigg]\, ,\nonumber
\eea
where we have introduced a concise notation 
\bea
h_n=4\pi ng \, .
\eea

We proceed integrating by parts and noting that boundary terms always vanish we arrive at the following expression 
\bea
\Delta S_{p,q}(g)&=&-2ig\, \sfrac{(p-q-1)}{(p+q-1)!} \sum_{n=1}^{\infty}\int_1^{\infty}{\rm d}z \,  Q_n(z)(z-1)^{p+q}~{}_{2}F_{1}(1-z)\,  ,
\eea
where for conciseness we omitted the parameters of ${}_{2}F_{1}$ and introduce the following function
\bea\label{Q}
Q_n(z)=\sum_{k=0}^{\infty} \frac{(-h_n)^{k+1}}{k!} z^{p-\sfrac{1}{2}}\frac{d^{p-2}}{dz^{p-2}}z^{q-1}\frac{d^q}{dz^q}\,   z^{\frac{1}{2}(k+1)} \, .
\eea
In Appendix~\ref{app:derivation_of_Q} we show that $Q_n(z)$ has the following representation as a double sum  
\bea
\label{QDS}
&&Q_n(z)=e^{-h_n\sqrt{z}}\sum_{k=0}^q\sum_{m=0}^{p-k-2}(-1)^{1+k+p}2^{1-k-m-p}(h_n\sqrt{z})^{p+k-m}\times \\
&&~~~~~~~~~~~~~~~~~~~~~~~~~~~~~~~~~~~~~\times\frac{\sqrt{\pi }\, q!\,  \Gamma(p+m-k-1)}{h_n\, k!\, m!\,  (q-k)! \, \Gamma(p-m-k-1)\Gamma(\frac{3}{2}+k-q)}\, . \nonumber
\eea
Since 
\bea
(-1)^{p+k-m}h_n^{p+k-m}\frac{\pa^{p+k-m}}{\pa h_n^{p+k-m}}\, e^{-h_n\sqrt{z}}=(h_n\sqrt{z})^{p+k-m} \, e^{-h_n\sqrt{z}}\, ,
\eea
we have 
{\small
\bea
\hspace{-0.6cm}Q_n(z)&=&\frac{1}{h_n}\sum_{k=0}^q\sum_{m=0}^{p-k-2}
\frac{(-1)^{m+1}\sqrt{\pi }\, 2^{1-k-m-p}\, q!\,  \Gamma(p+m-k-1)}{k!\,  m!\, (q-k)! \, \Gamma(p-m-k-1)\Gamma(\frac{3}{2}+k-q)} \, \times \\
&&\hspace{5cm}~~~~~~~~~~~~\times h_n^{p+k-m}\frac{\pa^{p+k-m}}{\pa h_n^{p+k-m}}\, e^{-h_n\sqrt{z}}\, . \nonumber\eea
}
\normalsize

\noindent
Thus, we can represent $Q_n(z)$ as a certain differential operator acting on $e^{-h_n\sqrt{z}}$:
\bea
Q_n(z)=\hat{Q}_n  e^{-h_n\sqrt{z}} \, ,
\eea
with the whole $z$ dependence just sitting in the exponent. 
Then further computation reduces to the following integral
{\small
\bea
\label{diffop_on_f}
\Delta S_{p,q}(g)&=&-2ig\sfrac{(p-q-1)}{(p+q-1)!}\sum_{n=1}^{\infty}\hat{Q}_n\int_1^{\infty}{\rm d}z \,  e^{-h_n\sqrt{z}} (z-1)^{p+q}~{}_{2}F_{1}(a,b,c,1-z)\, , 
\eea
}
where $a=p+\sfrac{1}{2}$, $b=q+\sfrac{3}{2}$ and $c=p+q+1$.
Thus, we are led to compute the integral
\bea\la{int}
f(h)=\int_{1}^{\infty} {\rm d}z\, e^{-h\sqrt{z}}(z-1)^{c-1}~{}_{2}F_{1}(a,b,c,1-z)\, .
\eea
For generic values of $a,b,c$ this integral is given in \cite{PBM3}.
Keeping for the moment  $p$ and $q$ generic (non-integer), the answer is given by the following formula
{\small
\bea
\label{MI}
\hspace{-0.3cm}
\frac{f(h)}{h}&=&\tfrac{1}{2\pi}\Gamma(-p)\Gamma(-1-q)\Gamma(1+p+q)~{}_{1}F_{2}\Big(\{\tfrac{1}{2}\};\{1+p,2+q\};t\Big)  \\
&+&\frac{2\Gamma(2p-1)\Gamma(p-q-1)\Gamma(1+p+q)}{4^p \Gamma(\tfrac{1}{2}+p)\Gamma(-\tfrac{1}{2}+p)}\tfrac{1}{t^{p}}~{}_{1}F_{2}\Big(\{\tfrac{1}{2}-p\};\{1-p,2-p+q\}; t\Big) \nonumber\\
&+&\frac{2\Gamma(1-p+q)\Gamma(1+2q)\Gamma(1+p+q)}{4^{1+q} \Gamma(\tfrac{1}{2}+q)\Gamma(\tfrac{3}{2}+q)}\tfrac{1}{t^{1+q}}~{}_{1}F_{2}\Big(\{-\tfrac{1}{2}-q\};\{p-q,-q\}; t\Big)\, , \nonumber
\eea
}where we have introduced a concise notation $t=h^2/4$.
This formula can be obtained by using the Mellin transform technique, see {\it e.g.}  \cite{Mar}. 
While well-defined for generic $p$ and $q$, the above expression becomes nonsensical for $p$ and $q$ being positive integers. 
In the latter case the answer can still be found from (\ref{MI}) by using the continuity principle -- first one starts from generic 
$p,q$ close to integer values  by introducing a kind of regularisation and then takes a limit to these values. A regularisation parameter 
controls the apparent singularities which are supposed to cancel in the final expression.

To proceed, we introduce the following  shorthand notation 
\bea
H_1\equiv ~{}_{1}F_{2}\Big(\{\tfrac{1}{2}\};\{1+p,2+q\};t\Big)=\sum_{k=0}^\infty \frac{\G(\tfrac{1}{2}+k)\G(1+p)\G(2+q)}{\sqrt{\pi}\G(1+k)\G(1+k+p)\G(2+p+k)}t^k \, 
\eea
and consider the power series expansion for  the second hypergeometric function
{\small
\bea\nonumber
&&H_2\equiv t^{-p}~{}_{1}F_{2}\Big(\{\tfrac{1}{2}-p\};\{1-p,2-p+q\}; t\Big)=\\
&&~~~~~~~~~~~~~~~~~~~=\sum_{k=0}^{\infty}\frac{\Gamma(1-p)\Gamma(\tfrac{1}{2}+k-p)\Gamma(2-p+q)}{\Gamma(1+k)\Gamma(\tfrac{1}{2}-p)\Gamma(1+k-p)\Gamma(2+k-p+q)}t^{k-p}\, .
\eea
}
Denote by $\bp$ and $\bq$  positive integers to which $p$ and $q$ are close by.  Then this sum can be split into three parts
{\small
\bea
H_2&=&\frac{1}{\Gamma(\tfrac{1}{2}-p)}\sum_{k=0}^{\bp-\bq-2}\frac{\G(\tfrac{1}{2}+k-p)}{\G(1+k)}\frac{\G(p-k)}{\G(p)}\frac{\G(p-q-k-1)}{\G(p-q-1)}t^{k-p}+\\
&+&\frac{\G(2-p+q)}{\G(\tfrac{1}{2}-p)\G(p)}\sum_{k=0}^{\bq} \frac{(-1)^{k-1+\bp-\bq}\G(-\tfrac{1}{2}+k-p+\bp-\bq)\G(1-k+p-\bp+\bq)}{\G(k+\bp-\bq)\G(1+k+\bp-p+q-\bq)} 
t^{k-1-p+\bp-\bq}\, , \nonumber \\
&+&\frac{\G(1-p)\G(2-p+q)}{\G(\tfrac{1}{2}-p)}\sum_{k=0}^{\infty}\frac{\G(\tfrac{1}{2}+k-p+\bp)}{\G(1+k+\bp)\G(1+k-p+\bp)\G(2+k-p+q+\bp)}t^{k-p+\bp}\, .
\nonumber
\eea
}
Here to obtain the second line we made a shift of the original summation label $k$ as $k\to k+\bp-\bq-1$, while to get the third line we shifted
as $k\to k+\bp$. Note that the first line of $H_2$ is finite in the limit $p\to \bp$, $q\to \bq$, while the second and the third lines are ``linearly" and ``quadratically" divergent, respectively,
{\it cf.} the factors in front of the corresponding sums. 

Analogously, we consider 
\bea
&&H_3\equiv \tfrac{1}{t^{1+q}}~{}_{1}F_{2}\Big(\{-\tfrac{1}{2}-q\};\{p-q,-q\}; t\Big)=\nonumber \\
&&~~~~~~~~~~~~~~=\sum_{k=0}^{\infty} \frac{\G(-\tfrac{1}{2}+k-q)\G(p-q)\G(-q)}{\G(1+k)\G(-\frac{1}{2}-q)\G(k-q)\G(k+p-q)}t^{k-q-1}
\eea
and split the sum into two parts 
{\small
\bea
H_3&=&\frac{\G(p-q)}{\G(1+q)\G(-\tfrac{1}{2}-q)}\sum_{k=0}^{\bq} (-1)^k\frac{\G(-\tfrac{1}{2}+k-q)\G(1-k+q)}{\G(1+k)\G(k+p-q)} t^{k-q-1}+\nonumber \\
&+&\frac{\G(-q)\G(p-q)}{\G(-\tfrac{1}{2}-q)}\sum_{k=0}^{\infty}\frac{\G(\tfrac{1}{2}+k-q+\bq)}{\G(2+k+\bq)\Gamma(1+k-q+\bq)\Gamma(1+k+p-q+\bq)}t^{k-q+\bq}\, .
\eea
}
To obtain the second line we made a shift of the original summation label as $k\to k+\bq+1$. The first line in the expression above is finite in the limit 
$p\to \bp$, $q\to \bq$, while the second one is ``linearly" divergent. 

Now we put everything together and simplify the factors in front of the sums 
{\small
\bea
\frac{f(h)}{h}&=&-\frac{\sqrt{\pi}\, \Gamma(1+p+q)}{2\sin(\pi p)\sin(\pi q)}\sum_{k=0}^\infty \frac{\G(\tfrac{1}{2}+k)}{\G(1+k)\G(1+k+p)\G(2+p+k)}t^k \nonumber \\
&+&\cos(\pi p)\G(1+p+q)\sum_{k=0}^{\bp-\bq-2}\frac{\G(\tfrac{1}{2}+k-p)\G(p-k)\G(p-q-k-1)}{2\pi^{3/2}\G(1+k)}t^{k-p}\, \nonumber \\
&+&\frac{\cos(\pi p)\G(1+p+q)}{2\sqrt{\pi}\sin \pi(p-q)}\sum_{k=0}^{\bq} \frac{(-1)^{k+\bp-\bq}\G(-\tfrac{1}{2}+k-p+\bp-\bq)\G(1-k+p-\bp+\bq)}{\G(k+\bp-\bq)\G(1+k+\bp-p+q-\bq)} 
t^{k-1-p+\bp-\bq}\, , \nonumber \\
&-&\frac{\sqrt{\pi}\cot(\pi p)\Gamma(1+p+q)}{2\sin \pi(p-q)}\sum_{k=0}^{\infty}\frac{\G(\tfrac{1}{2}+k-p+\bp)}{\G(1+k+\bp)\G(1+k-p+\bp)\G(2+k-p+q+\bp)}t^{k-p+\bp} \label{div}\\
&-&\frac{\cos(\pi q)\G(1+p+q)}{2\sqrt{\pi}\sin\pi(p-q)}\sum_{k=0}^{\bq} (-1)^k\frac{\G(-\tfrac{1}{2}+k-q)\G(1-k+q)}{\G(1+k)\G(k+p-q)} t^{k-q-1} \nonumber \\
&+&\frac{\sqrt{\pi}\cot(\pi p)\Gamma(1+p+q)}{2\sin\pi(p-q)}\sum_{k=0}^{\infty}\frac{\G(\tfrac{1}{2}+k-q+\bq)}{\G(2+k+\bq)\Gamma(1+k-q+\bq)\Gamma(1+k+p-q+\bq)}t^{k-q+\bq}\, .\nonumber
\eea
}

The second line in this expression is finite (it comes from the first line of $H_2$) and we can therefore put there $p=\bar{p}$, $q=\bar{q}$. This gives the 
first contribution $I_1$ to $f(h)$ corresponding to integer values of $p,q$
\bea
I_1=(-1)^p\G(1+p+q)\sum_{k=0}^{p-q-2}\frac{\G(\tfrac{1}{2}+k-p)\G(p-k)\G(p-q-k-1)}{2\pi^{3/2}\G(1+k)}t^{k-p}\, .\eea
Obviously, $I_1$ contains inverse powers of $t$ from $t^{-p}$ up to $t^{-q-2}$. 

The rest of (\ref{div}) is divergent. To proceed, we introduce the following regularisation 
\bea\label{reg}
p=\bp+\tfrac{1}{2}\eps\, , ~~~~q=\bq-\tfrac{1}{2}\eps\, .
\eea
To take the limit, we need the formulae 
\bea
\sin\pi(\eps+m)=\sin(\pi \eps)(-1)^{m}\, ,~~~~\cot\pi (\pm\tfrac{\eps}{2}+ m)=\pm \cot\tfrac{\pi}{2}\eps\, ,
\eea
valid for any integer $m$. The second contribution to $f(h)$ comes therefore from finite sums 
{\small
\bea
I_2&=&\lim_{\eps\to 0} 
\Bigg[\frac{\cos(\pi p)\G(1+p+q)}{2\sqrt{\pi}\sin \pi(p-q)}\sum_{k=0}^{\bq} \frac{(-1)^{k+\bp-\bq}\G(-\tfrac{1}{2}+k-p+\bp-\bq)\G(1-k+p-\bp+\bq)}{\G(k+\bp-\bq)\G(1+k+\bp-p+q-\bq)} 
t^{k-1-p+\bp-\bq}\,   \nonumber \\
&&~~~-\frac{\cos(\pi q)\G(1+p+q)}{2\sqrt{\pi}\sin\pi(p-q)}\sum_{k=0}^{\bq} (-1)^k\frac{\G(-\tfrac{1}{2}+k-q)\G(1-k+q)}{\G(1+k)\G(k+p-q)} t^{k-q-1} \Bigg]\, .
\eea
}
Substituting here the formulae (\ref{reg}) and taking the limit $\eps\to 0$, we find 
{\small
\bea
&&I_2=-\G(1+p+q)\sum_{k=0}^q (-1)^{k+p}\frac{\G(-\tfrac{1}{2}+k-q)\G(1-k+q)}{2\pi^{3/2}\Gamma(1+k)\Gamma(k+p-q)}t^{k-q-1}\times \nonumber \\
&&~~~~~~\times \Big[\log t +\psi(-\tfrac{1}{2}+k-q)-\psi(1+k)-\psi(k+p-q)-\psi(1-k+q)\Big]\, ,
\eea
}
where after the computation we replaced $\bp\to p$ and $\bq\to q$.
Note that this term contains inverse powers of $t$ from $t^{-q-1}$ up to $t^{-1}$. 

Finally, the third contribution comes from infinite sums 
{\small
\bea
&&I_3=\Gamma(1+p+q)\lim_{\eps\to 0}\Bigg[-\frac{\sqrt{\pi}\, }{2\sin(\pi p)\sin(\pi q)}\sum_{k=0}^\infty \frac{\G(\tfrac{1}{2}+k)}{\G(1+k)\G(1+k+p)\G(2+p+k)}t^k \nonumber \\
&&~~~~~~~~~~~~~~~-\frac{\sqrt{\pi}\cot(\pi p)}{2\sin \pi(p-q)}\sum_{k=0}^{\infty}\frac{\G(\tfrac{1}{2}+k-p+\bp)}{\G(1+k+\bp)\G(1+k-p+\bp)\G(2+k-p+q+\bp)}t^{k-p+\bp} \nonumber\\
&&~~~~~~~~~~~~~~~+\frac{\sqrt{\pi}\cot(\pi p)}{2\sin\pi(p-q)}\sum_{k=0}^{\infty}\frac{\G(\tfrac{1}{2}+k-q+\bq)}{\G(2+k+\bq)\Gamma(1+k-q+\bq)\Gamma(1+k+p-q+\bq)}t^{k-q+\bq}\Bigg]\, .\nonumber
\eea
}
The expression $I_3$ delivers the most complicated contribution which upon taking the limit and renaming $\bp\to p$ and $\bq\to q$ reads
{\small
\bea
I_3&=&-(\log t)^2 \sum_{k=0}^{\infty}\frac{(-1)^{p-q}\G(\tfrac{1}{2}+k)\Gamma(1+p+q)}{4\pi^{3/2} \G(1+k)\G(1+k+p)\G(2+k+q)}t^k+\nonumber \\
&&~-\log t \, \, \sum_{k=0}^{\infty} \frac{(-1)^{p-q}\G(\tfrac{1}{2}+k)\G(1+p+q)}{2\pi^{3/2}\G(1+k)\G(1+k+p)\G(2+k+q)}t^k \times \nonumber \\
&&~~~~~~~~~~~~~ \times \Big[ \psi(\tfrac{1}{2}+k)-\psi(1+k)-\psi(1+k+p)-\psi(2+k+q)\Big]- \label{I3}\\
&&\hspace{1cm} - \sum_{k=0}^{\infty} \frac{(-1)^{p-q}\G(\tfrac{1}{2}+k)\G(1+p+q)}{4\pi^{3/2}\G(1+k)\G(1+k+p)\G(2+k+q)}t^k \times \nonumber \\
&&\hspace{2cm} \times \Bigg[\Big( \psi(1+k)+\psi(1+k+p)+\psi(2+k+p)-\psi(\tfrac{1}{2}+k)\Big)^2+ \nonumber \\
&&\hspace{2.5cm} +\psi^{(1)}(\tfrac{1}{2}+k)- \psi^{(1)}(1+k)-\psi^{(1)}(1+k+p)+\psi^{(1)}(2+k+p)\Bigg] \nonumber \, .
\eea
}
In this way we have found that the original integral is given by the sum of three terms
\bea
f(h)=h(I_1+I_2+I_3)\, .
\eea
In fact, the whole expression $I_3$ can be written as 
{\small
\bea\label{epsI3}
I_3&=&\frac{d^2}{d\eps^2} \sum_{k=0}^{\infty}\frac{(-1)^{p+q+1}\G(\tfrac{1}{2}+k+\eps)\Gamma(1+p+q)}{4\pi^{3/2} \G(1+k+\eps)\G(1+k+p+\eps)\G(2+k+q+\eps)}t^{k+\eps}\Big|_{\eps=0}=\\
&=&\frac{d^2}{d\eps^2}\frac{ t^{\eps}(-1)^{p+q+1}\G(\tfrac{1}{2}+\eps)\G(1+p+q)}{4\pi^{3/2}\G(1+\eps)\G(1+\eps+p)\G(2+\eps+q)}
~{}_{2}F_{3}\Big(\{1,\tfrac{1}{2}+\eps\};\{1+\eps,1+\eps+p,2+\eps+q\};t\Big)\Big|_{\eps=0}\, . \nonumber
\eea
}

\section{Strong coupling expansion of the discontinuity}
\label{sec:Strong}
Here we show how to obtain an asymptotic expansion at large $g$ starting from the exact answer for the difference $\Delta S_{p,q}(g)$. 
To this end we have to analyse the expansion of $I_3$ when $t\to\infty$. The simplest way to proceed is to use the formula (\ref{epsI3}), where 
we keep $\eps$ finite and send $t\to \infty$. The corresponding expansion of $~{}_{2}F_{3}$ is known to be 
{\small
\bea
&&~{}_{2}F_{3}\Big(\{a_1,a_2\};\{b_1,b_2,b_3\};t\Big)={\mathscr F}_1+{\mathscr F}_2+{\mathscr F}_3\, , 
\eea
}
where 
{\small
\bea
\mathscr{F}_1&=&\frac{\G(b_1)\G(b_2)\G(b_3)\G(a_2-a_1)}{\G(a_2)\G(b_1-a_1)\G(b_2-a_1)\G(b_3-a_1)}\times \nonumber\\
&&\times ~(-t)^{-a_1}{}_{4}F_{1}\Big(\{a_1,a_1-b_1+1,a_1-b_2+1,a_1-b_3+1\};\{a_1-a_2+1\};\tfrac{1}{t}\Big)\, , 
\eea
\bea
\mathscr{F}_2&=&\frac{\G(b_1)\G(b_2)\G(b_3)\G(a_1-a_2)}{\G(a_1)\G(b_1-a_2)\G(b_2-a_2)\G(b_3-a_2)}\times \nonumber\\
&&\times ~(-t)^{-a_2}{}_{4}F_{1}\Big(\{a_2,a_2-b_1+1,a_2-b_2+1,a_2-b_3+1\};\{a_2-a_1+1\};\tfrac{1}{t}\Big)\, , 
\eea
}
and $\mathscr{F}_3$ will be discussed later.

For the case at hand we identify
\bea
\label{aabbb}
a_1=1\, , ~~~a_2=\tfrac{1}{2}+\eps\, , ~~~b_1=1+\eps\, , ~~~b_2=1+\eps+p\, , ~~~b_3=2+\eps+q\, .
\eea
We start with analysis of the contribution of $\mathscr{F}_1$ into $I_3$, which we denote as $I_3^{(1)}$.
We have 
{\small
\bea
I_3^{(1)}&=&\frac{d^2}{d\eps^2}\Bigg[\frac{ t^{\eps-1}(-1)^{p+q}}{4\pi^{3/2}}\frac{\G(1+p+q)\G(-\tfrac{1}{2}+\eps)\G(\tfrac{3}{2}-\eps)}{
\G(\eps)\G(1-\eps)\G(1-\eps-p)\G(-\eps-q)\G(p+\eps)\G(1+\eps+q) }\times \nonumber\\
&&~~~~~~\times \sum_{k=0}^{\infty} \frac{\G(1-\eps+k)\G(1-\eps+k-p)\G(-\eps+k-q)}{\G(\tfrac{3}{2}-\eps+k)}\frac{1}{t^k}\Bigg] \Bigg|_{\eps=0}\, . \eea
}
Further simplification gives
{\small
\bea\nonumber
I_3^{(1)}&=&\sum_{k=0}^{\infty}\frac{d^2}{d\eps^2}\Bigg[\frac{\G(1+p+q)\G(1-\eps+k)\G(1-\eps+k-p)\G(-\eps+k-q)\sin^2(\pi\eps)\tan(\pi\eps)}{4\pi^{7/2}\G(\tfrac{3}{2}-\eps+k)}\frac{1}{t^{k+1-\eps}}\Bigg] \Bigg|_{\eps=0}\, .
\eea
}
Now it is important to realise that the expression in the brackets above has different behaviour in the limit $\eps\to 0$ depending on the value of the summation variable $k$. If $k\geq p$ then due to the factor 
$\sin^2(\pi\eps)\tan(\pi\eps)$ the expansion starts from $\eps^3$ and therefore it does not produce any contribution at order $\eps^2$. This means that we can cut the infinite sum at $k=p-1$. Then we naturally spilt it 
into two parts 
{\small
\bea\nonumber
I_3^{(1)}&=&\sum_{k=q+1}^{p-1}\frac{d^2}{d\eps^2}\Bigg[\frac{\G(1+p+q)\G(1-\eps+k)\G(1-\eps+k-p)\G(-\eps+k-q)\sin^2(\pi\eps)\tan(\pi\eps)}{4\pi^{7/2}\G(\tfrac{3}{2}-\eps+k)}\frac{1}{t^{k+1-\eps}}\Bigg] \Bigg|_{\eps=0}\\
&+&\sum_{k=0}^{q}\frac{d^2}{d\eps^2}\Bigg[\frac{\G(1+p+q)\G(1-\eps+k)\G(1-\eps+k-p)\G(-\eps+k-q)\sin^2(\pi\eps)\tan(\pi\eps)}{4\pi^{7/2}\G(\tfrac{3}{2}-\eps+k)}\frac{1}{t^{k+1-\eps}}\Bigg] \Bigg|_{\eps=0}
\, .
\eea
}
Then we make a replacement in both sums
\bea
\G(1-\eps+k-p)\sin(\pi\eps)=\frac{\pi(-1)^{p+k}}{\G(\eps+p-k)}\, ,
\eea
and in the second one we also replace
\bea
\G(-\eps+k-q)\sin(\pi\eps)=\frac{\pi(-1)^{1+k+q}}{\Gamma(1+\eps+q-k)}\, .
\eea 
This gives 
{\small
\bea\nonumber
I_3^{(1)}&=&\sum_{k=q+1}^{p-1}\frac{d^2}{d\eps^2}\Bigg[\frac{(-1)^{p+k}\G(1+p+q)\G(1-\eps+k)\G(-\eps+k-q)\sin(\pi\eps)\tan(\pi\eps)}{4\pi^{5/2}\G(\tfrac{3}{2}-\eps+k)\G(\eps+p-k)}\frac{1}{t^{k+1-\eps}}\Bigg] \Bigg|_{\eps=0}\\
&+&\sum_{k=0}^{q}\frac{d^2}{d\eps^2}\Bigg[\frac{\G(1+p+q)\G(1-\eps+k)\tan(\pi\eps)}{4\pi^{3/2}\G(\tfrac{3}{2}-\eps+k)\G(\eps+p-k)\Gamma(1+\eps+q-k)}\frac{1}{t^{k+1-\eps}}\Bigg] \Bigg|_{\eps=0}
\, .
\eea
}
In the first sum $\sin(\pi\eps)\tan(\pi\eps)\sim \pi^2\eps^2$ in the limit $\eps\to 0$ which allows one to immediately find the corresponding contribution. To proliferate a comparison with the finite contributions 
delivered by $I_1$ and $I_2$, it is convenient to implement in the first sum the change of the summation variable $k\to -k+p-1$, while in the second one $k\to -k+q$, correspondingly.
This gives
{\small
\bea
\nonumber
I_3^{(1)}&=&-\G(1+p+q)\sum_{k=0}^{p-q-2}\frac{(-1)^{k}\G(p-k)\G(p-q-k-1)}{2\sqrt{\pi}\G(\tfrac{1}{2}-k+p)\G(1+k)}t^{k-p} \\
&+&\sum_{k=0}^{q}\frac{d^2}{d\eps^2}\Bigg[\frac{\G(1+p+q)\G(1-\eps-k+q)\tan(\pi\eps)}{4\pi^{3/2}\G(\tfrac{3}{2}-\eps-k+q)\G(\eps+p-q+k)\Gamma(1+\eps+k)}t ^{k-q-1+\eps}\Bigg] \Bigg|_{\eps=0}\, .\eea
}
Since 
{
\bea\label{idG}
\frac{1}{\G(\tfrac{1}{2}-k+p)}=\frac{1}{\pi}(-1)^{k+p}\G(\tfrac{1}{2}+k-p)\, ,
\eea
}
we observe that the first sum just becomes $-I_1$, while differentiation over $\eps$ in the second one leaves us with the following answer
{\small
\bea\nonumber
I_3^{(1)}&=&-I_1-\G(1+p+q)\sum_{k=0}^q\frac{(-1)^{p+q}\G(1-k+q)}{2\sqrt{\pi}\G(1+k)\G(k+p-q)\G(\tfrac{3}{2}-k+q)}t^{k-q-1} \times\\
&\times & \Big[\log t -\psi(1+k)-\psi(k+p-q)-\psi(1-k+q)+\psi(\tfrac{3}{2}-k+q)\Big]\, .
\eea
}
Now taking into account eq.(\ref{idG}) as well as the fact that $\psi(\tfrac{3}{2}-k+q)=\psi(-\tfrac{1}{2}+k-q)$, we see that the second sum is nothing else but $-I_2$. Thus, we have found, that 
\bea
I_3^{(1)}=-I_1-I_2 \, ,
\eea
that is in the strong coupling expansion the contribution of $I_3^{(1)}$ cancels out against the sum $I_1+I_2$.

Now we analyse the contribution of the terms $\mathscr{F}_2$, which we denote as $I_3^{(2)}$. We have 
{\small
\bea
I_3^{(2)}=\frac{i(-1)^{p+q}\G(1+p+q)}{4\pi \sqrt{t}\G(\tfrac{1}{2}+p)\G(\tfrac{3}{2}+q)}~
{}_{3}F_{0}\Big(\{\tfrac{1}{2},\tfrac{1}{2}-p,-\tfrac{1}{2}-q\};\{0\}, \tfrac{1}{t}\Big)\frac{d^2}{d\eps^2}\frac{e^{-i\pi \eps}}{\cos(\pi \eps)}\Big|_{\eps=0}=0\, . 
\eea
}

Finally, the contribution $\mathscr{F}_3$ is given by the following formula
{\small
\bea
\mathscr{F}_3=\frac{\Gamma(b_1)\Gamma(b_2)\Gamma(b_3)}{2\sqrt{\pi}\Gamma(a_1)\Gamma(a_2)}t^{\tfrac{\nu}{2}}
\Bigg(e^{i\pi\nu-2\sqrt{t}}\sum_{\ell=0}^\infty (-1)^\ell \frac{2^{-\ell}c_\ell}{(\sqrt{t})^\ell}+e^{2\sqrt{t}}\sum_{\ell=0}^{\infty}\frac{2^{-\ell}c_\ell}{(\sqrt{t})^\ell}\Bigg)\, ,
\eea
}
\normalsize

\vskip -0.1cm
\noindent
where $\nu=a_1+a_2-b_1-b_2-b_2+\tfrac{1}{2}=-2-p-q-2\eps$. Determination of the asymptotic coefficients $c_{\ell}$ represents a rather non-trivial task
which we undertake in Appendix~\ref{app:cl}. There we show that the coefficients $c_{\ell}$ do not depend on $\eps$ and are given by the following explicit formula 
{\small
\bea
\label{coeff_cl}
c_{\ell}=\frac{\Gamma(2+\ell+p+q)_3F_2\left( \left\{ -\ell, \sfrac{1}{2}+p, \sfrac{3}{2}+q \right\}, \left\{ 1-\sfrac{\ell}{2}+\sfrac{p}{2}+\sfrac{q}{2},  \sfrac{3}{2}-\sfrac{\ell}{2}+\sfrac{p}{2}+\sfrac{q}{2} \right\}, 1 \right) }{(-2)^\ell \Gamma(\ell+1)\Gamma(2-\ell+p+q)}\, .
\eea
}
\normalsize

\vskip -0.3cm
\noindent
Hence we have the following contribution of $\mathscr{F}_1$ which we denote $I_3^{(3)}$,
{\small
\bea\nonumber
I_3^{(3)}=\frac{d^2}{d\eps^2}\frac{(-1)^{p+q+1}\G(1+p+q)}{8\pi^{2} t (\sqrt{t})^{p+q}}\Bigg((-1)^{p+q}e^{-2\pi i \eps}e^{-2\sqrt{t}}
\sum_{\ell=0}^\infty (-1)^\ell \frac{2^{-\ell}c_\ell}{(\sqrt{t})^\ell}+e^{2\sqrt{t}}\sum_{\ell=0}^{\infty}\frac{2^{-\ell}c_\ell}{(\sqrt{t})^\ell}\Bigg)\Bigg|_{\eps=0}\, .
\eea
}
Differentiating over $\eps$ and taking the limit $\eps\to 0$ leaves us with the following expression
\bea
I_3^{(3)}=\frac{\G(1+p+q)}{2 t (\sqrt{t})^{p+q}} e^{-2\sqrt{t}}\sum_{k=0}^\infty (-1)^\ell \frac{2^{-\ell}c_\ell}{(\sqrt{t})^\ell}\, .
\eea
Note that the growing exponent $e^{2\sqrt{t}}$ does not enter the asymptotic expansion. Recalling that $t=h^2/4$ we arrive at the following strong coupling asymptotic expansion of the integral (\ref{int})
\bea
f(h)=e^{-h}\, \frac{2^{p+q+1}\G(p+q+1)}{h^{p+q+1}}\sum_{\ell=0}^{\infty}(-1)^\ell\frac{c_\ell}{h^\ell}\, .
\eea
With this expression at hand we can now find the asymptotic expansion of $\Delta S_{p,q}$. According to eq.(\ref{diffop_on_f}) we have 
{\small
\bea
\Delta S_{p,q}(g)&=&-i g\, (p-q-1)(p+q)\sum_{k=0}^q\sum_{m=0}^{p-k-2}
\frac{(-1)^{m+1}\sqrt{\pi }\, 2^{3-k-m+q}\, q!\,  \Gamma(p+m-k-1)}{k!\,  m!\, (q-k)! \, \Gamma(p-m-k-1)\Gamma(\frac{3}{2}+k-q)} \times \nonumber\\
&\times &\sum_{\ell=0}^{\infty}(-1)^\ell c_{\ell}\, \sum_{n=1}^{\infty} h_n^{p+k-m-1}\frac{\pa^{p+k-m}}{\pa h_n^{p+k-m}}\Bigg[\frac{e^{-h_n}}{h_n^{\ell+p+q+1}}\Bigg]\,  ,
\eea
}
\normalsize

\vskip -0.3cm
\noindent
Performing differentiations we  get 
{\small
\bea
\Delta S_{p,q}(g)&=&i g\, (p-q-1)(p+q)\sum_{k=0}^q\sum_{m=0}^{p-k-2}
\frac{(-1)^{p+k}\sqrt{\pi }\, 2^{3-k-m+q}\, q!\,  \Gamma(p+m-k-1)}{k!\,  m!\, (q-k)! \, \Gamma(p-m-k-1)\Gamma(\frac{3}{2}+k-q)} \times\\
&\times &\sum_{\ell=0}^{\infty}(-1)^\ell c_{\ell}\, \sum_{s=q+m-k}^{p+q}\frac{\Gamma(p-m+k+1)\G(1+k+p+\ell+s-m)}{\Gamma(1+\ell+p+q)\Gamma(1+s+k-m-q)\Gamma(1+p+q-s)}\sum_{n=1}^{\infty} \frac{e^{-h_n}}{h_n^{2+\ell+s}} \nonumber\,  .
\eea
}
\normalsize

\noindent
Due to the gamma function standing in the middle of the denominator in the second line of the above formula, 
the sum over $s$ can be extended down to zero. Next we introduce a ``loop" parameter $L=\ell+s+3$ and change the order of summation arranging the sum over $L$ to precede the one over $\ell$:
{\small
\bea
&&\Delta S_{p,q}(g)=i g\, (p-q-1)(p+q)\sum_{k=0}^q\sum_{m=0}^{p-k-2}
\frac{(-1)^{p+k}\sqrt{\pi }\, 2^{3-k-m+q}\, q!\,  \Gamma(p+m-k-1)}{k!\,  m!\, (q-k)! \, \Gamma(p-m-k-1)\Gamma(\frac{3}{2}+k-q)} \times\\
&&\nonumber\times \Bigg[ \sum_{L=3}^{p+q+2}\, \sum_{\ell=0}^{L-3}\frac{(-1)^\ell c_{\ell}\, \Gamma(p-m+k+1)\G(k+p+L-m-2)}{\Gamma(1+\ell+p+q)\Gamma(L-2-\ell+k-m-q)\Gamma(4+p+q-L+\ell)}\sum_{n=1}^{\infty} \frac{e^{-h_n}}{h_n^{L-1}} 
 \\ 
&&+
\sum_{L=p+q+3}^{\infty}\, \sum_{\ell=L-3-p-q}^{L-3}\frac{(-1)^\ell c_{\ell}\, \Gamma(p-m+k+1)\G(k+p+L-m-2)}{\Gamma(1+\ell+p+q)\Gamma(L-2-\ell+k-m-q)\Gamma(4+p+q-L+\ell)}\sum_{n=1}^{\infty} \frac{e^{-h_n}}{h_n^{L-1}} \Bigg]
\nonumber\, . 
\eea
}
\normalsize

\noindent
Here in the last line the lower integration bound  $L-3-p-q\geq 0$ of the variable $\ell$ can be extended down to zero without changing the answer because of the gamma function $\Gamma(4+p+q-L+\ell)$.
This allows one to combine two sums over $L$ and obtain a formula 
{\small
\bea
&&\Delta S_{p,q}(g)=i g\, (p-q-1)(p+q)\sum_{L=3}^{\infty} \frac{{\rm Li}_{L-1}(e^{-4\pi g})}{(4\pi g)^{L-1}}\times  \nonumber\\
&&\times \, \sum_{\ell=0}^{L-3} \frac{(-1)^\ell c_{\ell}}{\Gamma(1+\ell+p+q)\Gamma(4+p+q-L+\ell)\Gamma(L-2-\ell+k-m-q)} \label{DS_complex} \\
&&\times \, \sum_{k=0}^q\sum_{m=0}^{p-k-2}
\frac{(-1)^{p+k}\sqrt{\pi }\, 2^{3-k-m+q}\, q!\,  \Gamma(p+m-k-1)}{k!\,  m!\, (q-k)! \, \Gamma(p-m-k-1)\Gamma(\frac{3}{2}+k-q)\Gamma(p-m+k+1)\G(k+p+L-m-2)}\nonumber\,  ,
\eea
}
\normalsize

\vskip -0.3cm
\noindent
where we have taken into account that
\bea
\sum_{n=1}^{\infty} \frac{e^{-h_n}}{h_n^{L-1}}=\frac{{\rm Li}_{L-1}(e^{-4\pi g})}{(4\pi g)^{L-1}}\, .
\eea
In Appendix~\ref{app:cl} by using the explicit form (\ref{coeff_cl}) of the coefficients $c_{\ell}$ we bring  the expression for discontinuity $\Delta S_{p,q}(g)$ found above to 
the following form 
 \bea\label{eq:GSpert}
\Delta S_{p,q}(g) =(4 i g)(p-q-1)(p+q) \sum_{L=3}^\infty \frac{{\rm Li}_{L-1}\big(e^{-4\pi g}\big)}{(4\pi g)^{L-1}} c_L(p,q)\,,
\eea
where the coefficients $c_L(p,q)$ are given by
\bea
c_L(p,q) = &&\label{eq:cLpq}\sum_{k=0}^{L-3} \frac{\Gamma(p+\frac{1}{2}+k)\Gamma(q+\frac{3}{2}+k)}{\Gamma(p+\frac{1}{2})\Gamma(q+\frac{3}{2}) \Gamma(k+1)} \times\\
&&\times\notag\sum_{n=0}^{k+p+q}\frac{(-1)^n2^{2n+3} (n+1)}{\sqrt{\pi}}\frac{\Gamma(p+\frac{1}{2}+n)\Gamma(q+\frac{3}{2}+n)}{\Gamma(p+q+1+k-n)\Gamma(n+\frac{3}{2})\Gamma(2n+5-L)}\,.
\eea 
In Appendix~\ref{app:cl}  we also provide an alternative method to compute the discontinuity  $\Delta S_{p,q}(g)$ and find the same expression (\ref{eq:DSpert}).


 \section{Dispersion relation and the non-perturbative sector}
 \label{sec:DispRel}
 Having computed the discontinuities of the modified Borel transform across the two Stokes directions $0$ and $\pi$, we can obtain the asymptotic expansion of the perturbative coefficients $c_{p,q}^{(n)}$ for $n\gg1$, via a standard dispersion-like type of argument \cite{Bender:1973,Collins:1977dw}.
 The way to proceed is to consider the Cauchy integral for the analytic continuation (\ref{eq:cPert}) of the purely perturbative series
 $c^P_{p,q}(g)$:
 \bea
 c^P_{p,q}(g) =\frac{1}{2\pi i} \oint {\rm d} w\,\frac{c^P_{p,q}(w)}{w-g}\,,
 \eea
 where the contour is around the complex point $g$.
 
\begin{figure} \begin{center}
 \includegraphics[scale=0.4]{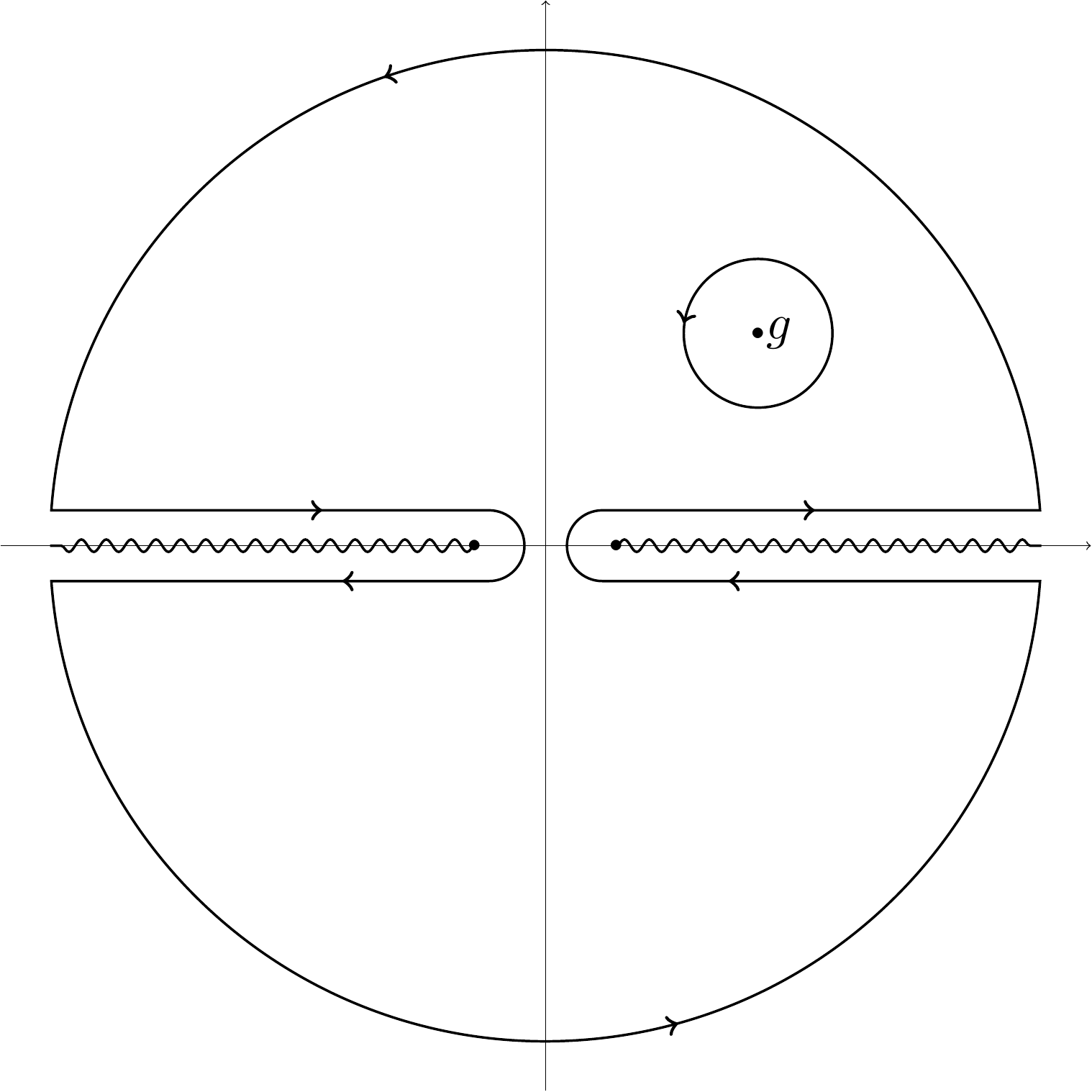}
 \caption{The Cauchy contour around the point $g$ can be closed outward as a sum over Hankel contours.}\label{fig:Cauchy}
 \end{center}
 \end{figure}

 Making use of
 \bea
 \frac{1}{w-g} = -\sum_{n=0}^\infty w^n g^{-n-1}\,,
 \eea
 valid for $g\to \infty$, we can read the perturbative coefficients $c_{p,q}^{(n)}$  
 from the contour integral
 \begin{align}
 c_{p,q}^{(n)} &\notag= -\frac{1}{2\pi i} \oint {\rm d} w\, c^P_{p,q}(w) \,w^{n-2}\\
 &= -\frac{1}{2\pi i} \int_0^\infty {\rm d}w\,  {\rm Disc}_0\,c^P_{p,q}(w)\,w^{n-2}
 -\frac{1}{2\pi i} \int_0^{-\infty} {\rm d}w\,  {\rm Disc}_\pi\,c^P_{p,q}(w)\,w^{n-2}\,,
 \end{align}
where we pushed the contour of integration to infinity as depicted in Figure~\ref{fig:Cauchy}, under that assumption that the residue at infinity of $c^P_{p,q}$ vanishes.

\begin{figure} \begin{center}
 \includegraphics[scale=0.4]{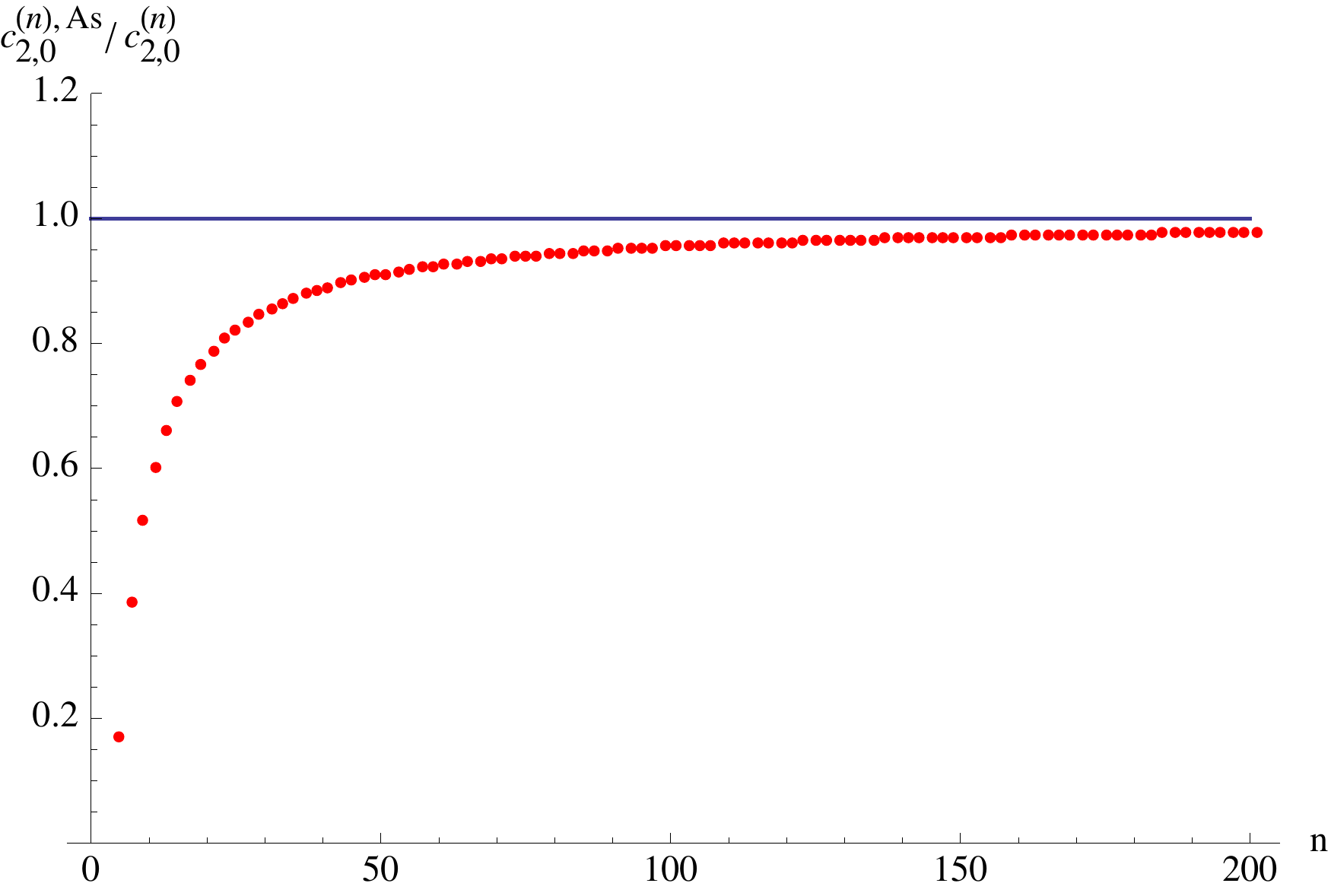}
 \caption{Ratio between the approximated coefficients $c_{2,0}^{(n)\tiny{\mbox{As}}}$ (\ref{eq:CoefAppLead}), valid for $n\gg1$, and the exact coefficients $c_{2,0}^{(n)}$ (\ref{c_pq}).\label{fig:LargeOrder}}
 \end{center}
 \end{figure}

 We know the discontinuities across the singular directions $0$ and $\pi$ from (\ref{eq:Disc0}-\ref{eq:DiscPi})
 so
 \bea\notag
c_{p,q}^{(n)} =  \frac{1}{2\pi i}  \int_0^\infty {\rm d}w\,\Delta S_{p,q}(w) w^{n-2}+
\frac{1}{2\pi i} \int_0^{-\infty} {\rm d}w\,\Delta S_{p,q}(-w) w^{n-2}\,.
 \eea
 To compute these two integral we make use of the perturbative expansion (\ref{eq:GSpert}), and, by analyzing loop order, $L$, by loop order we simply need to evaluate
 \bea
 \int_0^\infty {\rm d}w  \frac{w \,{\rm Li}_{L-1}\big(e^{-4\pi w}\big)}{(4\pi w)^{L-1}} w^{n-2}\,,
 \eea
 that, for $n\gg1$, gives 
 \bea
 \frac{ \zeta(n) \Gamma(n + 1 - L) }{(4\pi )^n}\,.
 \eea
 So, by putting everything together, we obtain the asymptotic expansion valid for $n\gg1$
 \bea
 c_{p,q}^{(n)} &\sim & \frac{2\left(1-(-1)^n\right)}{\pi} \frac{ \zeta(n)\Gamma(n-2)}{(4\pi)^n} (p+q)(p-q-1)\times \nonumber \\
 &\times&\left( c_3(p,q)+ \frac{c_4(p,q)}{n-3}+ \frac{c_5(p,q)}{(n-3)(n-4)}+ {\cal O}(n^{-3})  \right)\, ,\label{eq:CoefApp}
 \eea
 where the first three coefficients are 
 \bea
c_3(p,q)&=&4 (-1)^{(p-q)}\, , \nonumber \\
 c_4(p,q)&=&\label{eq:cLpqex}4 (-1)^{(p-q)}\times \big(2p(p-1)+\sfrac{1}{2}(2q+1)^2\big)\, , \\
  c_5(p,q)&=&4 (-1)^{(p-q)}\times \frac{1}{8}\big(-3+4p(p-1)+4q(q+1) \big)\big(4p(p-1)+(2q+1)^2\big) \, .\nonumber\eea
 Note that the $n$ even coefficients completely disappear from this analysis because, as explained before, the $c_{p,q}^{(n)}$ with $n$ even are non-vanishing only for a finite number of terms. 
The large order behaviour of the perturbative coefficients captures precisely the lower order perturbative coefficients on top of the non-perturbative contributions in the transseries (\ref{eq:TS}), {\textit{i.e.}} the coefficients for the strong coupling expansion of ${\tilde{\Phi}}^{\text{\tiny{ NP}}}_{p,q}(g)$. In Figure~\ref{fig:LargeOrder} we show how well, at large $n$, the perturbative coefficients $c_{p,q}^{(n)}$ can be approximated by even their leading asymptotic expansion 
\bea\label{eq:CoefAppLead}
c_{p,q}^{(n)\tiny{\mbox{As}}} = \frac{2\left(1-(-1)^n\right)}{\pi} \frac{ \zeta(n)\Gamma(n-2)}{(4\pi)^n} (p+q)(p-q-1)c_3(p,q)\,.
\eea
 
The formula  (\ref{eq:CoefApp}) allows us to obtain an explicit formula for the polynomials $c_L(p,q)$ by comparing the large $n$ asymptotic expansion of the coefficients $c_{p,q}^{(n)}$ with the right hand side of (\ref{eq:CoefApp}). Hence, 
we need to asymptotically expand formula (\ref{c_pq}) for large $n$. To this end we consider the ratio between $c_{p,q}^{(n)}$ and its leading asymptotic coefficient which for $n$ odd takes the form
\bea
R^{(n)}(p,q)&=&c_{p,q}^{(n)}  \Big/ c_{p,q}^{(n)\tiny{\mbox{As}}}=\frac{c_{p,q}^{(n)}\,\pi (4\pi)^n }{4 \zeta(n)\Gamma(n-2) (p+q)(p-q-1) c_3(p,q) }\sim \nonumber\\
&\sim& \label{eq:Rexp}1+ \frac{c_4(p,q)}{c_3(p,q)}\frac{1}{(n-3)}+ \frac{c_5(p,q)}{c_3(p,q)}\frac{1}{(n-3)(n-4)}+ {\cal O}(n^{-3})  \, .\eea
In what follows it appears advantageous to use the change of variables $n=2(m+1)$ where $m$ is half-integer and replace $q\to q-1$. Then for $R^{(n)}(p,q-1)$ we get 
\bea\label{Rpq}
R^{(n)}(p,q-1)=\frac{2^{4m}m}{\pi}\, B(m+1-p,m+p)B(m+1-q,m+q)\, ,
\eea 
where $B(a,b)$ is the Euler beta integral
\bea
B(a,b)=\int_0^1 {\rm d}v\, v^{b-1}(1-v)^{a-1}\, .
\eea
We observe that in the formula (\ref{Rpq}) contribution of $p$ and $q$ completely factorises and comes in a symmetric fashion. Therefore, our task now is to find an asymptotic expansion of the Euler integral 
when $m\to \infty$. First we compute the integral by means of the saddle point method.  Consider 
\bea
B\equiv B(m+1-p,m+p)&=&\int_0^1 {\rm d}v\, v^{m+p-1}(1-v)^{m-p}= \nonumber\\
&=& \int_0^1 {\rm d}v\, v^{p-1}(1-v)^{-p}e^{m\log(v(1-v))}\, . \eea
For large $m$ the dominant contribution to this integral comes from the critical point $v=\sfrac{1}{2}$ for which the ``action" is $\log(v(1-v))\vert_{v=1/2}=-\log 4$.
This motivates to perform the following change of integration variable
\bea
t=\log(4)-\log(v(1-v))
\eea
 which converts the integral to 
 \bea
B=2^{-2m}\int_0^\infty {\rm d}t \frac{e^{-mt}}{2\sqrt{e^t-1}}\Bigg[ \Big(e^{t/2}+\sqrt{e^t-1}\Big)^{2p-1}+\Big(e^{t/2}-\sqrt{e^t-1}\Big)^{2p-1}\Bigg]\, . \eea
Now using binomial expansions twice we rewrite the integrand as a double sum 
 \bea
 B&=& 2^{-2m}\sum_{s=0}^{p-1}\sum_{r=0}^s \binom{2p-1}{2s}\binom{s}{r}(-1)^r  \int_0^\infty {\rm d}t \frac{e^{(-m+p-r-1/2)t}}{\sqrt{e^t-1}} \, .\eea
 Evaluating this integral in the regime $m\gg 1$ and further performing one summation we arrive at 
 \bea
 B=-2^{-2m}\sqrt{\pi}\, \sum_{r=0}^{p-1}(-1)^{p-r}4^r \frac{(2p-1)\G(p+r)}{\G(2r+2)\G(p-r)}\times \frac{\G(m-r)}{\G(m+\frac{1}{2}-r)}\, .
 \eea
 The ratio of two gamma function has the known asymptotic expansion in the limit $m\to \infty$, namely
 \bea
 \frac{\G(m-r)}{\G(m+\frac{1}{2}-r)}\sim m^{-1/2}\sum_{l=0}^{\infty} (-1)^l (1/2)_l\frac{B_l^{(1/2)}(-r)}{l!}\frac{1}{m^l}\, ,
  \eea
  where $B_l^{(1/2)}(-r)$ are the generalised Bernoulli polynomials also known as Norlund polynomials, see {\it e.g.} \cite{Temme}. 
 
 Using this result we can obtain the asymptotic expansion of the Euler beta for $m\gg1$:
 \bea
  B(m+1-p,m+p) &\sim& -2^{-2m}\sqrt{\frac{\pi}{m}} \times \sum_{l=0}^\infty \frac{1}{m^l} d_l(p)\,,
   \eea
  where
  \bea
  d_l(p) = (-1)^{p+l}\frac{(2p-1)\,(1/2)_l }{l!} \sum_{r=0}^{p-1} (-1)^{r} 4^r B^{(1/2)}_l(-r) \frac{\G(p+r)}{\G(2r+2)\G(p-r)}\,.
  \eea
  
 The function $R_n(p,q-1)$ can be expanded for large $n$, using the variable $n=2(m+1)$, via the convolution of the above coefficients
 \bea
 R_n(p,q-1) \sim \sum_{l=0}^\infty \frac{1}{m^l} \left(\sum_{k=0}^l d_k(p)\, d_{l-k}(q)\right)\,.
 \eea
 This is not quite the expansion we sought for, as shown in equation (\ref{eq:Rexp}) we want to express this ratio as
  \bea
 R_n(p,q-1) \sim  1+ \frac{c_4(p,q-1)}{c_3(p,q-1)}\frac{1}{2m-1}+ \frac{c_5(p,q-1)}{c_3(p,q-1)}\frac{1}{(2m-1)(2m-2)}+ {\cal O}(m^{-3})\,.\nonumber
 \eea
 We can easily relate one expansion to the other via
 \bea\label{eqcLpq2}
 c_L(p,q) = -4\,(-1)^{p+q}\sum_{l=0}^{L-3} {\rm S}^{(l)}_{L-3} \,2^l \left(\sum_{k=0}^l d_k(p)\, d_{l-k}(q+1)\right)\,,
 \eea
 where $S^{(l)}_L$ denotes the Stirling number of the first kind.
 
 As we will shortly see, these coefficients $c_L(p,q)$, polynomials in $p$ and $q$, will give rise to important non-perturbative contribution to the dressing phase (\ref{eq:dressing}).
  
 \section{Non-perturbative contributions to the dressing phase}
 \label{sec:NPdress}
 
  We have shown in Section~\ref{sec:Resum}, that the purely perturbative asymptotic power series expansion (\ref{c_rs}), is not enough to reconstruct the coefficients of the BES dressing phase (\ref{eq:BES}). We need to replace the perturbative expansion by the transseries representation (\ref{eq:TS}) whose Borel-Ecalle resummation (\ref{eq:TSresum}) matches precisely the non-perturbative result (\ref{eq:BES}). In this section we see the effects that our transseries expansion produces to the dressing phase.
 
 \subsection{Effects of the non-perturbative sector to the dressing phase}
 Our replacement from the perturbative power series (\ref{c_rs}) to the transseries (\ref{eq:TS}) is not without consequences. In \cite{Beisert:2006ib} the authors showed that if we restrict the sum (\ref{c_rs}) to even $n$, we obtain a strong coupling solution to the crossing symmetry equation (\ref{eq:crossing}).
This particular solution does not have the correct weak coupling limit and for this reason the authors considered the analytic continuation of the series (\ref{c_rs}) by summing over all the integers $n$. This amounts to adding to the dressing phase a solution to the homogeneous crossing symmetry equation
 \bea
i\theta(x_j,x_k)+i \theta(1/x_j,x_k) = 0\,.\label{eq:Hcrossing}
\eea

The BES coefficients (\ref{eq:BES}) proposed in \cite{Beisert:2006ez} thus interpolates between the formal power series expansion (\ref{c_rs}) at strong coupling and the correct gauge theory weak coupling limit.
The crucial point is that this integral representation for the BES coefficients is not quite equivalent to the formal power series (\ref{c_rs}), but rather it is obtained via the Borel-Ecalle resummation of the transseries (\ref{eq:TS}).
This means that the non-perturbative terms we added must lead to additional contributions to the dressing phase, solutions to the homogeneous crossing symmetry equation (\ref{eq:Hcrossing}).

Let us compute this additional non-perturbative contributions to the dressing phase.
Since we know from \cite{Beisert:2006ib} that the formal power series (\ref{c_rs}) solves the full crossing symmetry equation (\ref{eq:crossing}) we can just focus on the purely non-perturbative terms (\ref{eq:GSpert}) of our transseries ansatz (\ref{eq:TS}).
The non-perturbative contributions to the function $\chi(x_1,x_2)$, given by equation (\ref{eq:Chi}) and written using $p,q$ variables is
\bea
\chi_{NP}(x_1,x_2) = \frak{s}  \sum_{p=2}^\infty\sum_{q=0}^{p-2} \frac{ i \Delta S_{p,q}(g)}{( p - q+1) (p + q)}\frac{1}{ x_1^{p-q -1} x_2^{p + q}}\,,\label{eq:ChiNP}
\eea
where $\frak{s} $ is once again the transseries parameter discussed in Section~\ref{subsec:disc}, {\textit{i.e.}} $\frak{s}  = -i/2$ for $0<\mbox{Arg} \,g <\pi/2$ and  $\frak{s}  = +i/2$ for $-\pi/2<\mbox{Arg}\, g <0$.

We do not know how to perform this double sum (\ref{eq:ChiNP}) using the exact integral representation (\ref{DSQ}) for $\Delta S_{p,q}(g)$, but we can easily compute it loop order by loop order using the strong coupling expansion (\ref{eq:GSpert}).

Using (\ref{eq:GSpert}) we can rewrite the above equation in the form of the loop expansion
\bea
\chi_{NP}(x_1,x_2) = -4\,\frak{s}  \sum_{L=3}^\infty \frac{{\rm Li}_{L-1}\big(e^{-4\pi g}\big)}{(4\pi g)^{L-1}} \chi_{NP}^{(L)}(x_1,x_2) \,,\label{eq:ChiNPSum}
\eea
with the definition
\bea
\chi_{NP}^{(L)}(x_1,x_2)= \sum_{p=2}^\infty\sum_{q=0}^{p-2} \frac{ c_L(p,q)}{ x_1^{p-q -1} x_2^{p + q}}\,,\label{eq:ChiNPL}
\eea
where the coefficients $c_L(p,q)$ are given for example by (\ref{eq:cLpq}) (see also the other equivalent forms presented in equation (\ref{eqcLpq2}) and Appendix~\ref{app:cl}).
Note that the coupling constant, $g$, only appears in front of the series (\ref{eq:ChiNPSum}).
It is surprising that the $L^{th}$ loop contribution to $\chi_{NP}$ coming from {\it{all}} instantons sectors can be fully resummed giving rise to the exponentially suppressed factor ${\rm Li}_{L-1}\big(e^{-4\pi g}\big)$.

From the explicit coefficients (\ref{eq:cLpq})-(\ref{eq:cLpqex}) it is a straightforward calculation to obtain the first few
\begin{align}
\chi_{NP}^{(3)}(x_1,x_2) &\notag =   \frac{4 x_2}{(1 + x_1 x_2) (  x_2^2-1)}\,,\\
\chi_{NP}^{(4)}(x_1,x_2) &\notag = \frac{2x_2 (1 - 6 x_1 x_2 + 14 x_2^2 + x_1^2 x_2^2 + 36 x_1 x_2^3 + x_2^4 + 6 x_1^2 x_2^4 + 
 2 x_1 x_2^5 + 9 x_1^2 x_2^6)}{ (1 + 
   x_1 x_2)^3 ( x_2^2-1)^3}   \,,\\
   \chi_{NP}^{(5)}(x_1,x_2) &\label{eq:chiNPex}=\frac{x_2 P(x_1,x_2)}{ (1 + 
   x_1 x_2)^5 ( x_2^2-1)^5}\,,
\end{align}
where $P(x_1,x_2)$ is a certain polynomial of degree $4$ and $12$ respectively in $x_1$ and $x_2$.
The reader can easily develop higher order contributions to $\chi_{NP}^{(L)}$ from the formula (\ref{eq:ChiNPL}).

At this point it is simply a matter of calculation to plug these non-perturbative contributions $\chi_{NP}^{(L)}$  into the dressing phase (\ref{eq:thetachi}) and show that these new terms are solutions to the homogeneous crossing symmetry equations (\ref{eq:Hcrossing}). Note that the full series (\ref{eq:ChiNPSum}) is a solution to the homogenous equation because every order in the $g^{-1}$ expansion solves (\ref{eq:Hcrossing}): {\textit{i.e.}} the coefficient $\chi_{NP}^{(L)}(x_1,x_2)$ of the $g^{-L+1}$ term is already on its own a solution to the crossing symmetry equation coming from the resummation of infinitely many instanton sectors. 

Note that the first non-perturbative contribution is given by $\chi_{NP}^{(3)}$, which corresponds to a three-loop perturbative correction $g^{-2}$, on top of a non-perturbative background. As mentioned in the Introduction, the vanishing of the tree level, one- and two-loops contributions might be explained by a protection mechanism based on vanishing of the zero mode factors, forcing perturbation theory on top of these mysterious non-perturbative saddles to start from three-loops .

 We claim that the complete non-perturbative correction (\ref{eq:ChiNP}) to the dressing phase, since it is a formal sum (\ref{eq:ChiNPSum}) of homogeneous solutions, gives also rise to a solution to the homogeneous crossing symmetry equation (\ref{eq:Hcrossing}), very likely {\it not} of the simple rational form in $x_1,x_2$ as the coefficients $ \chi_{NP}^{(L)}$ just encountered.

\subsection{Generating solutions to the homogenous crossing symmetry equation}

From the large order behaviour (\ref{eq:CoefApp}) of the perturbative coefficients (\ref{c_rsn}) we can construct a generating functional to obtain solutions to the homogeneous crossing symmetry equations (\ref{eq:Hcrossing}).
In \cite{Beisert:2006ib} the authors noticed that the perturbative coefficient $c^{(n)}_{p,q}$, with $n$ odd, $n>1$, generates a contribution to the dressing phase that solves (\ref{eq:Hcrossing}).
Similarly, for $n\gg1$, we can consider the asymptotic expansion (\ref{eq:CoefApp}) and, as we have just seen, 
for each loop order $L$, the perturbative coefficient $c_L(p,q)$ yields once again solutions to the homogeneous crossing symmetry equation. 

Thus we can consider, similarly to (\ref{eq:Chi}), the expression
\bea\label{eq:gfinitial}
\sum_{p=2}^\infty\sum_{q=0}^{p-2}\frac{ c^{(2z+1)}_{p,q} }{  (p-q-1)(p+q)}\frac{1}{x_1^{p-q -1} x_2^{p + q}}\,.
\eea
When $n=2z+1$ is an odd integer, this function reproduces the known perturbative contributions to the dressing phase. Viceversa, when $z\gg1$, we know from (\ref{eq:CoefApp}) that the perturbative coefficients $c^{(2z+1)}_{p,q}$ can be written as an asymptotic expansion in $z^{-1}$ in terms of the non-perturbative sector's coefficients $c_L(p,q)$ (\ref{eqcLpq2}). Thanks to the analysis of the previous section, each one of these terms will produce a solution to (\ref{eq:Hcrossing}) and equation (\ref{eq:gfinitial}) will basically sum up all of these contributions and it will still solve the homogeneous problem since it is a linear problem.
Hence equation (\ref{eq:gfinitial}) is somehow interpolating between the perturbative and the non-perturbative solutions to  (\ref{eq:Hcrossing}). 

Discarding from equation (\ref{eq:gfinitial}) an overall factor which is only $z$-dependent, we consider the generating functional for homogenous solutions to the crossing symmetry equation (\ref{eq:Hcrossing}) given by
\bea
\Xi(z;x_1,x_2) = \sum_{p=2}^\infty \sum_{q=0}^{p-2} \frac{(-1)^{p+q}}{x_1^{p-q -1} x_2^{p + q}}
  \left( z-1/2 \right)_ p  \left( z+1/2\right)_ {-p}  \left( z+1/2\right)_ q \left( z-1/2 \right)_ {-q} \,.
 \eea
 
The sum over $q$ can be easily performed giving
\bea
\Xi(z;x_1,x_2) = \sum_{p=2}^\infty \left(I_1 +I_2\right)\,,
\eea 
with 
\bea
I_1 =(-1)^p\,
        \left(z-1/2 \right)_ p  \left( z+1/2\right)_ {-p}  \frac{\phantom{i}_2F_1\left(
        1, \frac{1}{2}+z; \frac{3}{2} - z\left\vert\, \frac{x1}{x2}\right.\right)}{x_1^p x_2^{p-1}}\,,
\eea
and 
\bea
I_2 = 
        \left[\left(z-1/2 \right)_ p  \left( z+1/2\right)_ {-p} \right]^2 \frac{\phantom{i}_2F_1\left(
        1, p-\frac{1}{2}+z; p+\frac{1}{2} - z\left\vert\, \frac{x_1}{x_2}\right.\right)}{x_2^{2p-1}}\,.
\eea
 
For $I_1$ the sum over $p$ is straightforward
\bea
S_1 =\label{eq:S1} \sum_{p=2}^\infty I_1 = \frac{ (1 + 2 z)_2F_1\left(1, \frac{3}{2}+ z; \frac{5}{2} - z\left\vert \frac{1}{
  x_1x_2}\right.\right)}{(2z-3)x_1 x_2} \frac{ _2F_1\left(1, \frac{1}{2}+ z; \frac{3}{2} - z\left\vert \frac{x_1}{
  x_2}\right.\right)}{ x_2}\,.
\eea 
 The first term in $S_1$ is obviously symmetric in $x_1\leftrightarrow x_2$, while for $z\in\mathbb{N}$ one can easily show using the inversion formula for the hypergeometric function\footnote{We simply used equation (15.8.2) of \cite{NIST} for the case at hand.} that also the second fraction is symmetric. This means that, for $z\in\mathbb{N}$, the contribution of $S_1$ to the dressing phase (\ref{eq:thetachi}) is actually zero.
 
 The second contribution to $\Xi$ comes from
 \bea
 S_2= \sum_{p=2}^\infty I_2  = \sum_{p=2}^\infty  \left[\left(z-1/2 \right)_ p  \left( z+1/2\right)_ {-p} \right]^2 \frac{\phantom{i}_2F_1\left(
        1, p-\frac{1}{2}+z; p+\frac{1}{2} - z\left\vert\, \frac{x_1}{x_2}\right.\right)}{x_2^{2p-1}}\,.
 \eea
 This sum is trickier than $S_1$ because the index of summation $p$ appears in the parameters of the hypergeometric function. We notice that, for $z\in\mathbb{N}$, the difference between the two parameters $c$ and $b$ of the hypergeometric is  $ p+1/2 - z - ( p-1/2+z) = 2z-1\in\mathbb{N}$. This allows us to use the reduction formula, see {\it e.g.} \cite{PBM3},
 \bea\notag
 _2F_1\left(
        1, p-\frac{1}{2}+z; p+\frac{1}{2} - z\left\vert\, \frac{x_1}{x_2}\right.\right) =\sum_{k=0}^{2z-1} \binom{2 z - 1}{ k} 
        \frac{k!}{
  \left( p - z+1/2 \right)_k}  \left(\frac{x_1}{x_2-x_1}\right)^k\frac{x_2}{x_2-x_1}\,.
\eea
 The sum over $p$ can be now performed
 \bea
 S_2&=&\label{eq:S2}\frac{\left[(1 + 2 z) \,\Gamma(3/2 - z)\right]^2 }{4 x_2^2 ( x_2-x_1)} \times\\
 &\,&\notag\times \sum_{k=0}^{2z-1}   \left(\frac{x_1}{x_2-x_1} \right)^k   \frac{\Gamma(2z)}{\Gamma(2z-k)} \,_3{\tilde{F}}_2\left(1, z+\frac{3}{2}, 
    z+\frac{3}{2}; \frac{5}{2} - z, \frac{5}{2} + k - z\left\vert \frac{1}{x_2^2}\right.\right)\,,
 \eea
 where $_3{\tilde{F}}_2$ denotes the generalized hypergeometric function regularized.
 
 For $z\in\mathbb{N}$, our generating functional $\Xi(z;x_1,x_2)$ produces only rational functions of $x_1,x_2$. In particular, using the explicit formulas (\ref{eq:S1})-(\ref{eq:S2}), we can easily check that $\Xi(1;x_1,x_2)$ coincides precisely (modulo an overall numerical factor) with the three world sheet loops contribution $\chi^{(3)}(x_1,x_2)$ presented in equation (5.6) of \cite{Beisert:2006ib}. 
 
 Similarly, from our studies of large order behaviour (\ref{eq:CoefApp}), we expect the following behaviour of the generating functional for $z\gg1$
 \bea
 \Xi(z;x_1,x_2)\sim \chi_{NP}^{(3)}(x_1,x_2) +\frac{\chi_{NP}^{(4)}(x_1,x_2) }{(2z-2)}+\frac{\chi_{NP}^{(5)}(x_1,x_2) }{(2z-2)(2z-3)}+O(z^{-3})\,,
 \eea
where the rational function $\chi_{NP}^{(L)}(x_1,x_2)$ are precisely the non-perturbative contributions to the dressing phase (\ref{eq:chiNPex}) previously computed.
 
It would be interesting to obtain an analytic expression for  $ \Xi(z;x_1,x_2)$ for arbitrary values of $z$ and show that it solves the homogenous crossing symmetry equation.

\section{Acknowledgements}
We would like to thank Benjamin Basso, Lorenzo Bianchi, Davide Fioravanti, Marco Rossi and Ricardo Schiappa for useful discussions. We specially thank Sergey Frolov for his careful read of our manuscript. The work of G.A. and S.S. is supported by the German Science Foundation (DFG) under the
Collaborative Research Center (SFB) 676 Particles, Strings and the Early Universe and the Research Training Group 1670. D.D. is grateful to Hamburg University and DESY for their hospitality and to the Collaborative Research Center SFB 676 for financial support during the final stages of this project.

\appendix
\section{Derivation of the discontinuity of $\Omega$}
\label{app:discon}
To obtain the discontinuity of $\Omega$, , {\it cf.} (\ref{eq:Omega}), we first apply the reduction technique \cite{PBM3}  which allows one to reduce our particular ${}_{4}F_{3}$ to a multiple derivative of ${}_{2}F_{1}$ with respect to the variable $z$ by means of the following formula
\bea\label{reduction}
\Omega&=&\sum_{k=0}^{p-2}\sum_{m=0}^q \binom{p-2}{k}\binom{q}{m}\frac{1}{(5/2)_k(3/2)_m} z^m\frac{d^m}{dz^m}z^k\frac{d^{k}}{dz^{k}}f=\\
\nonumber
&=&\sum_{k=0}^{p-2}\sum_{m=0}^q \sum_{s=0}^m\binom{p-2}{k}\binom{q}{m}\binom{m}{s}\frac{1}{(5/2)_k(3/2)_m}\frac{\Gamma(k+1)}{\Gamma(1-m+k+s)}z^{k+s}\frac{d^{k+s}f}{dz^{k+s}}\, ,
\eea
where we have introduced a concise notation 
\bea
f(z): ={}_{2}F_{1}\big(\sfrac{3}{2}-p,\sfrac{1}{2}-q, 2, z\big)\, .
\eea
First the sum over $m$ can be straightforwardly taken leaving behind 
\bea
\label{reduction0}
\Omega=\sum_{k=0}^{p-2}\sum_{s=0}^q \frac{3\times  2^{2(k+s)} \Gamma(-1+p)\Gamma(1+q) {}_{2}F_{1}(-k,s-q, 3/2+s,1)}{(3+2k)\Gamma(2+2k)\Gamma(p-k-1)\Gamma(1+q-s)\Gamma(2+2s)}z^{k+s}\frac{d^{k+s}f}{dz^{k+s}}\, .  \nonumber 
\eea
And further summation gives 
\bea  \label{diff4F3}\Omega&=&\frac{3\pi}{8}\frac{\Gamma(p-1)}{\Gamma(q+\sfrac{3}{2})}\times \\
&&~~~~~
\times \sum_{\ell=0}^{p+q-2}\frac{~{}_{3}\tilde{F}_{2}\left( \left\{ \sfrac{3}{2}+\ell, 2+\ell-p-q, -q    \right\}, \left\{ 1+\ell-q,  \sfrac{5}{2}+\ell-q \right\}, 1 \right)}{\Gamma(p+q-\ell-1)}    z^{\ell}\frac{d^{\ell}f}{dz^{\ell}}\, ,\nonumber
\eea
where $~{}_{3}\tilde{F}_{2}$ stands for the regularised hypergeometric function, which is given in fact by the finite sum
\bea
&&\frac{~{}_{3}\tilde{F}_{2}\left( \left\{ \sfrac{3}{2}+\ell, 2+\ell-p-q, -q    \right\}, \left\{ 1+\ell-q,  \sfrac{5}{2}+\ell-q \right\}, 1 \right)}{\Gamma(p+q-\ell-1)} =\\
&&~~~~=\sum_{r=0}^q \frac{\Gamma(\sfrac{3}{2}+\ell+r)q!}{\Gamma(\sfrac{3}{2}+\ell)r!(q-r)!}\frac{1}{\Gamma(1+\ell-q+r)\Gamma(\sfrac{5}{2}+\ell-q+r)\Gamma(p+q-1-\ell-r)}\, . \nonumber
\eea

The function $f$ has a branch cut on the interval $[1,\infty)$ and the corresponding discontinuity is known to be 
\bea
{\rm Disc}f(z)=2\pi i\frac{(z-1)^{p+q}~{}_{2}F_{1}(\sfrac{1}{2}+p,\sfrac{3}{2}+q,p+q+1,1-z)}{\Gamma(\sfrac{3}{2}-p)\Gamma(\sfrac{1}{2}-q)\Gamma(1+p+q)}\, .\eea
Using the series representation for ${}_{2}F_{1}$ we find from this formula 
\bea
z^{\ell}\frac{d^{\ell}{\rm Disc}(f)}{dz^{\ell}}=-\frac{8i}{\pi} \sum_{n=0}^{\infty}\frac{(-1)^{n+p+q}z^\ell(z-1)^{n+p+q-\ell}\Gamma(\sfrac{1}{2}+n+p)\Gamma(\sfrac{3}{2}+n+q)}{(2p-1)(2q+1)\Gamma(n+1)\Gamma(p+q+n+1-\ell)}\, .
\eea
Plugging everything together we get 
{\small
\bea\nonumber 
{\rm Disc}(\Omega)&=&-3i\frac{\Gamma(p-1)}{\Gamma(q+\sfrac{3}{2})}\sum_{n=0}^{\infty}\sum_{\ell=0}^{p+q-2}\sum_{r=0}^q \frac{(-1)^{n+p+q}z^\ell(z-1)^{n+p+q-\ell}\Gamma(\sfrac{1}{2}+n+p)\Gamma(\sfrac{3}{2}+n+q)}{(2p-1)(2q+1)\Gamma(n+1)\Gamma(p+q+n+1-\ell)}\times \\
&\times& \frac{\Gamma(\sfrac{3}{2}+\ell+r)q!}{\Gamma(\sfrac{3}{2}+\ell)r!(q-r)!}\frac{1}{\Gamma(1+\ell-q+r)\Gamma(\sfrac{5}{2}+\ell-q+r)\Gamma(p+q-1-\ell-r)}\, .
\eea
}
\normalsize
\vskip -0.3cm
\noindent
Because of $\Gamma(1+\ell-q+r)$ in the denominator, the sum over $\ell$ can be restricted to runs from $q-r$ to $p+q-2$. We therefore make a change of variable $\ell=q-r+s$, so that $s$ runs from $0$ to $p+r-2$.
Then we get 
{\small
\bea\nonumber 
&&{\rm Disc} (\Omega)=-3i\sum_{n=0}^{\infty}\frac{(-1)^{n+p+q}\Gamma(p-1)\Gamma(\sfrac{3}{2}+n+q)\Gamma(\sfrac{1}{2}+n+p)}{\Gamma(1+n)\Gamma(q+\sfrac{3}{2})(2p-1)(2q+1)}\sum_{r=0}^q \frac{q!}{r!(q-r)!}\times\\
&&~~~~~~\times \sum_{s=0}^{p-2+r}\frac{ \Gamma(\sfrac{3}{2}+q+s)\, (z-1)^{r-s+n+p}z^{q-r+s}}{\Gamma(1+s)\Gamma(\sfrac{5}{2}+s)\Gamma(p-1-s)\Gamma(\sfrac{3}{2}+q-r+s)\Gamma(1+r-s+n+p)}\, .
\eea
}
\normalsize
\vskip -0.3cm
\noindent
Here in the denominator of the last sum the term $\Gamma(p-1-s)$  cuts the summation range for $s$ at $p-2$. Therefore, we can change the order of summation in $r$ and $s$ and write
\bea\nonumber 
&&{\rm Disc} (\Omega)=-\sum_{n=0}^{\infty}\frac{3i(-1)^{n+p+q}\Gamma(p-1)}{\Gamma(q+\sfrac{3}{2})(2p-1)(2q+1)} \sum_{s=0}^{p-2}\frac{q!\, \Gamma(\sfrac{3}{2}+q+s)\Gamma(\sfrac{3}{2}+n+q)\Gamma(\sfrac{1}{2}+n+p)}{\Gamma(1+n)\Gamma(1+s)\Gamma(\sfrac{5}{2}+s)\Gamma(p-1-s)} \times \\
&&~~~~~~~~\times \sum_{r=0}^q \frac{(z-1)^{r-s+n+p}z^{q-r+s}}{r!(q-r)!\Gamma(\sfrac{3}{2}+q-r+s)\Gamma(1+r-s+n+p)}\, .
\eea
The sum over $r$ yields 
\bea\nonumber
&&\mathcal{R}\equiv \sum_{r=0}^q \frac{(z-1)^{r-s+n+p}z^{q-r+s}}{r!(q-r)!\Gamma(\sfrac{3}{2}+q-r+s)\Gamma(1+r-s+n+p)}=\\
&&~~~~~~~~~~~~~~~~~=\frac{(z-1)^{n+p-s}z^{q+s} {}_{2}F_{1}(-q,-\sfrac{1}{2}-q-s, 1+n+p-s,\frac{z-1}{z})}{\Gamma(1+q)\Gamma(1+n+p-s)\Gamma(\sfrac{3}{2}+q+s)}\, .
\eea
Next with the help of the well-known transformation formula 
 \bea\label{tran_F}
{}_{2}F_{1}(a,b,c;z)=(1-z)^{-a}{}_{2}F_{1}(a,c-b,c;\frac{z}{z-1})
 \eea
we can write 
{\small
\bea\nonumber
{}_{2}F_{1}(-q,-\sfrac{1}{2}-q-s, 1+n+p-s,\frac{z-1}{z})=z^{-q}{}_{2}F_{1}(-q,\sfrac{3}{2}+n+p+q, 1+n+p-s,1-z)\, ,\eea
}
so that 
\bea
\mathcal{R}=\frac{(z-1)^{n+p-s}z^{s}{}_{2}F_{1}(-q,\sfrac{3}{2}+n+p+q, 1+n+p-s,1-z)}{\Gamma(1+q)\Gamma(1+n+p-s)\Gamma(\sfrac{3}{2}+q+s)}\, .\eea
Next, the following identity holds
\bea\nonumber
&&(z-1)^{n+p-s}z^{s}{}_{2}F_{1}(-q,\sfrac{3}{2}+n+p+q, 1+n+p-s,1-z)=\\
&&~~~~~~~~~~~=\frac{(-1)^{n+p+q-s}\Gamma(1+n+p-s)}{\Gamma(1+n+p+q-s)}\frac{1}{\sqrt{z}}\frac{d^q}{dz^q} \Big[(1-z)^{q+n+p-s}z^{\sfrac{1}{2}+q+s}\Big]\, .\eea
Hence, we get 
\bea
\nonumber 
{\rm Disc} (\Omega)&=&-\frac{3i}{\sqrt{z}}\frac{d^q}{dz^q}\frac{\Gamma(p-1)}{\Gamma(q+\sfrac{3}{2})(2p-1)(2q+1)}\sum_{n=0}^{\infty}\frac{ \Gamma(\sfrac{1}{2}+n+p)\Gamma(\sfrac{3}{2}+n+q)}{\Gamma(1+n)} \times \\
&\times&\sum_{s=0}^{p-2}\frac{(-1)^s\,  (1-z)^{q+n+p-s}z^{\sfrac{1}{2}+q+s}}{\Gamma(1+s)\Gamma(\sfrac{5}{2}+s)\Gamma(p-1-s)\Gamma(1+n+p+q-s)}\, .\eea
Further, summing over $s$ results into 
\bea
\nonumber
&&{\mathcal S}=\sum_{s=0}^{p-2}\frac{(-1)^s\,  (1-z)^{q+n+p-s}z^{\sfrac{1}{2}+q+s}}{\Gamma(1+s)\Gamma(\sfrac{5}{2}+s)\Gamma(p-1-s)\Gamma(1+n+p+q-s)}=\\
&&~~~~~~~~~=\frac{4}{3\sqrt{\pi}}\frac{(1-z)^{n+p+q}z^{\sfrac{1}{2}+q}
{}_{2}F_{1}(2-p,-n-p-q, \sfrac{5}{2},\frac{z}{z-1})}{\Gamma(p-1)\Gamma(1+n+p+q)}=\\
\nonumber
&&~~~~~~~~~=\frac{4}{3\sqrt{\pi}}\frac{(1-z)^{n+q+2}z^{\sfrac{1}{2}+q}{}_{2}F_{1}(2-p,\sfrac{5}{2}+n+p+q, \sfrac{5}{2},z)}{\Gamma(p-1)\Gamma(1+n+p+q)}  \, , \eea
where to obtain the last expression we again used the transformation formula  (\ref{tran_F}). Now, taking into account that 
\bea\nonumber
(1-z)^{n+q+2}z^{\sfrac{1}{2}+q}{}_{2}F_{1}(2-p,\sfrac{5}{2}+n+p+q, \sfrac{5}{2},z)=\frac{3\sqrt{\pi}z^{q-1}}{4\Gamma(\sfrac{1}{2}+p)}\frac{d^{p-2}}{dz^{p-2}}\Big[z^{p-\sfrac{1}{2}}(1-z)^{n+p+q}\Big]\, ,
\eea
the expression for $\mathcal{S}$ acquires the form
\bea
{\mathcal S}=\frac{z^{q-1}}{\Gamma(\sfrac{1}{2}+p)\Gamma(p-1)\Gamma(1+n+p+q)}\frac{d^{p-2}}{dz^{p-2}}\Big[z^{p-\sfrac{1}{2}}(1-z)^{n+p+q}\Big]\, .\eea
Thus, for the discontinuity we have 
\bea 
&&{\rm Disc} (\Omega)=-\frac{3i}{\sqrt{z}}\frac{d^q}{dz^q}z^{q-1}\frac{d^{p-2}}{dz^{p-2}}z^{p-\sfrac{1}{2}}(1-z)^{p+q}\times\\
\nonumber
&&\times
\frac{1}{(2p-1)(2q+1)(p+q)!}\sum_{n=0}^{\infty}\frac{\Gamma(\sfrac{3}{2}+q+n)}{\Gamma(q+\sfrac{3}{2})}\frac{ \Gamma(\sfrac{1}{2}+p+n)}{\Gamma(\sfrac{1}{2}+p)}\frac{\Gamma(1+p+q)}{\Gamma(1+p+q+n)}\frac{(1-z)^n}{n!}\, .
\eea
Summing up we finally get the desired formula (\ref{disc_phin_hat}).

\section{From the Borel image $\hat{\phi}_{p,q}$ to its representation $\hat{\Phi}_{p,q}$}\label{app:phi_to_Phi}
\subsection{First proof}
The main ingredient of the formula (\ref{phihat}) is its non-polynomial part represented by the hypergeometric function $\Omega(z)$, {\it cf.} (\ref{eq:Omega}),
where we have introduced a variable $z=x^2$.
To proceed, we will use representation (\ref{diff4F3}), where we analytically continue the function $f$ in the complex plane for the values $|\mbox{Arg}\,(-z)|<\pi$.
The corresponding formula is well known and reads
{\small
\bea\nonumber
f(z)&=&\frac{i}{\Gamma(\sfrac{1}{2}-q)}\Bigg[\frac{(-1)^{q+1} z^{q-\sfrac{1}{2}} }{\Gamma(p+\sfrac{1}{2}) }\sum_{n=0}^{\infty} \frac{ (\sfrac{3}{2}-p)_{n+p-q-1} (\sfrac{1}{2}-p)_{n+p-q-1} }{n!(n+p-q-1)!}z^{-n}\Big(\log(-z)+h(p,q,n)\Big)+\nonumber\\
&&~~~~~~~~~~~~+ (-1)^p z^{p-\sfrac{3}{2}}\sum_{n=0}^{p-q-2}\frac{(\sfrac{3}{2}-p)_n\Gamma(p-q-1-n)}{n!\Gamma(\sfrac{1}{2}+p-n)}z^{-n}\Bigg]\, ,
\eea
}
where 
\bea
h(p,q,n)=\psi(p-q+n)+\psi(n+1)-\psi(\sfrac{1}{2}-q+n)-\psi(\sfrac{3}{2}+q-n)\, .
\eea
Obviously, the function $f(z)$ has a cut on the real axis. Taking into account that $p$ and $q$ are positive integers it is elementary to find the real part for $f(z)$ for $z$ positive.  Using the fact that 
$\log(-z)=\log|z|+i\pi$, we find that
\bea
\Re\, f(z)= \frac{(-1)^{p+q}}{\Gamma(\sfrac{3}{2}-p)\Gamma(\sfrac{1}{2}-q)}\sum_{n=0}^{\infty} \frac{ \Gamma(\sfrac{1}{2}+n-q)\Gamma(-\sfrac{1}{2}+n-q) }{n!\, (n+p-q-1)!}z^{q-n-\sfrac{1}{2}}\, , ~~~~z>0\, .
\eea
As a next step we compute 
\bea
  z^{\ell}\frac{d^{\ell}(\Re  f)}{dz^{\ell}}=\pi\frac{ (-1)^p}{\Gamma(\sfrac{3}{2}-p)\Gamma(\sfrac{1}{2}-q)}\sum_{n=0}^{\infty} \frac{(-1)^n\Gamma(n-q-\sfrac{1}{2})}{n! \, \Gamma(n+p-q)\Gamma(\sfrac{1}{2}-\ell-n+q)}z^{q-n-\sfrac{1}{2}} \, . \eea
Substituting this result into the real part of (\ref{diff4F3}) and replacing the regularised hypergeometric function via its normal counterpart, we obtain 
\bea  
\label{diff4F30} 
\Re\, \Omega =\frac{ (-1)^p}{\Gamma(\sfrac{3}{2}-p)\Gamma(\sfrac{1}{2}-q)}\sum_{n=0}^{\infty} \frac{(-1)^n\Gamma(n-q-\sfrac{1}{2})}{n! \, \Gamma(n+p-q)}\, S(n) \, z^{q-n-\sfrac{1}{2}}  \, , \nonumber
\eea
where we need to compute the following sum
\bea\label{Sn}
S(n) =\varkappa\sum_{\ell=0}^{p+q-2} \frac{~{}_{3}F_{2}\left( \left\{ \sfrac{3}{2}+\ell, 2+\ell-p-q, -q    \right\}, \left\{ 1+\ell-q,  \sfrac{5}{2}+\ell-q \right\}, 1 \right)}
{\Gamma(1+\ell-q)\Gamma(\sfrac{5}{2}+\ell-q)\Gamma(\sfrac{1}{2}-\ell-n+q)\Gamma(p+q-1-\ell)} \, ,\eea
where the coefficient $\varkappa$ is chosen for convenience to be 
\bea
\varkappa=\frac{3\pi^2}{8}\frac{\Gamma(p-1)}{\Gamma(q+\sfrac{3}{2})}\,  .
\eea
Here for the hypergeometric function we can substitute its definition
\bea
&&~{}_{3}F_{2}\left( \left\{ \sfrac{3}{2}+\ell, 2+\ell-p-q, -q    \right\}, \left\{ 1+\ell-q,  \sfrac{5}{2}+\ell-q \right\}, 1 \right)=\\
\nonumber
&&~=
\sum_{r=0}^q \frac{(-1)^r}{r!}\frac{\Gamma(r-q)}{\Gamma(-q)}\frac{\Gamma(\sfrac{3}{2}+\ell+r)}{\Gamma(\frac{3}{2}+\ell)}\frac{\Gamma(1+\ell-q)}{\Gamma(1+\ell-q+r)}\frac{\Gamma(\sfrac{5}{2}+\ell-q)}{\Gamma(\sfrac{5}{2}+\ell-q+r)}
\frac{\, \Gamma(p+q-1-\ell)}{\Gamma(p+q-1-\ell-r)}\, .
\eea
After the change of order of summation eq.(\ref{Sn}) acquires the form
\bea\label{Sn1}
&&S(n) =\varkappa \sum_{r=0}^q  (-1)^r \frac{\Gamma(r-q)}{r!\Gamma(-q)}\times \\
\nonumber
&&\times 
\sum_{\ell=q-r}^{p+q-2-r}\frac{\Gamma(\sfrac{3}{2}+\ell+r)}{\Gamma(\sfrac{3}{2}+\ell)\Gamma(\sfrac{1}{2}-\ell-n+q)\Gamma(1+\ell-q+r)\Gamma(\sfrac{5}{2}+\ell-q+r)\Gamma(p+q-1-\ell-r)}\, ,
\eea
where the restrictions on the summation variable $\ell$ are clear from the arguments of the $\Gamma$-functions entering the denominators of the second sum.
We further shift the sum variable $\ell$ as $\ell=q-r+s$,  and get the double sum, which we write  in the following order
\bea\nonumber
&&S(n) =\frac{\varkappa}{\pi} (-1)^n \sum_{s=0}^{p-2}\frac{(-1)^{s}\Gamma(\sfrac{3}{2}+q+s)}{ s!\Gamma(\sfrac{5}{2}+s)\Gamma(p-1-s)} \sum_{r=0}^q\frac{\Gamma(\sfrac{1}{2}+n+s-r)}{\Gamma(\sfrac{3}{2}+q+s-r)}\frac{q!\, (-1)^r  }{r! (q-r)!}
\, .
\eea
The internal sum is given by 
\bea
\nonumber
\sum_{r=0}^q\frac{\Gamma(\sfrac{1}{2}+n+s-r)}{\Gamma(\sfrac{3}{2}+q+s-r)}\frac{q!\, (-1)^r }{r! (q-r)!}= \frac{\Gamma(\sfrac{1}{2}+n+s)}{\Gamma(\sfrac{3}{2}+q+s)}\frac{\Gamma(\sfrac{1}{2}-n-s)}{\Gamma(\sfrac{1}{2}+q-n-s)}\frac{\Gamma(2q+1-n)}{\Gamma(q+1-n)}  \, .
\eea
Finally, to perform the last sum over $r$ we have to carefully distinguish two cases: $q\neq 0$ and $q=0$. We treat these cases in turn.
\begin{enumerate}[label=\it{\arabic{*}})~]
\item Case $q\neq 0$. We have 
\bea
S(n)=\frac{3\pi^2}{8}\frac{1}{\Gamma(\sfrac{1}{2}+p)\Gamma(\frac{3}{2}+q)}\left\{ \begin{tabular}{l}  $ \frac{\Gamma(p+q-n)\Gamma(1+2q-n)}{\Gamma(\sfrac{1}{2}+q-n)\Gamma(1+q-n)\Gamma(2+q-n)}$\, ,~~~~~$0\leq n\leq q$\, ;  \\
0  ,~~~~~$q+1\leq n\leq p+q-1$\, ;  \\
$(-1)^{p+q}  \frac{\Gamma(n-1-q)\Gamma(n-q)}{\Gamma(n-2q)\Gamma(1+n-p-q)\Gamma(\sfrac{1}{2}-n+q)} $\, ,  ~~~$n\geq p+q$.
\end{tabular}
\right.  
\eea
Now we are ready to compute the real part of $\Omega$ 
\bea
&&\Re\Omega =\frac{3}{2}\frac{(-1)^q}{(1-2p)(1+2q)}\times \\
&& \times \Bigg[z^{-\sfrac{1}{2}}\sum_{n=0}^{q} \frac{(-1)^n\Gamma(n-q-\sfrac{1}{2})}{n! \, \Gamma(n+p-q)}\, \frac{\Gamma(p+q-n)\Gamma(1+2q-n)}{\Gamma(\sfrac{1}{2}+q-n)\Gamma(1+q-n)\Gamma(2+q-n)} \, z^{q-n}  + \nonumber \\
&&~~~~(-1)^{p+q}  z^{-\sfrac{1}{2}}\sum_{n=p+q}^{\infty} \frac{(-1)^n\Gamma(n-q-\sfrac{1}{2})}{n! \, \Gamma(n+p-q)}\,\frac{\Gamma(n-1-q)\Gamma(n-q)}{\Gamma(n-2q)\Gamma(1+n-p-q)\Gamma(\sfrac{1}{2}-n+q)} \, z^{q-n}\Bigg]\, , \nonumber \eea
Changing the summation indices appropriately, we find the final answer 
{\small
\bea\label{RO1}
&&\Re\Omega(z) =\frac{1}{(2p-1)(2q+1)}\Bigg[3\,  z^{-\sfrac{1}{2}} ~{}_{4}F_{3}\left( \left\{ 1-p, p,-q,1+q \right\}, \left\{ \sfrac{1}{2}, \sfrac{3}{2}, 2\right\}, z \right) \\
&&- z^{-\sfrac{1}{2}-p}\frac{3\times 2^{3-4p}(-1)^{p+q} \Gamma(2p-2)}{\Gamma(p-q)\Gamma(p+q+1)}~{}_{4}F_{3}\left( \left\{ p-1, p-\sfrac{1}{2},p,p+\sfrac{1}{2} \right\}, \left\{ 2p, p-q, p+q+1\right\}; \sfrac{1}{z} \right)
\Bigg]\, . \nonumber\eea
}

Now we recall that function $\Re \, \hat{\phi}_{p,q}$ reads as 
{\small
\bea \label{phihat2} 
   \Re\, \hat{\phi}_{p,q}(x) &=& \frac{4}{3} (p-q-1)(p+q)x^2 \times \\
   &\times&\Big[3~{}_{4}F_{3}\left( \left\{ 1-p, p, -q, 1+q \right\}, \left\{ \sfrac{1}{2}, \sfrac{3}{2}, 2 \right\}, x^2 \right) -  (2p-1)(2q+1)\, x~\Re\, \Omega(x^2) \Big], \nonumber
\eea
}
Substituting here (\ref{RO1}) we find that the polynomial part cancels out completely and we are left with the desired result (\ref{main_coin}).

\item Case $q=0$.  In this situation we have 
\bea
S_{q=0}(n)=\frac{3\pi^2}{8}\frac{1}{\Gamma(\sfrac{1}{2}+p)\Gamma(\frac{3}{2})}\left\{ \begin{tabular}{l}  $ \frac{\Gamma(p-n)}{\Gamma(\sfrac{1}{2}-n)\Gamma(2-n)}$\, ,~~~~~$n=0,1$;  \\
0  ,~~~~~$2\leq n\leq p-1$\, ;  \\
$(-1)^{p}  \frac{\Gamma(n-1)}{\Gamma(1+n-p)\Gamma(\sfrac{1}{2}-n)} $\, ,  ~~~$n\geq p$.
\end{tabular}
\right.  
\eea
We therefore find 
\bea
\Re\Omega &=&\frac{3}{2}\frac{1}{1-2p} \Bigg[z^{-\sfrac{1}{2}}\sum_{n=0}^{1} \frac{(-1)^n\Gamma(n-\sfrac{1}{2})}{n! \, \Gamma(n+p)}\, \frac{\Gamma(p-n)}{\Gamma(\sfrac{1}{2}-n)\Gamma(2-n)} \, z^{-n}  + \nonumber \\
&&~~~~~~~~~~~~~~~~+(-1)^{p}  z^{-\sfrac{1}{2}}\sum_{n=p}^{\infty} \frac{(-1)^n\Gamma(n-\sfrac{1}{2})}{n! \, \Gamma(n+p)}\,\frac{\Gamma(n-1)}{\Gamma(1+n-p)\Gamma(\sfrac{1}{2}-n)} \, z^{-n}\Bigg]\, , \nonumber \eea
which gives 
\bea
\Re\Omega &=&\frac{1}{2p-1} \Bigg[3z^{-\sfrac{1}{2}}-\frac{3}{4p(p-1)}z^{-\sfrac{3}{2}} \\
&&- z^{-\sfrac{1}{2}-p}\frac{3\times 2^{3-4p}(-1)^{p} \Gamma(2p-2)}{\Gamma(p)\Gamma(p+1)}~{}_{3}F_{2}\left( \left\{ p-1, p-\sfrac{1}{2},p+\sfrac{1}{2} \right\}, \left\{ 2p, p+1\right\}; \sfrac{1}{z} \right)
\Bigg]\, . \nonumber\eea
Then we specify (\ref{phihat2}) for $q=0$ and get 
\bea \label{phihat1} 
&&   \Re\, \hat{\phi}_{p,q}(x) = \frac{4}{3} p(p-1)x^2 \times \nonumber\\
 &&~~~\times\Bigg[3- \, x~\Big[3x^{-1}-\frac{3}{4p(p-1)}x^{-3} \\
&&~~~ - x^{-1-2p}\frac{3\times 2^{3-4p}(-1)^{p} \Gamma(2p-2)}{\Gamma(p)\Gamma(p+1)}~{}_{3}F_{2}\left( \left\{ p-1, p-\sfrac{1}{2},p+\sfrac{1}{2} \right\}, \left\{ 2p, p+1\right\}; \sfrac{1}{x^2} \right)
\Big] \Bigg], \nonumber
\eea
which finally boils down to 
\bea
  \Re\, \hat{\phi}_{p,q}(x) &=&1+ (-1)^{p}2^{5-4p} p(p-1)\frac{\Gamma(2p-2)}{\Gamma(p)\Gamma(p+1)}\times \\
                                        &\times & x^{2-2p}~{}_{3}F_{2}\left( \left\{ p-1, p-\sfrac{1}{2},p+\sfrac{1}{2} \right\}, \left\{ 2p, p+1\right\}; \sfrac{1}{x^2} \right) \, . \nonumber
\eea
\end{enumerate}
Thus, we have proved that in all the cases $\Re\, \hat{\phi}_{p,q}$ is equivalent to eq.(\ref{BES_c}).

\subsection{Second proof}

Another approach is based on the Mellin-Barnes integral representation \cite{NIST} for the hypergeometric function \eqref{eq:Omega}:
\small{
\bea\label{eq:Mellin}
  \frac{\prod_{k=1}^4\G(a_k)}{\prod_{k=1}^3 \G(b_k)} {}_{4}F_{3}(\mathbf{a}, \mathbf{b}, z) = \frac{1}{2\pi i} \int\limits_{-i\infty}^{i\infty}  \frac{\prod_{i=1}^4 \G(a_i + s)}{\prod_{i=1}^3 \G(b_i + s)} \G(-s) (-z)^s \mathrm{d}s,
\eea
}
\normalsize
where the integration contour separates the poles of $\G(a_k+s)$, k=1,...,4, from those of $\G(-s)$. The right-hand side of \eqref{eq:Mellin} provides the analytic continuation of the left-hand side from the open unit disk to the sector $|\mbox{Arg}\,(1-z)| < \pi$. In our case parameters are: $b_1 = \frac{3}{2}$, $b_2 = 2$, $b_3 = \frac{5}{2}$ and $a_{i+1} = a_i + m_i$, $i=1,2,3$, with $a_1 = \frac{3}{2} - p$, $m_1 = p-q-1$, $m_2=2q+1$, $m_3 = p-q-1$. Since all $m_i \in \mathbb{N}$, the function of integration in \eqref{eq:Mellin} has the following poles:
\begin{itemize}
 \item first order poles $s=-a_1-k$, $k=0,1,2,...,m_1-1$;
 \item second order poles $s=-a_2-k$, $k=0,1,2,...,m_2-1$;
 \item third order poles $s=-a_3-k$, $k=0,1,2,...,m_3-1$;
 \item fourth order poles $s=-a_4-k$, $k=0,1,2,...,\infty$.
\end{itemize}
The method is thus build on the further calculation of the residues of the integrand at each pole. The full expression for the residues includes a huge amount of summands. Therefore, we present only the needed real part which comes from the terms of the form $(-z)^{1/2+n}\log(-z)$, $n\in \mathbb{N}$.
\subsubsection*{The $1^{\text{st}}$ order poles}
In this case the residue is given by 
\small{
\bea
 \text{Res}_1 = (-1)^k (-z)^{-\frac{3}{2}+p-k}\frac{\G(\frac{3}{2}+k-p)\G(2p-k-1)\G(p-q-1-k)\G(p+q-k)}{k! \G(p-k)\G(p-k+\frac{1}{2})\G(p-k+1)}
\eea
}
\normalsize
and is purely imaginary. Thus, $\Re\, \text{Res}_1 = 0$.

\subsubsection*{The $2^{\text{nd}}$ order poles}
At the $k-$th pole, $0\leq k \leq 2q$, we have the following expression for the real part of the residue:
\small{
\bea
 \Re\, \text{Res}_2 = (i\pi)(-1)^{p-q+1} (-z)^{-\frac{1}{2}-k+q}\frac{\G(\frac{1}{2}+k-q)\G(p+q-k)\G(2q-k+1)}{k!(p-q+k-1)!\G(q-k+1)\G(\frac{3}{2}+q-k)\G(q-k+2)}
\eea
}
\normalsize
Note, that the residue is non-zero only for $0 \leq k \leq q$. After changing factorials by Gamma functions, transforming $k\to q-s$, $0 \leq s \leq q$, and using $\G(1-z)=\pi/(\G(z)\sin(\pi z))$, one gets
\small{
\bea
  \Re\, \text{Res}_2 = 2 (-1)^{p-q+1} \pi z^{-\frac{1}{2}}\frac{(1-p)_s (p)_s (-q)_s (1+q)_s}{(1/2)_s (3/2)_s (2)_s}\frac{z^s}{\G(1+s)}
\eea
}
\normalsize
Summing over all the second order poles ($0 \leq s \leq q$), we arrive at
\small{
\bea
  \Re\, \text{Res}_2 = 2 (-1)^{p-q+1} \pi z^{-\frac{1}{2}} {}_{4}F_{3}\left( \left\{ 1-p, p, -q, 1+q \right\}, \left\{ \sfrac{1}{2}, \sfrac{3}{2}, 2 \right\}, z \right)
\eea
}\normalsize
Putting all the coefficients from \eqref{eq:Mellin} and \eqref{phihat}, one gets the polynomial part in $\Re\,\hat{\p}_{p,q}(x)$ with a minus sign. This means that the first term in \eqref{phihat} is cancelled by this residue term.   

\subsubsection*{The $3^{\text{rd}}$ order poles}
Here $0\leq k \leq p-q-2$ and the the real part of the residue at the $k-$th pole is given by
\small{
\bea
 \Re\, \text{Res}_3&=&  \frac{(i \pi)(-1)^{p+q-1} (-z)^{-\frac{3}{2}-k-q}}{\G(-k-q)\G(1-k-q)} \frac{\G(-1-k+p-q)\G(\frac{3}{2}+k+q)}{k!(k+p+q)!(1+k+2q)!\G(\frac{1}{2}-k-q)}\times\nonumber\\
 &\times& \left(\log(z)+\psi(1+k)-\psi(-k-q)-\psi(\frac{1}{2}-k-q)-\psi(1-k-q)+\right.\\
 &+&\left.\psi(-1-k+p-q)-\psi(\frac{3}{2}+k+q)+\psi(1+k+p+q)+\psi(2+k+2q)\right).\nonumber
\eea
}
\normalsize

\vskip -0.3cm
\noindent
Note that  $1/\infty^2$ behavior of $1/(\G(-k-q)\G(1-k-q))$ is not compensated by the rest part. This leads to $\Re\, \text{Res}_3=0$.

\subsubsection*{The $4^{\text{th}}$ order poles }

In this case the expression for the real part of the residue $\Re\, \text{Res}_4$ is huge. Nevertheless, again thanks to the $\G(1-k-p)\G(2-k-p)$ factor in the denominator most of the terms vanish. The rest is given by
\small{
\bea
 \Re\, \text{Res}_4 &=& (i\pi)(-z)^{-\frac{1}{2}-k-p} \frac{\G(\frac{1}{2}+k+p)}{\G(\frac{3}{2}-k-p)} \frac{1}{2k!(-1+k+2p)!(-1+k+p-q)!(k+p+q)!}\times\nonumber\\
 &\times&\frac{(\psi(1-k-p)+\psi(2-k-p))^2-(\psi'(1-k-p)+\psi'(2-k-p))}{\G(1-k-p)\G(2-k-p)}.
\eea
}
\normalsize

\vskip -0.3cm
\noindent
This can be simplified with the help of the following identities: $\psi(1-z)=\pi \cot(\pi z)+ \psi(z)$, $\psi'(1-z) = \pi^2/\sin^2(\pi z) - \psi'(z)$ and $\G(1-z)=\pi/(\G(z)\sin(\pi z))$:
{\small{
\bea
 \Re\, \text{Res}_4 = z^{-\frac{1}{2}-p}\frac{\G(k+p-1)\G(k+p-\frac{1}{2})\G(k+p+\frac{1}{2})\G(k+p)}{\G(k+2p)\G(k+p-q)\G(k+p+q+1)} \frac{z^{-k}}{k!}
\eea
}}
Summing over all the forth order poles ($0 \leq k < \infty$), taking into account all the coefficients from \eqref{eq:Mellin} and \eqref{phihat}, one comes to the desired expression for $\Re\,\hat{\Phi}_{p,q}$ which proves \eqref{main_coin} for $q\neq 0$.

\subsubsection*{$q=0$}

Strictly speaking, the direct substitution $q=0$ into the results obtained above leads to the wrong answer. In this case one has to perform the same procedure from the very beginning. This happens because \eqref{eq:Omega} reduces to ${}_3F_2(\{\sfrac{3}{2}-p,\sfrac{1}{2}, p+\sfrac{1}{2}\}, \{2, \sfrac{5}{2}\}, z)$ and now we have a different system of poles. The final result gives an additional factor 1 which completely corresponds to \eqref{main_coin}.

\section{Derivation of $Q$}\label{app:derivation_of_Q}
In Section~\ref{sec:disc} we introduced the following function 
\bea\label{c1}
Q(z)=\sum_{k=0}^{\infty} \frac{(-h)^{k+1}}{k!} z^{p-\sfrac{1}{2}}\frac{d^{p-2}}{dz^{p-2}}z^{q-1}\frac{d^q}{dz^q}\,   z^{\frac{1}{2}(k+1)} \, .
\eea
Performing straightforward differentiations and then summation over $k$ we arrive at the following formula 
\bea
\label{c2}
Q(z)=-\frac{\pi h z {}_{1}F_{2}(\{\sfrac{3}{2}\},\{\sfrac{5}{2}-p, \sfrac{3}{2}-q\}, \sfrac{h^2z}{4})}{2\Gamma(\sfrac{5}{2}-p)\Gamma(\sfrac{3}{2}-q)}+\frac{h^2 z^{3/2} {}_{2}F_{3}(\{1,2\},\{\sfrac{3}{2},3-p,2-q\}, \sfrac{h^2z}{4})}{\Gamma(3-p)\Gamma(2-q)}\, .\eea
Further one can show that for $p\geq q+2$, $q\geq 0$, the following identity takes place 
\bea\nonumber
\frac{h^2 z^{3/2} {}_{2}F_{3}(\{1,2\},\{\sfrac{3}{2},3-p,2-q\}, \sfrac{h^2z}{4})}{\Gamma(3-p)\Gamma(2-q)}=-\frac{4\sqrt{\pi}\Big(\frac{h^2 z}{4}\Big)^{p-\sfrac{1}{2}}}{h}\frac{\Gamma(p){}_{1}F_{2}(\{p\},\{p-\sfrac{1}{2}, p-q\}, \sfrac{h^2z}{4})}{\Gamma(p-\sfrac{1}{2})\Gamma(p-q)}\, .
\eea
Thus, $Q(z)$ is essentially written as the sum of two ${}_{1}F_{2}$ functions that both have the same characteristic feature. Namely, if the upper parameter is $\rho+q$, where $q\geq 0$, then among the lower parameters there is $\rho$.
In this situation we can apply the following reduction formula, where ${}_{1}F_{2}$ gets replaced by a finite sum of ${}_{0}F_{1}$, the latter being expressed via the modified Bessel function of the first kind $I_{\nu}$,
\bea
\label{c2}
\begin{aligned}
Q(z)&=-\frac{\pi}{h}\sum_{k=0}^q 2^{\sfrac{1}{2}-k-p}(h\sqrt{z})^{\sfrac{1}{2}+k+p}\frac{\Gamma(1+q)}{\Gamma(1+k)\Gamma(1+q-k)\Gamma(\sfrac{3}{2}+k-q)}\, I_{\sfrac{3}{2}+k-p}(h\sqrt{z})\\
&+\frac{\sqrt{\pi}}{h}\sum_{k=0}^q 2^{\sfrac{3}{2}-k-p}(h\sqrt{z})^{\sfrac{1}{2}+k+p}\frac{\Gamma(1+q)\Gamma(p)}{\Gamma(1+k)\Gamma(1+q-k)\Gamma(p+k-q)}\, I_{-\sfrac{3}{2}+k+p}(h\sqrt{z})\, .
\end{aligned}
\eea
Note that due to our restrictions on the range of $p$ and $q$ the index $\nu$ of the first Bessel function $I_{\sfrac{3}{2}+k-p}$ is always negative, while the index of the second one, $I_{-\sfrac{3}{2}+k+p}$, is always positive.
Moreover, the index always takes half-integer values which means that $I_{\nu}$ can be written via elementary functions. Indeed, let us introduce the following auxiliary functions 
\bea
\begin{aligned}
{\mathscr I}^e_n(x)&=\sum_{m=0}^{\big[\sfrac{n}{2}\big]} \frac{\Gamma(1+n+2m)}{(2m)! \, \Gamma(1+n-2m)} (2x)^{-2m}\, , \\
{\mathscr I}^o_n(x)&=\sum_{m=0}^{\big[\sfrac{n-1}{2}\big]} \frac{\Gamma(2+n+2m)}{(2m+1)! \, \Gamma(n-2m)} (2x)^{-2m-1}\, , \\
{\mathscr I}_n(x)&=\sum_{m=0}^{n} \frac{\Gamma(n+m+1)}{m! \, \Gamma(n+1-m)} (2x)^{-m}=e^x \sqrt{\frac{2x}{\pi}}\, K_{\sfrac{1}{2}+n}(x)\, , \end{aligned}
\eea
where $[n]\equiv {\rm Floor}(n)$ and $K_{\nu}(x)$ is the Macdonald function. Then for an integer $n\geq 0$ we have 
\bea
\label{mBrep}
\begin{aligned}
I_{\sfrac{1}{2}+n}(x)&=e^{-x}(-1)^{n+1}\frac{{\mathscr I}^e_n(x)+ {\mathscr I}^o_n(x)   }{\sqrt{2\pi x}}+e^x \frac{{\mathscr I}^e_n(x)-{\mathscr I}^o_n(x)   }{\sqrt{2\pi x}}\, , \\
I_{-\sfrac{1}{2}-n}(x)&=e^{-x}(-1)^{n+1}\frac{{\mathscr I}^e_n(x)+ {\mathscr I}^o_n(x)  - 2 {\mathscr I}_n(x) }{\sqrt{2\pi x}}+e^x \frac{{\mathscr I}^e_n(x)-{\mathscr I}^o_n(x)   }{\sqrt{2\pi x}}\, .
\end{aligned}
\eea
First, we note that each individual $I_{\nu}$ involves the growing exponent, $e^x$, as $x\to \infty$. On the other hand, as is obvious from (\ref{DSQ}), these terms cannot appear in the final answer for $Q(z)$.
Thus, upon summing up they all must cancel. Second, concerning the terms with the damping exponent $e^{-x}$, our numerical analysis shows that the terms involving the functions ${\mathscr I}^e_n(x)$
and ${\mathscr I}^o_n(x)$ all cancel in the sum, so that the only contribution left comes from ${\mathscr I}_n(x)$. In this way we find 
\bea\nonumber
Q(z)=\frac{\pi}{h}\sum_{k=0}^q (h\sqrt{z})^{\sfrac{1}{2}+k+p}\frac{2^{\sfrac{3}{2}-k-p}\Gamma(1+q)}{\Gamma(1+k)\Gamma(1+q-k)\Gamma(\sfrac{3}{2}+k-q)}\, \frac{e^{-h\sqrt{z}}(-1)^{p+k+1}{\mathscr I}_{p-k-2}(h\sqrt{z})}{\sqrt{2\pi h\sqrt{z}}}\, ,\eea
where upon substituting the series representation for ${\mathscr I}_{p-k-2}(h\sqrt{z})$ and replacing $h\to h_n$, we obtain the desired formula (\ref{QDS}). Note also that in terms of $K_{\nu}$, the above formula reads as
\bea
Q(z)=\frac{1}{h}\sum_{k=0}^q(-1)^{p+k+1}2^{\frac{3}{2}-k-p}(h\sqrt{z})^{\frac{1}{2}+p+k}\frac{K_{p-k-\frac{3}{2}}(h\sqrt{z})\Gamma(1+q)}{ \Gamma(1+k)\Gamma(1+q-k)\Gamma(\frac{3}{2}+k-q)}\, .\nonumber
\eea

\section{Details for the construction of the asymptotic expansion for  $\Delta S_{p,q}(g)$}
\label{app:cl}
Here we resolve several technical issues concerning construction of the asymptotic expansion of the discontinuity $\Delta S_{p,q}(g)$
and also present an alternative method to derive the same asymptotic expansion.

\subsection{Solution of the difference equation for $c_\ell$}
The coefficients $c_{\ell}$ which arise in the asymptotic expansion of the function ${}_{2}F_{3}$ can be determined recurrently by using Riney's method.
The corresponding recurrence formula reads \cite{NIST}
\bea
c_0=1\, , ~~~~c_\ell=-\frac{1}{4\ell}\sum_{k=0}^{\ell-1}c_k e_{\ell, k}\, ,
\eea
where 
\bea
e_{\ell, k}=\sum_{j=1}^{4} \frac{\G(1-\nu-2b_j+\ell+2)}{\G(1-\nu-2b_j+k)}\frac{\displaystyle{\prod_{i=1}^2}{\textstyle{ (a_{i}-b_j)}  }}{\displaystyle{\prod_{i=1,i\neq j}^4}{\textstyle{(b_i-b_j)}}}\, , ~~~~b_4\equiv 1\, .
\eea
Here the coefficients $a_1,\ldots ,b_3$ are given by (\ref{aabbb}) and $\nu=a_1+a_2-b_1-b_2-b_2+\tfrac{1}{2}=-2-p-q-2\eps$. 
For our purposes, however,  this recurrence formula is not enough as we need to determine these coefficients in the closed form,
{\it i.e.} without referring to any recurrence procedure. 

\smallskip

We start our analysis with some observations. First,
computing $e_{\ell,k}$ explicitly 
we note that they do not depend on $\eps$ and, as a consequence, $c_{\ell}$ are also $\eps$ independent. Second, the coefficients $c_{\ell}$
satisfy certain difference equations. To understand this issue, consider the differential equation for the function ${}_{2}F_{3}$:
\bea
\nonumber
\Big(\vartheta(\vartheta+b_1-1)(\vartheta+b_2-1)(\vartheta+b_3-1)-t(\vartheta+a_1)(\vartheta+a_2)\Big){}_{2}F_{3}(\{a_1,a_2,a_3\},\{b_1,b_2\},t)=0\, ,
\eea
where  $\vartheta=t\frac{d}{dt}$. To derive the difference equations for $c_{\ell}$, it is enough to substitute in this equation the part of the asymptotic expansion for ${}_{2}F_{3}$
which contains either damping or growing exponent 
\bea
\mathscr{F}_3^{-} = t^{\tfrac{\nu}{2} }\, e^{i\pi\nu-2\sqrt{t}}\,  \sum_{\ell=0}^\infty \frac{ (-1)^\ell c_\ell}{ 2^{\ell} \, t^{\ell/2}   }\, , ~~~~~~
\mathscr{F}_3^{+} = t^{\tfrac{\nu}{2} }\, e^{2\sqrt{t}}  \,            \sum_{\ell=0}^{\infty}            \frac{c_\ell}{2^{\ell} \,  t^{\ell/2}   } \, .
\eea
Both lead to the same difference equations, so it is enough to consider only  one of them. Substituting for instance $\mathscr{F}_3^{-}$, we get an expression which contains 
$\eps$-independent term and the term proportional to $\eps$. Since the original equation is valued for arbitrary $\eps$ these terms  must separately vanish. Collecting terms 
proportional to $t^{\ell/2}$ in the first, $\eps$-independent, term we get the following difference equation:
\bea
&&(-1+\ell+p-q)(1+\ell-p+q)(1+\ell+p+q)^2c_{\ell-1}- \nonumber\\
&&~~-\big(1+4\ell^3-2p^3+2p^2(1+q)-2q(1+q)(2+q)+2p(2+q)^2+ \nonumber \\
&&~~~~~~~~~~~~~~~~+6\ell^2(2+p+q)+2\ell(5+3q+p(7+4q)) \big)c_{\ell}+ \label{diff_eq_1}\\
&&~~+\big(11+5\ell^2+7p-(p-q)^2+3q+\ell(15+4(p+q))\big)c_{\ell+1}-2(2+\ell)c_{\ell+2}=0  \nonumber \, .
\eea
From the second term proportional to $\eps$ we find a simpler difference equation, namely,
{\small
\bea\label{diff_eq}
&&(-1+\ell+p-q)(1+\ell-p+q)(1+\ell+p+q)c_{\ell-1}+\\
&&~~~~~~~~~~~~~~+(-3\ell^2-\ell(5+2p+2q)+(p-q)^2-3p+q-1)c_\ell+2(1+\ell)c_{\ell+1}=0\, . \nonumber
\eea 
}
In fact, the second equation (\ref{diff_eq}) implies the first. Shifting in eq.(\ref{diff_eq}) the variable $\ell\to \ell+1$, solving for $c_{\ell+2}$
and plugging this solution into (\ref{diff_eq_1}), we observe that the last equation factorises and it contains the left hand side of eq.(\ref{diff_eq}) as a factor.
Thus, fulfilment of (\ref{diff_eq}) implies the fulfilment of (\ref{diff_eq_1}).

\smallskip

Now we explain how to find a closed formula for the coefficients $c_{\ell}$. It is not difficult to see that these coefficients must arise in the large $s$-expansion 
of the following ratio of the gamma functions
\bea\label{expE}
\frac{\G(s+\frac{1}{2})}{\Gamma(s+1)\Gamma(s+p+1)\Gamma(s+q+2)}\sim \sum_{\ell=0}^{\infty}
\frac{2^{2s+5/2+p+q}}{(2\pi)^{1/2}\G(2s+3+p+q+j)}\, c_{\ell}\, .
\eea
Indeed, according to the discussion in chapter 2.2.2. by \cite{Paris_Kaminski},  the numbers $c_{\ell}$ can be computed from the following recursion formula 
\bea
c_{\ell}=-\frac{1}{4\ell}\sum_{k=0}^{\ell-1}c_k E_{\ell,k}\, ,
\eea
where 
\bea
E_{\ell,k}=\sum_{j=1}^3 \frac{\Gamma(5+p+q-2b_j)}{\G(3+p+q-2b_j)}   \frac{ (1/2-b_j) }{ \prod_{i=1}^{'3} {\textstyle (b_i-b_j)}    }\, .
\eea
Here $b_1=1+p$, $b_2=2+q$, $b_3=1$, and   
 the prime signifies omission of the term with $i=j$. Computing recurrently the first  few coefficients 
\bea
c_0&=&1\, , \nonumber \\
c_1&=&\sfrac{1}{2}(1-p^2-q(1+q)+p(3+2q)) \, , \nonumber \\
c_2&=&\sfrac{1}{8}(9+p^4+(q-1)q^2(3+q)-2p^3(3+2q)- \\
&&~~~~~~~~~~~~~~~
-2p(3+2q)(-4+q+q^2)+p^2(5+2q(7+3q))) \, , \nonumber\\
&& \ldots \, ,    \nonumber
\eea
we verify that they form a sequence satisfying the difference equation (\ref{diff_eq}). It is however unknown how to 
produce an expansion of the left hand side of eq.(\ref{expE}) in a way which would allow one to read off the closed 
formula for an arbitrary coefficient $c_{\ell}$. What is however known in the closed form is the following asymptotic expansion 
\bea
\frac{\G(s+\frac{1}{2})}{\Gamma(s+1)\Gamma(s+p+1)\Gamma(s+q+2)}\sim \sum_{j=0}^{\infty}\frac{\upsilon(j)}{\Gamma(s+5/2+p+q+j)\Gamma(s+1)}
\eea
with the coefficients 
\bea\upsilon(j)=\frac{(1/2+p)_j(3/2+q)_j}{j!} \, ,
\eea
see formula (2.2.39) in \cite{Paris_Kaminski}. At this point it is natural to  use 
the large $s$ asymptotic expansion of the inverse product of two gamma functions, {\it cf.} eq.(2.2.34) in \cite{Paris_Kaminski}, 
\bea
\frac{1}{\Gamma(s+5/2+p+q+j)\Gamma(s+1)}\sim \frac{2^{2s+5/2+p+q+j}}{(2\pi)^{1/2}}\sum_{k=0}^{\infty}\frac{\sigma(k,j)}{\Gamma(2s+3+p+q+j+k)}
\eea
where 
\bea
\sigma(k,j)&=&\frac{(-2)^{-k}}{k!}\prod_{r=1}^k \Big( (\sfrac{3}{2}+p+q+j)^2-(r-\sfrac{1}{2})^2\Big)= \nonumber \\
&=&\frac{(-1-j-p-q)_k \, (2+j+p+q)_k}{2^k\, k!}\, .
\eea
Thus, we arrive at the following double sum representation 
{\small
\bea
\frac{\G(s+\frac{1}{2})}{\Gamma(s+1)\Gamma(s+p+1)\Gamma(s+q+2)}\sim \sum_{j=0}^{\infty}\sum_{k=0}^{\infty} \frac{2^{2s+5/2+p+q}}{(2\pi)^{1/2}} \frac{2^j \upsilon(j)\sigma(k,j)}{\Gamma(2s+3+p+q+j+k)}\, . \eea
}
\normalsize

\vskip -0.1cm
\noindent
This expansion is to be compared with (\ref{expE}). To this end we make a change of the summation variables $j+k=\ell$ and get 
{\small
\bea
\frac{\G(s+\frac{1}{2})}{\Gamma(s+1)\Gamma(s+p+1)\Gamma(s+q+2)}&\sim& \nonumber \\
&&\hspace{-3cm} \sim \sum_{\ell=0}^{\infty}\frac{2^{2s+5/2+p+q}}{(2\pi)^{1/2}\Gamma(2s+3+p+q+\ell)}
\sum_{j=0}^{\ell} \, 2^j \upsilon(j)\sigma(\ell-j,j)\, . \eea
}
\normalsize

\vskip -0.3cm
\noindent
In this way  we get an explicit formula for the coefficients $c_{\ell}$:
\bea
c_{\ell}=\sum_{j=0}^{\ell} \, 2^j \upsilon(j)\sigma(\ell-j,j)\, .
\eea
Substituting here the corresponding coefficients and performing the summation over $j$ we arrive at the following result
{\small
\bea
c_{\ell}&=&(-1)^\ell \frac{\Gamma(2+\ell+p+q)}{\G(\frac{1}{2}+p)\G(\frac{3}{2}+q)}
\sum_{j=0}^{\ell}\frac{(-1)^{j}\, 2^{2j-1}\, \G(\frac{1}{2}+j+p)\Gamma(\frac{3}{2}+j+q)}
{\Gamma(1+\ell-j)\Gamma(1+j)\G(2+2j-\ell+p+q)}=
\label{sum_rep_cl}
\\
\nonumber
&=&\frac{\Gamma(2+\ell+p+q)_3F_2\left( \left\{ -\ell, \sfrac{1}{2}+p, \sfrac{3}{2}+q \right\}, \left\{ 1-\sfrac{\ell}{2}+\sfrac{p}{2}+\sfrac{q}{2},  \sfrac{3}{2}-\sfrac{\ell}{2}+\sfrac{p}{2}+\sfrac{q}{2} \right\}, 1 \right) }{(-2)^\ell \Gamma(\ell+1)\Gamma(2-\ell+p+q)} \, .
\eea
}
\normalsize

\vskip -0.1cm
\noindent
One can now directly verify that the coefficients $c_{\ell}$ given by the formula above satisfy the difference equation (\ref{diff_eq}) and coincide with those found through the recursion formula.

\subsection{Simplifying the expression for $\Delta S_{p,q}(g)$}
We start with the expression (\ref{DS_complex}) for $\Delta S_{p,q}(g)$, isolate
the sum over $\ell$ and substitute there the coefficients $c_\ell$
in the form of the sum (\ref{sum_rep_cl}). We get the following double sum
\bea
{\mathcal W}(k,m)&=&\sum_{\ell=0}^{L-3}\sum_{j=0}^{\ell} \frac{\Gamma(2+\ell+p+q)}{\G(\frac{1}{2}+p)\G(\frac{3}{2}+q)}
\frac{(-1)^{j}\, 2^{2j-1}\, \G(\frac{1}{2}+j+p)\Gamma(\frac{3}{2}+j+q)}
{\Gamma(1+\ell-j)\Gamma(1+j)\G(2+2j-\ell+p+q)}\times \nonumber \\
&\times& \frac{1}{\Gamma(1+\ell+p+q)\Gamma(4+p+q-L+\ell)\Gamma(L-2-\ell+k-m-q)}\, .
\eea
\normalsize

\vskip -0.0 cm
\noindent
Now we isolate from $\Delta S_{p,q}(g)$  the sum over $\ell$ and substitute there $c_\ell$
in the form of the sum (\ref{sum_rep_cl}). We get the following double sum
\bea
{\mathcal W}(k,m)&=&\sum_{\ell=0}^{L-3}\sum_{j=0}^{\ell} \frac{\Gamma(2+\ell+p+q)}{\G(\frac{1}{2}+p)\G(\frac{3}{2}+q)}
\frac{(-1)^{j}\, 2^{2j-1}\, \G(\frac{1}{2}+j+p)\Gamma(\frac{3}{2}+j+q)}
{\Gamma(1+\ell-j)\Gamma(1+j)\G(2+2j-\ell+p+q)}\times \nonumber \\
&\times& \frac{1}{\Gamma(1+\ell+p+q)\Gamma(4+p+q-L+\ell)\Gamma(L-2-\ell+k-m-q)}\, .
\eea
As a next step we interchange the order of summation and get 
\bea
{\mathcal W}(k,m)&=&\sum_{j=0}^{L-3}\sum_{\ell=j}^{L-3} \frac{\Gamma(2+\ell+p+q)}{\G(\frac{1}{2}+p)\G(\frac{3}{2}+q)}
\frac{(-1)^{j}\, 2^{2j-1}\, \G(\frac{1}{2}+j+p)\Gamma(\frac{3}{2}+j+q)}
{\Gamma(1+\ell-j)\Gamma(1+j)\G(2+2j-\ell+p+q)}\times \nonumber \\
&\times& \frac{1}{\Gamma(1+\ell+p+q)\Gamma(4+p+q-L+\ell)\Gamma(L-2-\ell+k-m-q)}\, .
\eea
Taking the sum over $\ell$ we obtain
{\footnotesize
\bea
&&{\mathcal W}(k,m)=\sum_{j=0}^{L-3}\frac{(-1)^j2^{j-1}\G(\frac{1}{2}+j+p)\Gamma(\frac{3}{2}+j+q)}{\G(\frac{1}{2}+p)\G(\frac{3}{2}+q)\G(1+j)}\times \\
&&\Bigg[ 2\frac{ _3F_2\left( \left\{ -1-j-p-q, 3+j-k-L+m+q,2+j+p+q \right\}, \left\{ 1+j+p+q,  4+j+p+q-L \right\}, 1/2 \right)   }{\G(-2-j+k+L-m-q)\G(1+j+p+q)\G(4+j+p+q-L)} \nonumber\\
&&- 2^{j-L+3}\frac{\G(L+p+q) }{\G(2+p+q)\G(L+p+q-1)}\times \nonumber \\
&&\times \frac{_4F_3\left( \left\{ 1 , -2j+L-3-p-q, 1-k+m+q,  L+p+q \right\}, \left\{-1-j+L, 2+p+q,  L-1+p+q\right\}, 1/2 \right)}{\G(-1-j+L)\G(k-m-q)\G(4+2j-L+p+q)}
\Bigg]\, . \nonumber
\eea
}
\normalsize

\noindent
Note that for the allowed values of $k$ and $m$ the function $_4F_3$ is always finite, but the gamma function $\G(k-m-q)$ which divides it is always infinite because $k\leq q$ and $m\geq 0$.
Thus, the term of $\mathcal W$ containing $_4F_3$ does not contribute to the discontinuity $\Delta S_{p,q}$ and we can safely omit it. We therefore consider 
{\footnotesize
\bea
&&{\mathcal W}(s,m)=\sum_{k=0}^{L-3}\frac{(-1)^k 2^{k }\G(\frac{1}{2}+k+p)\Gamma(\frac{3}{2}+k+q)}{\G(\frac{1}{2}+p)\G(\frac{3}{2}+q)\G(1+k)}\times \\
&&\times \frac{ _3F_2\left( \left\{ -1-k-p-q, 3-L-s+k+m+q,2+k+p+q \right\}, \left\{ 1+k+p+q,  4+k+p+q-L \right\}, 1/2 \right)   }{\G(-2+L+s-k-m-q)\G(1+k+p+q)\G(4+k+p+q-L)} \, , \nonumber
\eea
}
\normalsize

\noindent
where for further clarity we replace the index $k$ by $s$ and $j$ by $k$. Further,
the hypergeometric function
{\footnotesize
$$
{\mathcal V}= _3F_2\left( \left\{ -1-k-p-q, 3-L-s+k+m+q,2+k+p+q \right\}, \left\{ 1+k+p+q,  4+k+p+q-L \right\}, 1/2 \right)
$$
}
\normalsize

\vspace{-0.3cm}
\noindent
featuring in the last formula can be reduced, namely,
{\small 
\bea
{\mathcal V}&=&2\, _2F_1\left( -1-k-p-q,3+k-L+m+q-s,4+k+p+q-L,1/2  \right)- \\
&&~~~-~_2F_1\left( -k-p-q,3+k-L+m+q-s,4+k+p+q-L,1/2  \right)\, . \nonumber
\eea
}
To each of these two $_2F_1$'s we apply an identity
$$
_2F_1(a,b,c,z)=(1-z)^{-a}_2F_1(a,c-b,c,z/(z-1))\, 
$$
and get
{\small 
\bea
{\mathcal V}&=&2^{-k-p-q}~\Big[~_2F_1\left( -1-k-p-q,1-m+p+s,4+k+p+q-L,-1  \right)- \nonumber \\
&&~~~~~~~~~~~~~~ -~_2F_1\left( -k-p-q,1-m+p+s,4+k+p+q-L,-1  \right)\Big]=  \\
&=&2^{-k-p-q}\frac{1-m+p+s}{4+k+p+q-L}~_2F_1\left(-k-p-q,2-m+p+s,5+k+p+q-L,-1  \right)\, . \nonumber
\eea
}
Therefore,
{\small
\bea
{\mathcal W}(s,m)&=&\sum_{k=0}^{L-3}\frac{(-1)^k 2^{-p-q}\G(\frac{1}{2}+k+p)\Gamma(\frac{3}{2}+k+q)}{\G(\frac{1}{2}+p)\G(\frac{3}{2}+q)\G(1+k)}(p-m+s+1)\times \\
&\times &\frac{ _2F_1\left(-k-p-q,2-m+p+s,5+k+p+q-L,-1  \right)   }{\G(-2+L+s-k-m-q)\G(1+k+p+q)\G(5+k+p+q-L)} \, . \nonumber
\eea
}
\normalsize

\noindent
For the discontinuity we therefore find
{\small
\bea
\Delta S_{p,q}(g)&=&i g\, (p-q-1)(p+q)\sum_{L=3}^{\infty} \frac{{\rm Li}_{L-1}(e^{-4\pi g})}{(4\pi g)^{L-1}} \sum_{s=0}^q\sum_{m=0}^{p-s-2}{\mathcal W}(s,m)\nonumber \\
&\times &\,
\sqrt{\pi}\,\frac{(-1)^{p+s}\, 2^{3-s-m+q}\, q!\,  \Gamma(p+m-s-1)\Gamma(p-m+s+1)\G(-2+s+p+L-m)}{s!\,  m!\, (q-s)! \, \Gamma(p-m-s-1)\Gamma(\frac{3}{2}+s-q)}
\nonumber\,  
\eea
}
\normalsize

\noindent
or explicitly
{\small
\bea
&&\label{eq:DSQ1}\Delta S_{p,q}(g)=i g\, (p-q-1)(p+q)\sum_{L=3}^{\infty} \frac{{\rm Li}_{L-1}(e^{-4\pi g})}{(4\pi g)^{L-1}} \sum_{k=0}^{L-3}\frac{\G(\frac{1}{2}+k+p)\Gamma(\frac{3}{2}+k+q)}{\G(\frac{1}{2}+p)\G(\frac{3}{2}+q)\G(1+k)}\times  \\
&&\times \sum_{s=0}^q\sum_{m=0}^{p-s-2}\sqrt{\pi}\, \frac{(-1)^{p+s+k}\, 2^{3-s-m-p}\, q!\,  \Gamma(p+m-s-1)\Gamma(p-m+s+2)\G(-2+s+p+L-m)}{s!\,  m!\, (q-s)! \, \Gamma(p-m-s-1)\Gamma(\frac{3}{2}+s-q)}\times\nonumber \\
&&~~~~~~~~~~~~~~\times \frac{ _2F_1\left(-k-p-q,p-m+s+2,5+k+p+q-L,-1  \right)   }{\G(-2+L+s-k-m-q)\G(1+k+p+q)\G(5+k+p+q-L)} \, . \nonumber
\eea
}
\normalsize

\noindent
We note as an interesting fact that the hypergeometric function entering in the last expression is expressible via the following Jacobi polynomial
\bea
&&~_2F_1\left(-k-p-q,p-m+s+2,5+k+p+q-L,-1  \right)\\
&&~~~~~~~~~~~~~~~~=\frac{(k+p+q)!(-2)^{k+p+q}}{(5+k-L+p+q,k+p+q)}J_{k+p+q}^{(-2-k+m-2p-q-s,-2+k-L+p+q)}(0)\, . \nonumber
\eea
Expanding $~_2F_1$ into the hypergeometric series, one comes to another representation
{\small
\begin{align}\label{eq:DSQ1}
\Delta S_{p,q}(g)=&4 i g\, (p-q-1)(p+q)\sum_{L=3}^{\infty} \frac{{\rm Li}_{L-1}(e^{-4\pi g})}{(4\pi g)^{L-1}} C_L(p,q)\, ,
\end{align}
}
\normalsize

\noindent
where the coefficients $C_L(p,q)$ are
{\small
\begin{align}\label{eq:DSQ12}C_L(p,q) &=\sum_{k=0}^{L-3}\frac{\G(\frac{1}{2}+k+p)\Gamma(\frac{3}{2}+k+q)}{\G(\frac{1}{2}+p)\G(\frac{3}{2}+q)\G(1+k)}\times\, \nonumber  \\
&\times \sum_{\tau=0}^{p+q-2} (-1)^{k+\tau} 2^{-1-\tau} \sqrt{\pi} \frac{\G(1+q)\G(4+\tau)\G(\tau+L)}{\G(p+q+k+1)\G(p+q-\tau-1)\G(L+\tau-k-p-q)} \\
&~~~~\times{}_2\tilde{F}_1(-k-p-q,4+\tau;5+k-L+p+q;-1)\nonumber\\ 
&~~~~\times{}_3\tilde{F}_2\Big(\big\{\tfrac{1}{2}(4-2p+\tau), \tfrac{1}{2}(5-2p+\tau), 2-p-q+\tau\big\}, \big\{ 3-p+\tau, \tfrac{7}{2}-p-q+\tau \big\}, 1\Big)  \, . \nonumber
\end{align}
}
where $``\sim"$ denotes that the hypergeometric function is regularised.

In Appendix~\ref{D3} we provide an alternative but simpler expression for $C_L(p,q)$.

\subsection{Alternative derivation of the asymptotic expansion for $ \Delta S_{p,q}(g)$}\label{D3}

An alternative method to compute the discontinuity in (\ref{eq:disc}), is by using the Gauss series expansion for the hypergeometric function
\bea
_2F_1(\sfrac{1}{2}+p,\sfrac{3}{2}+q,p+q+1,1-z) =\sum_{k=0}^\infty \frac{(\frac{1}{2}+p)_k (\frac{3}{2}+q)_k}{(p+q+1)_k\Gamma(k+1)} (1-z)^k\,.
\eea
We need just to compute the following integral
\bea
f= \int_1^\infty {\rm d}z\,(-h_n\sqrt{z}) e^{-h_n\sqrt{z}}  \frac{d^q}{dz^q} \Big(z^{q-1}\frac{d^{p-2}}{dz^{p-2}}\Big[z^{p-\sfrac{1}{2}}(1-z)^{p+q+k}\Big]\Big).
\eea
First we use the binomial expansion to expand $(1-z)^{p+q+k}$ and rewrite
\bea
f\nonumber=\sum_{s=0}^{p+q+k} \frac{(-1)^s\Gamma(p+q+1+k)}{\Gamma(s+1)\Gamma(p+q+1+k-s)} \int_1^\infty {\rm d}z\,(-h_n\sqrt{z}) e^{-h_n\sqrt{z}}  \frac{d^q}{dz^q} \Big(z^{q-1}\frac{d^{p-2}}{dz^{p-2}}\, z^{p+s-\sfrac{1}{2}}\Big).
\eea
Now we compute the derivatives by using $d^{m} z^n = \Gamma(n+1)/\Gamma(n+1-m) z^{n-m}$, so the integral takes the form
\bea
f&=&\nonumber\sum_{s=0}^{p+q+k}(-1)^s \frac{\Gamma(p+q+1+k)}{\Gamma(s+1)\Gamma(p+q+1+k-s)}\times  \\
&&~~~~~~~~\times\frac{\Gamma(p+s+\frac{1}{2})\Gamma(q+s+\frac{3}{2})}{\Gamma(s+\frac{3}{2})\Gamma(s+\frac{5}{2})} \int_1^\infty {\rm d}z\,(-h_n) e^{-h_n\sqrt{z}} z^{s+1}\,.
\eea
We pass back to the original variable $x^2=z$ and note that
\bea
&&\nonumber 2\int_1^\infty {\rm d}x\,(-h_n) e^{-h_n x} x^{2s+3} = 2h_n \frac{d^{2s+3}}{d h_n^{2s+3}} \int_1^\infty {\rm d}x\,
 e^{-h_n x}=\\
&&~~~~~~~~~~~~~~=\nonumber 2 h_n \frac{d^{2s+3}}{d h_n^{2s+3}} \Big( \frac{e^{-h_n} }{h_n}\Big) = (-2) \sum_{l=0}^{2s+3}\frac{\Gamma(2s+4)}{\Gamma(2s+4-l)} \frac{e^{-h_n}}{h_n^l}\,. 
\eea
Here the summation range of $l$ can be extended all the way to infinity thanks to $\Gamma(2s+4-l)$ in the denominator.
The sum over $n$ in $\Delta S_{p,q}$ is now trivial and gives
\bea
\sum_{n=1}^\infty \frac{e^{-h_n}}{h_n^l} = \frac{{\rm Li}_l\big(e^{-4\pi g}\big)}{(4\pi g)^l}\,.
\eea
By putting everything together we obtain
\bea
&&\Delta S_{p,q}(g) =\label{eq:disc2S} (4 i g)(p-q-1)(p+q) \sum_{l=0}^\infty \frac{{\rm Li}_l\big(e^{-4\pi g}\big)}{(4\pi g)^l}\sum_{k=0}^\infty \frac{\Gamma(p+\frac{1}{2}+k)\Gamma(q+\frac{3}{2}+k)}{\Gamma(p+\frac{1}{2})\Gamma(q+\frac{3}{2})\Gamma(k+1)}\times \nonumber \\
&&\times  \sum_{s=0}^{p+q+k}
\frac{(-1)^s}{\Gamma(s+1)\Gamma(p+q+1+k-s)}  \frac{\Gamma(p+s+\frac{1}{2})\Gamma(q+s+\frac{3}{2})}{\Gamma(s+\frac{3}{2})\Gamma(s+\frac{5}{2})}\frac{\Gamma(2s+4)}{\Gamma(2s+4-l)}\,.
\eea
The sum over $s$ can be extended all the way to infinity thanks to $\Gamma(p+q+1+k-s)$ in the denominator and this can be performed:
\bea
\sum_{s=0}^{\infty}
&&\frac{(-1)^s}{\Gamma(s+1)\Gamma(p+q+1+k-s)}  \frac{\Gamma(p+s+\frac{1}{2})\Gamma(q+s+\frac{3}{2})}{\Gamma(s+\frac{3}{2})\Gamma(s+\frac{5}{2})}\frac{\Gamma(2s+4)}{\Gamma(2s+4-l)} =\\
&&\nonumber =2^l\frac{\Gamma(p+\frac{1}{2})\Gamma(q+\frac{3}{2})}{\Gamma(p+q+k+1)}~_4\tilde{F}_3\Big(\{2,p+\frac{1}{2},q+\frac{3}{2},-p-q-k\},\{\frac{3}{2},2-\frac{l}{2},\frac{5}{2}-\frac{l}{2}\};1\Big)\,,
\eea
where $~_4\tilde{F}_3$ is the regularized generalized hypergeometric function.

It can be shown, {\it c.f.}  \cite{PBM3},  that $~_4\tilde{F}_3\Big(\{2,p+\frac{1}{2},q+\frac{3}{2},-p-q-k\},\{\frac{3}{2},2-\frac{l}{2},\frac{5}{2}-\frac{l}{2}\};1\Big)$ vanishes for $k+2>l$. This implies that in (\ref{eq:disc2S}) the sum over $l$ actually starts from $l=2$, while $k$ runs from $0$ to $l-2$.
We can therefore shift $l=L-1$ and finally arrive at
\bea
&&\Delta S_{p,q}(g) =\label{eq:disc1S} (4 i g)(p-q-1)(p+q) \sum_{L=3}^\infty \frac{{\rm Li}_{L-1}\big(e^{-4\pi g}\big)}{(4\pi g)^{L-1}} 2^{L-1} \times\\
&&\times \nonumber\sum_{k=0}^{L-3} \frac{\Gamma(p+\frac{1}{2}+k)\Gamma(q+\frac{3}{2}+k)}{\Gamma(k+1)\Gamma(p+q+k+1)}~_4\tilde{F}_3\Big(\{2,p+\frac{1}{2},q+\frac{3}{2},-p-q-k\},\{\frac{3}{2},\frac{5-L}{2},\frac{6-L}{2}\};1\Big)\,.
\eea
Thus, the discontinuity takes the form
\bea\label{eq:DSpert}
\Delta S_{p,q}(g) =(4 i g)(p-q-1)(p+q) \sum_{L=3}^\infty \frac{{\rm Li}_{L-1}\big(e^{-4\pi g}\big)}{(4\pi g)^{L-1}} c_L(p,q)\,,
\eea
where the coefficients $c_L(p,q)$ are given by
\bea
c_L(p,q)& = &2^{L-1}\sum_{k=0}^{L-3} \frac{\Gamma(p+\frac{1}{2}+k)\Gamma(q+\frac{3}{2}+k)}{\Gamma(k+1)\Gamma(p+q+k+1)}\times \\
&&~~~~~\times~_4\tilde{F}_3\Big(\{2,p+\frac{1}{2},q+\frac{3}{2},-p-q-k\},\{\frac{3}{2},\frac{5-L}{2},\frac{6-L}{2}\};1\Big)\, ,\nonumber
\eea
or equivalently by expanding the hypergeometric function
\bea
c_L(p,q) = &&\label{eq:CLpq}\sum_{k=0}^{L-3} \frac{\Gamma(p+\frac{1}{2}+k)\Gamma(q+\frac{3}{2}+k)}{\Gamma(p+\frac{1}{2})\Gamma(q+\frac{3}{2}) \Gamma(k+1)} \times\\
&&\times\notag\sum_{n=0}^{k+p+q}\frac{(-1)^n 2^{2n+3} (n+1)}{\sqrt{\pi}}\frac{\Gamma(p+\frac{1}{2}+n)\Gamma(q+\frac{3}{2}+n)}{\Gamma(p+q+1+k-n)\Gamma(n+\frac{3}{2})\Gamma(2n+5-L)}\,.
\eea 
Note that these coefficients $c_L(p,q) $ entering  the expansion (\ref{eq:DSpert}) seem very different from the previously computed $C_L(p,q)$ given by (\ref{eq:DSQ12}), nonetheless we have numerically checked that the two expressions coincide $k$ by $k$ once we fix values for $p,q$ and $L$.

\end{document}